\documentclass[11pt]{article}

\usepackage{graphicx}
\usepackage{float}

\begin{document}

\title{Testing Relativistic Gravity with Radio Pulsars}

\author{Norbert Wex\\
\small Max-Planck-Institut f\"ur Radioastronomie\\[-0.7ex]
\small Auf dem H\"ugel 69, D-53121, Bonn, Germany}

\date{August 28, 2013}

\maketitle

\begin{abstract}
Before the 1970s, precision tests for gravity theories were constrained to the 
weak gravitational fields of the Solar system. Hence, only the weak-field 
slow-motion aspects of relativistic celestial mechanics could be investigated. 
Testing gravity beyond the first post-Newtonian contributions was for a long 
time out of reach.

The discovery of the first binary pulsar by Russell Hulse and Joseph Taylor in 
the summer of 1974 initiated a completely new field for testing the relativistic 
dynamics of gravitationally interacting bodies. For the first time the back 
reaction of gravitational wave emission on the binary motion could be studied. 
Furthermore, the Hulse-Taylor pulsar provided the first test bed for the 
orbital dynamics of strongly self-gravitating bodies.

To date there are a number of pulsars known, which can be utilized for precision 
test of gravity.  Depending on their orbital properties and their companion, 
these pulsars provide tests for various different aspects of relativistic 
dynamics. Besides tests of specific gravity theories, like general relativity or 
scalar-tensor gravity, there are pulsars that allow for generic constraints on 
potential deviations of gravity from general relativity in the quasi-stationary 
strong-field and the radiative regime.

This article presents a brief overview of this modern field of relativistic 
celestial mechanics, reviews some of the highlights of gravity tests with radio 
pulsars, and discusses their implications for gravitational physics and 
astronomy, including the upcoming gravitational wave astronomy.

\end{abstract}


\newpage

\tableofcontents

\newpage


\section{Introduction}

In about two years from now we will be celebrating the centenary of Einstein's 
general theory of relativity. On November 25th 1915 Einstein presented
his field equations of gravitation (without cosmological term) to the Prussian 
Academy of Science \cite{ein15b}. With this publication, general 
relativity (GR) was finally completed as a logically consistent physical
theory ({\it ``Damit ist endlich die allgemeine Relativit{\"a}tstheorie als 
logisches Geb{\"a}ude abgeschlossen.''}). Already one week before, based on the 
vacuum form of his field equations, Einstein was able to show that his theory of 
gravitation naturally explains the anomalous perihelion advance of the planet 
Mercury \cite{ein15a}. While in hindsight this can be seen as the first 
experimental test for GR, back in 1915 astronomers were still searching 
for a Newtonian explanation \cite{see15}. In his 1916 comprehensive summary 
of GR \cite{ein16a}, Einstein proposed three experimental tests:
\begin{itemize}
\item Gravitational redshift \index{Gravitational redshift}
(Einstein suggested to look for red-shift in the 
spectral lines of stars).
\item Light deflection \index{Light deflection}
(Einstein explicitly calculated the values for the Sun and Jupiter).
\item Perihelion precession \index{Perihelion precession}
of planetary orbits (Einstein emphasized the 
agreement of GR, with the observed perihelion precession of Mercury
with a reference to his calculations in \cite{ein15a}).
\end{itemize}
Gravitational redshift, a consequence of the equivalence principle, is 
common to all metric theories of gravity, and therefore in some respect its 
measurement has less discriminating power than the other two tests 
\cite{will93}. The first verification of gravitational light bending during the 
total eclipse on May 29th 1919 was far from being a high precision test, but 
clearly decided in favor of GR, against the Newtonian prediction, which is only
half the GR value \cite{ded20}. In the meantime this test has been greatly
improved, in the optical with the astrometric satellite HIPPARCOS \cite{fma97}, 
and in the radio with very long baseline interferometry 
\cite{lcs+95,sdlg04,fklb09}. The deflection predicted by GR has been verified 
with a precision of $1.5 \times 10^{-4}$. An even better test for the curvature 
of spacetime in the vicinity of the Sun is based on the 
Shapiro delay,\index{Shapiro delay} the 
so-called ``fourth test of GR'' \cite{sha64}. A measurement of the frequency 
shift of radio signals exchanged with the Cassini spacecraft lead to a $10^{-5}$ 
confirmation of GR \cite{bit03}. 
Apart from the four ``classical'' tests, GR has passed many other tests in the 
Solar system with flying colors: Lunar Laser Ranging tests for the
strong equivalence principle and the de-Sitter precession of the Moon's orbit
\cite{nor99}, the Gravity Probe B experiment for the relativistic spin 
precession of a gyroscope (geodetic and frame dragging) \cite{edp+11}, and the 
Lense-Thirring effect in satellite orbits \cite{cp04}, just to name a few.

GR, being a theory where fields travel with finite
speed, predicts the existence of gravitational waves that propagate
with the speed of light \cite{ein16b} and extract energy from 
(non-axisymmetric) material systems with accelerated masses \cite{ein18}. This 
is also true for a self-gravitating system, where the acceleration of the masses 
is driven by gravity itself, a question which was settled in a fully 
satisfactory manner only several decades after Einstein's pioneering papers (see 
\cite{ken07} for an excellent review). This fundamental property of GR could not 
be tested in the slow-motion environment of the Solar system, and the 
verification of the existence of gravitational waves had to wait until the 
discovery of the first binary pulsar in 1974 \cite{ht75}. Also, all the 
experiments in the Solar system can only test the weak-field aspects of gravity.
The spacetime of the Solar system is close to Minkowski space everywhere: 
To first order (in standard coordinates) the spatial components 
of the spacetime metric can be written as $g_{ij} = (1 - 2\Phi/c^2)\delta_{ij}$, 
where $\Phi$ denotes the Newtonian gravitational potential. At the surface of 
the Sun one finds $\Phi/c^2 \sim -2 \times 10^{-6}$, while at the surface of a neutron star $\Phi/c^2 \sim -0.2$. Consequently, gravity experiments 
with binary pulsars, not only yielded the first tests of the radiative 
properties of gravity, they also  took our gravity tests into a new regime of 
gravity.

\begin{figure}[H]
\centering
\includegraphics[width=140mm]{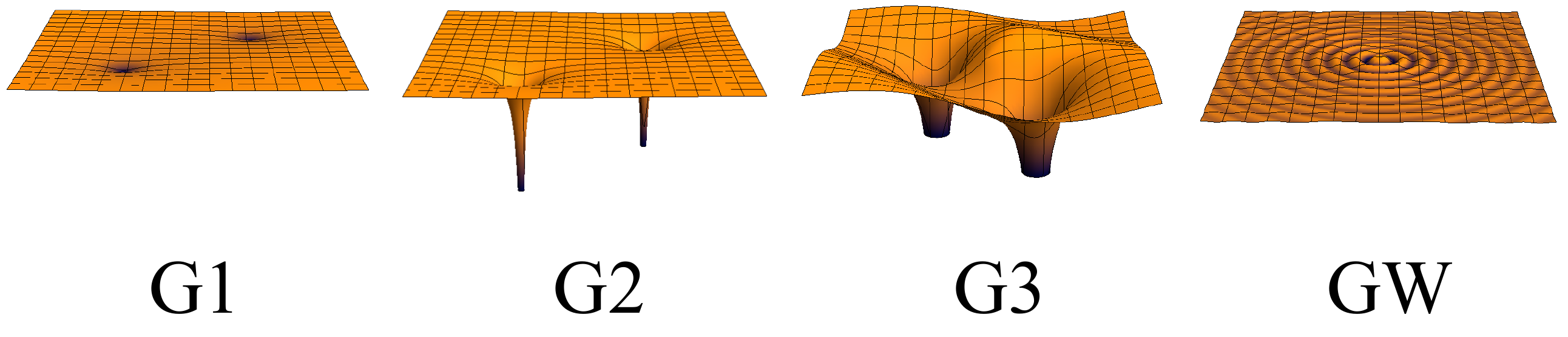}
\caption{Illustration of the different gravity regimes used in this 
review. 
\label{fig:gravregimes}}
\end{figure}

To categorize gravity tests with pulsars and to put them into context with other 
gravity tests it is useful to introduce the following four 
gravity regimes\index{gravity regimes}:
\begin{description}
\item[G1] {\it Quasi-stationary weak-field regime}: 
  \index{quasi-stationary weak-field regime}
  The motion of the masses
  is slow compared to the speed of light ($v \ll c$) and spacetime is only
  very weakly curved, i.e.~close to Minkowski spacetime 
  everywhere. This is, for instance, the case in the Solar system.
\item[G2] {\it Quasi-stationary strong-field regime}: 
  \index{quasi-stationary strong-field regime}
  The motion of the masses
  is slow compared to the speed of light ($v \ll c$), but one or more bodies
  of the system are strongly self-gravitating, i.e.~spacetime in their vicinity
  deviates significantly from Minkowski space. Prime examples here are binary
  pulsars, consisting of two well-separated neutron stars.
\item[G3] {\it Highly-dynamical strong-field regime}: 
  \index{highly-dynamical strong-field regime}
  Masses move at a significant fraction of the speed of light ($v \sim c$) 
  and spacetime is
  strongly curved and highly dynamical in the vicinity of the masses.
  This is the regime of merging neutron stars and black holes. 
\item[GW] {\it Radiation regime}: 
  \index{radiation regime}
  Synonym for the collection of the radiative 
  properties of gravity, most notably the generation of gravitational waves by 
  material sources, the propagation speed of gravitational waves, and their
  polarization properties.
\end{description}
Figure~\ref{fig:gravregimes} illustrates the different regimes. Gravity regime 
G1 is well tested in the Solar system. Binary pulsar experiments are presently 
our only precision experiments for gravity regime G2,
and the best tests for the radiative properties of gravity (regime GW)\footnote{
Gravitational wave damping has also been observed in a double white-dwarf system, which has an orbital period of just 13 minutes \cite{hkb12}.
This experiment combines gravity regimes G1 (note, $v/c \sim 3 \times 10^{-3}$)
and GW of figure~\ref{fig:gravregimes}.}. 
In the near future, gravitational wave detectors will allow a direct 
detection of gravitational waves (regime GW) and probe the strong and highly
dynamical spacetime of merging compact objects (regime G3). 
As we will discuss at the end of this review, pulsar timing arrays soon
should give us direct access to the nano-Hz gravitational wave band and 
probe the properties of these ultra-low-frequency gravitational waves 
(regime GW).


\subsection{Radio pulsars and pulsar timing}

\begin{figure}[h!]
\centering
\includegraphics[height=120mm]{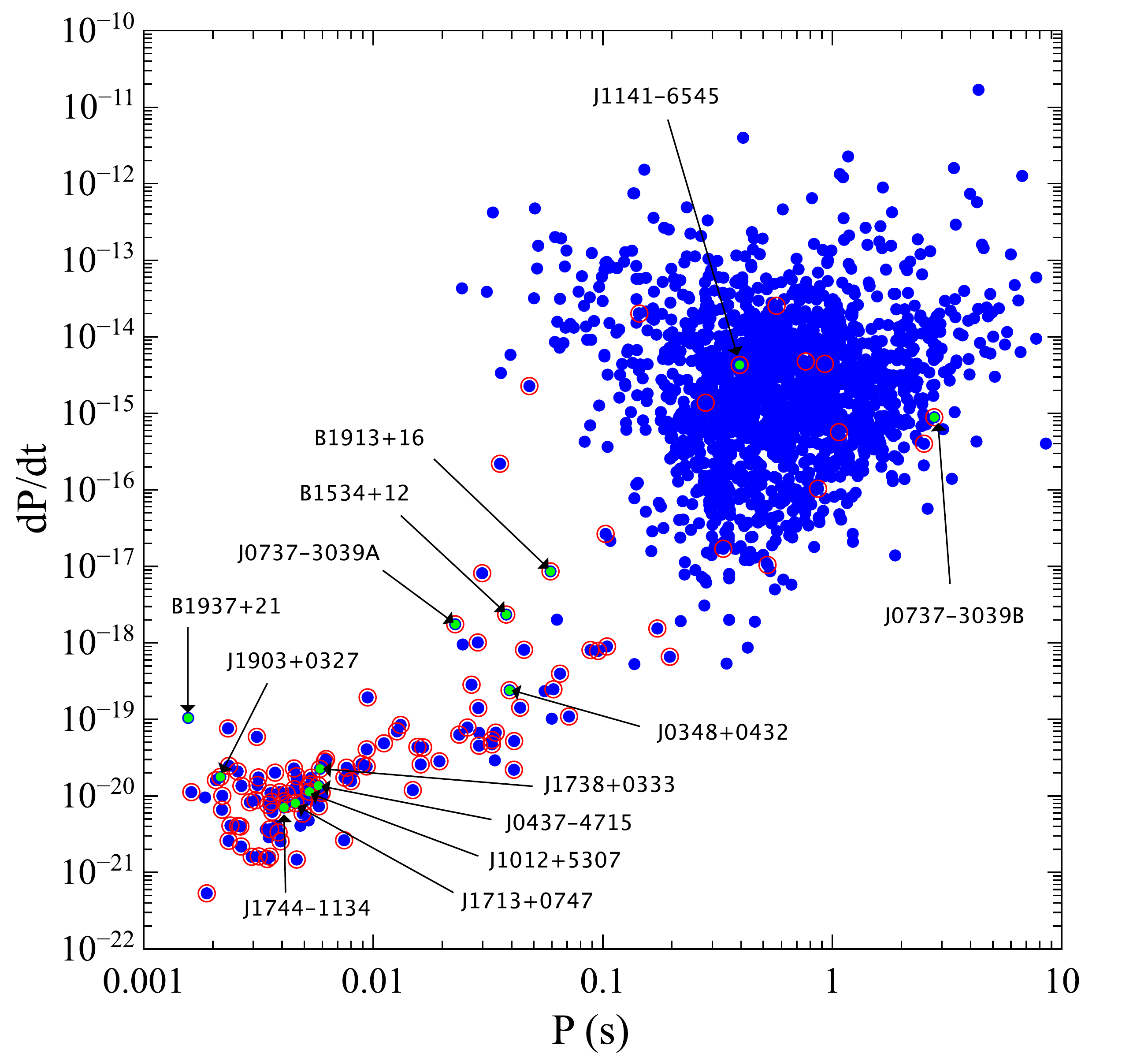}
\caption{The $P$-$\dot{P}$ diagram for radio pulsars. Binary pulsars are 
indicated by a red circle. Pulsars that play a particular role in this review 
are marked with a green dot and have their name as a label. The 
data are taken from the {\it ATNF Pulsar Catalogue} \cite{psrcat}. 
\label{fig:ppdot}}
\end{figure}

\begin{figure}[h]
\centering
\includegraphics[height=80mm]{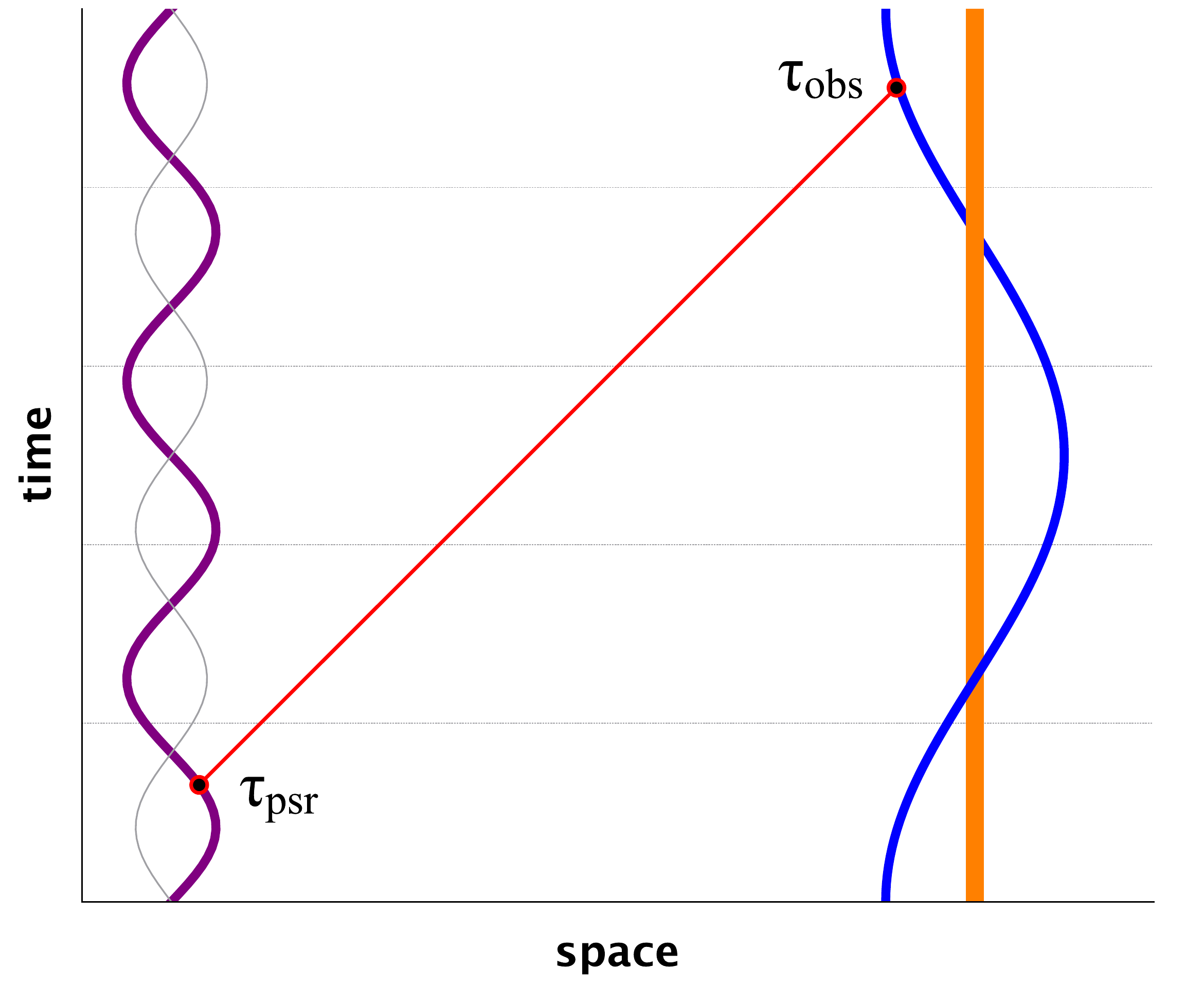}
\caption{Spacetime diagram illustration of pulsar timing. Pulsar timing 
connects the proper time of emission 
$\tau_{\rm psr}$, defined by the pulsar's intrinsic rotation, and the proper 
time of the observer on Earth $\tau_{\rm obs}$, measured by the atomic clock at 
the location of the radio telescope. The timing model, which expresses 
$\tau_{\rm obs}$ as a function of $\tau_{\rm psr}$, accounts for various 
``relativistic effects''
associated with the metric properties of the spacetime, i.e.~the world line
of the pulsar and the null-geodesic of the radio signal. In addition, it 
contains a number of terms related to the Earth motion and
relativistic corrections in the Solar system, like time dilation and 
signal propagation delays (see \cite{ehm06} for details).
\label{fig:spacetime}}
\end{figure}

Radio pulsars,\index{pulsars} 
i.e.~rotating neutron stars with coherent radio emission 
along their magnetic poles, were discovered in 1967 by Jocelyn Bell and Antony 
Hewish \cite{hbp+68}. Seven 
years later, Russell Hulse and Joseph Taylor discovered the first binary 
pulsar, a pulsar in orbit with a companion star \cite{ht75}. This discovery
marked the beginning of gravity tests with radio pulsars. Presently, more than
2000 radio pulsars are known, out of which about 10\% reside in binary systems 
\cite{psrcat}. The population of radio pulsars can be nicely presented in
a diagram that gives the two main characteristics of a pulsar: the rotational 
period $P$ and its temporal change $\dot{P}$ due to the loss of rotational 
energy (see figure~\ref{fig:ppdot}). Fast rotating pulsars with small $\dot{P}$ 
(millisecond pulsars) appear to be particularly stable in their rotation. On 
long time-scales, some of them rival the best atomic clocks in terms of 
stability \cite{lor05,hcm+12}. This property makes them ideal tools for 
precision astrometry, and hence (most) gravity tests with pulsars are simply 
clock comparison experiments to probe the spacetime of the binary pulsar, where 
the ``pulsar clock'' is read off by counting the pulses in the pulsar signal  
(see figure~\ref{fig:spacetime}). As a result, a wide range of relativistic 
effects related to orbital binary dynamics, time dilation and delays in the 
signal propagation can be tested. The technique used is the so-called 
{\it pulsar timing},\index{pulsar timing} 
which basically consists of measuring the exact arrival 
time of pulses at the radio telescope on Earth, and fitting an appropriate {\it 
timing model}\index{timing model} 
to these arrival times, to obtain a phase-connected solution. In 
the phase-connected approach lies the true strength of pulsar timing: the timing 
model has to account for every (observed) pulse over a time scale of several 
years, in some 
cases even several decades. This makes pulsar timing extremely sensitive to even 
tiny deviations in the model parameters, and therefore vastly superior to a simple measurement of Doppler-shifts in the pulse period. 
Table \ref{tab:bestof} illustrates the current precision capabilities of pulsar 
timing for various experiments, like mass determination, astrometry and gravity 
tests. We will not go into the details of pulsar observations and pulsar timing 
here, since there are numerous excellent reviews on these topics, for instance
\cite{sta03,lk04}, just to mention two. In this review we focus on the 
relativistic effects that play a role in pulsar timing observations, and how
pulsar timing can be used to test gravitational phenomena in generic as well
as theory-based frameworks.

\begin{table}[ht]
\caption{Examples of precision measurements using pulsar timing. 
  A number in bracket indicates
  the (one-sigma) uncertainty in the last digit of each value. The symbol
  $M_\odot$ stands for the Solar mass. (cf.~table~1 in \cite{kra12}).
  \label{tab:bestof} }
\vspace{1mm}
\centerline{
{\small
\begin{tabular}{lll}
\hline
Rotational period:	& 5.757451924362137(2)\,ms & \cite{vbv+08} \\
Orbital period:     & $0.102251562479(8)$\,d & (Kramer et al., in prep.) \\
Small eccentricity: & $(3.5 \pm 1.1) \times 10^{-7}$ & \cite{fwe+12} \\
Distance:           & $157(1)$\,pc	& \cite{vbv+08} \\
Proper motion:      & $140.915(1)$\,mas\,yr$^{-1}$ & \cite{vbv+08} \\
Masses of neutron stars: & $m_p = 1.4398(2) \,M_\odot$ & \cite{wnt10} \\ 
                         & $m_c = 1.3886(2) \,M_\odot$ & \cite{wnt10} \\ 
Mass of millisecond pulsar:	  & $1.667(7) \,M_\odot$ & \cite{fbw+11} \\
Mass of white-dwarf companion:    & $0.207(2) \,M_\odot$ & \cite{hbo06} \\
Mass of Jupiter and moons: & $9.547921(2) \times 10^{-4} \,M_\odot$ &      	\cite{chm+10} \\
Relativistic periastron advance:	 & $4.226598(5)$ deg yr$^{-1}$ & \cite{wnt10} \\
Gravitational wave damping: 	& 0.504(3)\,pico-Hz\,yr$^{-1}$ & (Kramer et al., in prep.) \\
GR validity (observed/GR):	   & $1.0000(5)$ & (Kramer et al., in prep.) \\
\hline
\end{tabular}
}}
\end{table}


\subsection{Binary pulsar motion in gravity theories}
\label{sec:EOM}

While in Newtonian gravity there is an exact solution to the 
equations of motion\index{equations of motion}
of two point masses that interact gravitationally, no such exact analytic 
solution is known in GR. In GR, the two-body problem has to be solved 
numerically or on the basis of approximation methods. A particularly well 
established and 
successful approximation scheme, to tackle the problem of motion of a system of 
well-separated bodies, is the 
{\em post-Newtonian approximation},\index{post-Newtonian approximation} 
which is 
based on the weak-field slow-motion assumption. However, to describe the 
motion and gravitational wave emission of binary pulsars, there are two 
main limitations of the post-Newtonian approximation that have to be overcome
(cf.~\cite{dam87}):
\begin{description} 
\item[A)] Near and inside the pulsar (and its companion, if it is also a
  neutron star) the gravitational field is strong and the weak-field assumption 
  no longer holds.
\item[B)] When it comes to generation of gravitational waves (of wavelength 
  $\lambda_{\rm GW}$) and their back-reaction 
  on the orbit (of size $r$ and period $P_b$), the post-Newtonian 
  approximation is only valid in the near zone ($r \ll \lambda_{\rm GW} = 
  c P_b/2$), and breaks down in the radiation zone ($r > \lambda_{\rm GW}$) 
  where gravitational waves propagate and boundary conditions are defined, 
  like the `no incoming radiation' condition. 
\end{description}
The discovery of the Hulse-Taylor pulsar was a particularly strong stimulus for 
the development of consistent approaches to compute 
the equations of motion for a binary system with strongly self-gravitating
bodies (gravity regime G2). As a result, by now there are fully self-consistent 
derivations for the gravitational wave emission and the damping of the orbit due to gravitational wave 
back-reaction for such systems. In fact, in GR, there are several independent 
approaches that lead to the same result, giving equations of motion for a binary 
system with non-rotating components that include terms up to 3.5 post-Newtonian 
order ($v^7/c^7$) \cite{bla06,fi07}. For the relative acceleration in the center-of-mass frame one finds the general form
\begin{eqnarray}
  \ddot{\bf r} &=& 
    -\frac{GM}{r^2}\left[
      (1 + A_2 + A_4 + A_5 + A_6 + A_7) \, \frac{\bf r}{r} \right.
      \nonumber\\
      && \left.\frac{}{} \qquad\qquad
      + (B_2 + B_4 + B_5 + B_6 + B_7) \, \dot{\bf r} \right] \;,
  \label{eq:eomstruc}
\end{eqnarray}
where the coefficients $A_k$ and $B_k$ are of order $c^{-k}$, and are functions 
of $r \equiv |{\bf r}|$, $\dot{r}$, $v \equiv |\dot{\bf r}|$, and the masses 
(see \cite{bla06} for explicit expressions). The quantity $M$ denotes the total
mass of the system. At this level of approximation, these 
equations of motion are also applicable to binaries containing strongly 
self-gravitating bodies, like neutron stars and black holes. This is a 
consequence of a remarkable property of Einstein's theory of gravity, the {\it 
effacement of the internal structure} 
\index{effacement of the internal structure}
\cite{dam83,dam87}: In GR, strong-field 
contributions are absorbed into the definition of the body's mass. 

In GR's post-Newtonian approximation scheme, gravitational wave damping enters 
for the first time at the 2.5 post-Newtonian level (order $v^5/c^5$), as a term 
in the equations of motion that is not invariant against time-reversal. The 
corresponding loss of orbital energy is given by the {\it quadrupole formula},
\index{quadrupole formula}
derived for the first time by Einstein within the linear approximation, 
for a material system where the gravitational interaction between the masses can 
be neglected \cite{ein18}. As it turns out, the quadrupole formula is also 
applicable for gravity regime G2 of figure~\ref{fig:gravregimes}, 
and therefore valid for binary pulsars as well (cf.~\cite{dam87}).

In alternative gravity theories, the gravitational wave back-reaction, 
generally, already enters
at the 1.5 post-Newtonian level (order $v^3/c^3$). This is the result of the 
emission of dipolar gravitational waves, and adds terms $A_3$ and $B_3$ to equation~(\ref{eq:eomstruc}) 
\cite{will93,gw02}. Furthermore, one does no longer have an effacement of the 
internal structure of a compact body, meaning that the orbital dynamics, in 
addition to the mass, depends on the ``sensitivity'' of the body, a quantity
that depends on its structure/compactness. Such 
modifications already enter at the ``Newtonian'' level, where the usual 
Newtonian gravitational constant $G$ is replaced by a (body-dependent) effective 
gravitational constant ${\cal G}$. For alternative gravity theories, it 
therefore generally
makes an important difference whether the pulsar companion is a compact neutron 
star or a much less compact white dwarf. In sum, alternative theories
of gravity generally predict deviations from GR in both the quasi-stationary and
the radiative properties of binary pulsars \cite{dt92,dam09}.

At the first post-Newtonian level, for fully conservative gravity theories 
without preferred location effects, one can construct a generic {\it modified 
Einstein-Infeld-Hoffmann Lagrangian}
\index{modified Einstein-Infeld-Hoffmann Lagrangian}
for a system of two gravitationally
interacting masses $m_p$ (pulsar) and $m_c$ (companion) at relative (coordinate) 
separation $r \equiv |{\bf x}_p - {\bf x}_c|$ and velocities ${\bf v}_p = 
\dot{\bf x}_p$ and 
${\bf v}_c = \dot{\bf x}_c$:
\begin{eqnarray}\label{eq:LO}
   L_{\rm O} &=& 
   -m_p c^2\left(1 - \frac{{\bf v}_p^2}{2c^2} - \frac{{\bf v}_p^4}{8c^4}\right) 
   -m_c c^2\left(1 - \frac{{\bf v}_c^2}{2c^2} - \frac{{\bf v}_c^4}{8c^4}\right)  
   \nonumber\\
   && + \frac{{\cal G}m_pm_c}{r} 
      \left[1  
      - \frac{{\bf v}_p \cdot {\bf v}_c}{2c^2} 
      - \frac{({\bf r}\cdot{\bf v}_p)({\bf r}\cdot{\bf v}_c)}{2c^2r^2} 
      + \varepsilon\,\frac{({\bf v}_p - {\bf v}_c)^2}{2c^2} \right]
  \nonumber\\
   && - \xi \, \frac{{\cal G}^2 M m_p m_c}{2c^2r^2} \;,
\end{eqnarray}
where $M \equiv m_p + m_c$. The body-dependent quantities ${\cal G}$, 
$\varepsilon$ and $\xi$ account for deviations from GR associated with the 
self-energy of the
individual masses \cite{will93,dt92}. In GR one simply finds ${\cal G} = G$, 
$\varepsilon = 3$, and $\xi = 1$. There are various analytical solutions to
the dynamics of (\ref{eq:LO}). The most widely used in pulsar astronomy is
the quasi-Keplerian parametrization by Damour and Deruelle \cite{dd85}. It forms
the basis of pulsar-timing models for relativistic binary pulsars, as we will 
discuss in more details in Section~\ref{sec:PPK}.

Beyond the first post-Newtonian level there is no fully generic framework for 
the gravitational dynamics of a binary system. However, one can find 
equations of motion valid for a general class of gravity theories, like in 
\cite{de96b} where a framework based on multi-scalar-tensor theories is 
introduced to discuss tests of relativistic gravity to the second 
post-Newtonian level, or in \cite{mw13} where the explicit equations of motion 
for non-spinning compact objects to 2.5 post-Newtonian order for a general class 
of scalar-tensor theories of gravity are given.


\subsection{Gravitational spin effects in binary pulsars}
\label{sec:geodprec}

In relativistic gravity theories, in general, the proper rotation of the 
bodies of a binary system directly affects their orbital and spin dynamics.
Equations of motion for spinning bodies in GR have been developed by numerous
authors, and in the meantime go way beyond the leading order contributions
(for reviews and references see, e.g., \cite{bo75,dam87,bru91,hss13}).
For present day pulsar-timing experiments it is sufficient to have a look at the
post-Newtonian leading order contributions. There one finds three 
contributions:\index{spin effects}
the spin-orbit (SO) interaction between the pulsar's spin ${\bf S}_p$ and the 
orbital angular momentum ${\bf L}$, the SO interaction between the companion's 
spin ${\bf S}_c$ and the orbital angular momentum, and finally the spin-spin 
interaction between the spin of the pulsar and the spin of the companion
\cite{bo75}.

Spin-spin interaction will remain negligible in binary pulsar experiments for 
the foreseeable future. They are many orders of magnitude below the second
post-Newtonian and spin-orbit effects \cite{wex95}, and 
many orders of magnitude below the measurement precision of present 
timing experiments. For this reason, we will not further discuss spin-spin 
effects here.

For a boost-invariant gravity theory, the (acceleration-dependent) Lagrangian 
for the spin-orbit interaction\index{spin-orbit interaction}
has the following general form (summation over
spatial indices $i,j$)
\begin{equation}\label{eq:LSO}
  L_{\rm SO}({\bf x}_A, {\bf v}_A, {\bf a}_A) =
    \frac{1}{c^2}\sum_A S_A^{ij} \left[
    \frac{1}{2}v_A^i a_A^j + \sum_{B \ne A} \frac{\Gamma_A^B m_B}{r_{AB}^3}
    (v_A^i - v_B^i)(x_A^j - x_B^j) \right] \;,
\end{equation}
where $S_A^{ij} \equiv \varepsilon^{ijk} S_A^k$ is the antisymmetric spin tensor
of body $A$ \cite{dam82,dam87,dt92}. The coupling function $\Gamma_A^B$ can 
also account 
for strong-field effects in the spin-orbit coupling. In GR $\Gamma_A^B = 2G$. 
For bodies with negligible gravitational self-energy, one finds in the framework 
of the {\it parametrized post-Newtonian (PPN) formalism}\footnote{The PPN 
formalism uses 10 parameters to parametrize in a generic way deviations from GR 
at the post-Newtonian level, within the class of metric gravity theories (see 
\cite{will93} for details).}
$\Gamma_A^B = (\gamma_{\rm PPN} + 1)G$, a quantity that is actually most 
tightly constrained by the light-bending and Shapiro-delay experiments in the 
Solar system, which test $\gamma_{\rm PPN}$ \cite{lcs+95,sdlg04,fklb09,bit03}.

In binary pulsars, spin-orbit coupling has two effects. On the one hand,
it adds spin-dependent terms to the equations of motion (\ref{eq:eomstruc}), 
which cause a Lense-Thirring precession of the orbit (for GR see 
\cite{bo75,ds88}). So far this contribution could not be tested in binary pulsar 
experiments. Prospects of its measurement will be discussed in the future outlook in Section~\ref{sec:summary}. On the other 
hand it leads to secular changes in the orientation of the spins of the two 
bodies (geodetic precession), most importantly the observed pulsar in a
pulsar binary \cite{dr74,bo75,ber75}. As we discuss in more 
details in Section~\ref{sec:gp}, a change in the rotational axis of the pulsar 
causes changes in 
the observed emission properties of the pulsar, as the line-of-sight gradually 
cuts through different regions of the magnetosphere.

As can be derived from (\ref{eq:LSO}), to first order in GR the geodetic 
precession \index{geodetic precession}
of the pulsar, averaged over one orbit, is given by 
($\hat{\bf L} \equiv {\bf L}/|{\bf L}|$)
\begin{equation}\label{eq:OmGP}
  \mathbf{\Omega}_p^{\rm SO} = \frac{n_b}{1 - e^2}
    \left(2 + \frac{3m_c}{2m_p}\right) \frac{m_p m_c}{M^2}
    \frac{V_b^2}{c^2} \, \hat{\bf L} \;,
\end{equation}
where $n_b \equiv 2\pi/P_b$ and $V_b \equiv (GM n_b)^{1/3}$.

It is expected that in alternative theories relativistic spin precession  
generally depends on self-gravitational effects, meaning, the actual precession 
may depend on the compactness of a self-gravitating body. For the class of 
theories that lead to the Lagrangian~(\ref{eq:LSO}), equation~(\ref{eq:OmGP}) 
modifies to
\begin{equation}\label{eq:OmGPag}
  \mathbf{\Omega}_p^{\rm SO} = \frac{n_b}{1 - e^2}
    \left[\frac{\Gamma_p^c}{\cal G} +
    \left(\frac{\Gamma_p^c}{\cal G} - \frac{1}{2}\right) \frac{m_c}{m_p}
    \right] \frac{m_p m_c}{M^2}
    \frac{{\cal V}_b^2}{c^2} \, \hat{\bf L} \;,
\end{equation}
where ${\cal V}_b \equiv ({\cal G}M n_b)^{1/3}$ is the strong-field
generalization of $V_b$. 

Effects from spin-induced quadrupole moments are negligible as well.
For double neutron-star systems they are many orders of magnitude below 
the second post-Newtonian and spin-orbit effects, due to the small extension of 
the bodies \cite{wex95}. If the companion is a more extended star, like a
white dwarf or a main-sequence star, the rotationally-induced quadrupole moment
might become important. A prime example is PSR~J0045$-$7319, where the 
quadrupole moment of the fast rotating companion causes a significant precession 
of the pulsar orbit \cite{kbm+96}. For all the binary pulsars discussed here, 
the quadrupole moments of pulsar and companion are (currently) negligible.

Finally, certain gravitational phenomena, not present in GR, can even lead to a 
spin precession of isolated pulsars, for instance, a violation of the local 
Lorentz invariance and a violation of the local position invariance in the 
gravitational sector, as we will discuss in more details in 
Sections~\ref{sec:LLI} and \ref{sec:LPI}.

 
\subsection{Phenomenological approach to relativistic effects in binary
pulsar observations}
\label{sec:PPK}

For binary pulsar experiments that test the quasi-stationary strong-field 
regime (G2) and the gravitational wave damping (GW), a phenomenological 
parametrization, the so-called `parametrized post-Keplerian' (PPK) formalism, 
\index{parametrized post-Keplerian formalism}
has been introduced by Damour \cite{dam88} and extended by Damour and Taylor 
\cite{dt92}. The PPK formalism parametrizes all the observable effects that 
can be extracted independently from binary pulsar timing and pulse-structure 
data. Consequently, the PPK formalism allows to obtain theory-independent 
information from binary pulsar observations by fitting for a set of 
{\it Keplerian} and {\it post-Keplerian parameters}. 
\index{Keplerian parameters}
\index{post-Keplerian parameters}

The description of the orbital motion is based on the quasi-Keplerian 
parametrization of Damour \& Deruelle, which is a solution to the first 
post-Newtonian equations of motion \cite{dd85,dd86}. The corresponding 
{\it Roemer delay} in the arrival time of the pulsar signals is
\index{Roemer delay}
\begin{equation}\label{eq:D_Roemer}
  \Delta_{\rm R} = x \sin\omega \left[\cos U - e(1 + \delta_r)\right] +
     x \cos\omega \left[1 - e^2(1 + \delta_\theta)^2\right]^{1/2} \sin U \;,
\end{equation}
where the eccentric anomaly $U$ is linked to the proper time of the pulsar $T$
via the Kepler equation
\index{Kepler equation}
\begin{equation}\label{eq:Kepler}
  U - e \sin U = 2\pi \left[\left(\frac{T - T_0}{P_b}\right)
       - \frac{\dot{P}_b}{2}\left(\frac{T - T_0}{P_b}\right)^2\right] \;.
\end{equation}
The five Keplerian parameters $P_b$, $e$, $x$, $\omega$, and $T_0$ denote the
orbital period, the orbital eccentricity, the projected semi-major axis of the 
pulsar orbit, the longitude of periastron, and the time of periastron passage, 
respectively.
The post-Keplerian parameter $\delta_r$ is not separately measurable, i.e.\ it 
can be absorbed into other timing parameters, and the post-Keplerian parameter 
$\delta_\theta$ has not been measured up to now in any of the binary pulsar 
systems. The relativistic precession of periastron 
\index{precession of periastron}
changes the the longitude of
periastron $\omega$ according to
\begin{equation}
  \omega = \omega_0 + \dot\omega \,\frac{P_b}{\pi}
    \arctan\left[\left(\frac{1+e}{1-e}\right)^{1/2} \tan\frac{U}{2}\right] \;,
\end{equation}
meaning, that averaged over a full orbit, the location of periastron shifts
by an angle $\dot\omega P_b$. The parameter $\dot\omega$ is the corresponding 
post-Keplerian parameter. A change in the orbital period, due to the 
emission of gravitational waves, is parametrized by the post-Keplerian parameter 
$\dot{P}_b$. Correspondingly, one has post-Keplerian parameters for the change
in the orbital eccentricity and the projected semi-major axis:
\begin{eqnarray}
  e &=& e_0 + \dot{e}(T - T_0) \;,\\
  x &=& x_0 + \dot{x}(T - T_0) \;.
\end{eqnarray}
Besides the Roemer delay $\Delta_{\rm R}$, there are two purely relativistic 
effects that play an important role in pulsar timing experiments. In an 
eccentric orbit, one has a changing time dilation of the ``pulsar clock''
due to a variation in the orbital velocity of the pulsar and a change of 
the gravitational redshift caused by the gravitational field of the companion. 
This so-called {\it Einstein delay} 
\index{Einstein delay} 
is a periodic effect, whose amplitude is 
given by the post-Keplerian parameter $\gamma$, and to first oder can be written 
as
\begin{equation}\label{eq:D_Einstein}
  \Delta_{\rm E} = \gamma \sin U \;.
\end{equation}
For sufficiently edge-on and/or eccentric orbits the propagation delay 
suffered by the pulsar signals in the gravitational field of the companion
becomes important. This so-called {\it Shapiro delay}, to first order, reads
\index{Shapiro delay}
\begin{equation}\label{eq:DeltaS}
  \Delta_{\rm S} = -2 r \ln\left[1 - e \cos U -
    s\sin\omega (\cos U - e) - s\cos\omega (1 - e^2)^{1/2}\sin U \right] \;,
\end{equation}
where the two post-Keplerian parameters $r$ and $s$ are called {\it range}
and {\it shape} of the Shapiro delay. The latter is linked to the inclination
of the orbit with respect to the line of sight, $i$, by $s = \sin i$. It is
important to note, that for $i \rightarrow 90^\circ$ 
equation~(\ref{eq:DeltaS}) breaks down and higher order corrections are
needed. But so far, equation~(\ref{eq:DeltaS}) is fully sufficient for the
timing observations of known pulsars \cite{mas11}.

Concerning the post-Keplerian parameters related to quasi-stationary effects, 
for the wide class of boost-invariant gravity theories one finds that 
they can be expressed as functions of the Keplerian parameters, the masses, and 
parameters generically accounting for gravitational self-field effects  
(cf.~equation~(\ref{eq:LO})) \cite{dt92,will93}:
\begin{eqnarray}
  \dot\omega &=& \frac{n_b}{1-e^2}
                 \left(\varepsilon - \frac{\xi}{2} + \frac{1}{2}\right)
                 \frac{{\cal V}_b^2}{c^2} \;, 
                 \label{eq:PKagOMDOT}\\
  \gamma     &=& \frac{e}{n_b} \left(\frac{G_{0c}}{\cal G} + {\cal K}_p^c +
                 \frac{m_c}{M} \right) \frac{m_c}{M}
                 \frac{{\cal V}_b^2}{c^2} \;, 
                 \label{eq:PKagGAMMA}\\
  r          &=& \frac{1 + \varepsilon_{0c}}{4} \, \frac{G_{0c}m_c}{c^3} \;, 
                 \label{eq:PKagM2}\\
  s          &=& x\,n_b \, \frac{M}{m_c}\,\frac{c}{{\cal V}_b}\;,
                 \label{eq:PKagSINI}
\end{eqnarray}
plus $\Omega^{\rm SO}$ from equation~(\ref{eq:OmGPag}). Here we have listed 
only those parameters that play a role in this review. For a complete
list and a more detailed discussion, the reader is referred to \cite{dt92}.
The quantities $G_{0c}$ and $\varepsilon_{0c}$ are related to the interaction of 
the companion with a test particle or a photon. 
The parameter ${\cal K}_p^c$ accounts for a possible change in the moment of 
inertia of the pulsar due to a change in the local gravitational constant. 
In GR one finds ${\cal G} = G_{0c} = G$,
$\varepsilon = \varepsilon_{0c} = 3$, $\xi = 1$ and ${\cal K}_p^c = 0$.
Consequently
\index{post-Keplerian parameters, in GR}
\begin{eqnarray}\label{eq:PKGR}
  \dot\omega^{\rm GR} &=& \frac{3n_b}{1-e^2} \, \frac{V_b^2}{c^2} \;, 
                          \label{eq:omdotGR}\\
  \gamma^{\rm GR}     &=& \frac{e}{n_b} \left(1 + \frac{m_c}{M} \right)  
                          \frac{m_c}{M} \, \frac{V_b^2}{c^2} \;, 
                          \label{eq:gammaGR}\\
  r^{\rm GR}          &=& \frac{G m_c}{c^3} \;, \\
  s^{\rm GR}          &=& x\,n_b \, \frac{M}{m_c} \, \frac{c}{V_b}\;.
\end{eqnarray}
These parameters are independent of the 
internal structure of the neutron star(s), due to the effacement of the internal 
structure, a property of GR \cite{dam83,dam87}.
For most alternative gravity theories this is not the case. For instance,
in the mono-scalar-tensor theories $T_1(\alpha_0,\beta_0)$ of \cite{de93,de96a}, 
one finds\footnote{The mono-scalar-tensor theories $T_1(\alpha_0,\beta_0)$ of 
\cite{de93,de96a} have a conformal coupling function $A(\varphi) = 
\alpha_0 (\varphi - \varphi_0) + \beta_0 (\varphi - \varphi_0)^2/2$.
The Jordan-Fierz-Brans-Dicke gravity
\index{Jordan-Fierz-Brans-Dicke gravity}
is the sub-class with 
$\beta_0 = 0$, and $\alpha_0^2 = (2\omega_{\rm BD} + 3)^{-1}$.}
\index{post-Keplerian parameters, in scalar-tensor gravity}
\begin{eqnarray}\label{eq:PKst1}
  \dot\omega^{T_1} &=& 
    \frac{n_b}{1-e^2}\left(\frac{3 -  \alpha_p\alpha_c}{1 + \alpha_p\alpha_c}
    - \frac{m_p\alpha_p^2\beta_c + m_c\alpha_c^2\beta_p}
           {2M(1 + \alpha_p\alpha_c)^2} \right)
    \frac{{\cal V}_b^2}{c^2} \;, \\
  \gamma^{T_1}     &=& \frac{e}{n_b} 
    \left(\frac{1 + k_p\alpha_c}{1 + \alpha_p\alpha_c} + \frac{m_c}{M}\right) 
    \frac{m_c}{M} \, \frac{{\cal V}_b^2}{c^2} \;, \\
  r^{T_1}          &=& \frac{G_\ast m_c}{c^3} \;, \\
  s^{T_1}          &=& x\,n_b\,\frac{M}{m_c}\,\frac{c}{{\cal V}_b}\;,
\end{eqnarray}
where ${\cal V}_b = [G_\ast(1 + \alpha_p\alpha_c)M n_b]^{1/3}$. The 
body-dependent quantities $\alpha_p$ and $\alpha_c$ denote the effective scalar 
coupling 
\index{effective scalar coupling}
of pulsar and companion respectively, and $\beta_A \equiv 
\partial\alpha_A/\partial\varphi_0$ where $\varphi_0$ denotes
the asymptotic value of the scalar field at spatial infinity. The 
quantity $k_p$ is related to the moment of inertia $I_p$ of the pulsar via
$k_p \equiv -\partial\ln I_p/\partial\varphi_0$. For a given equation of state,
the parameters $\alpha_A$, $\beta_A$, and $k_A$ depend on the fundamental 
constants of the theory, e.g.~$\alpha_0$ and $\beta_0$ in $T_1(\alpha_0,\beta_0)$, and the mass of the body. As we will demonstrate later,
these ``gravitational form factors'' can assume large values in the strong
gravitational fields of neutron stars. Depending on the value of $\beta_0$, 
this is even the case for a vanishingly small $\alpha_0$, where there are 
practically no measurable deviations from GR in the Solar system. In fact, even 
for $\alpha_0 = 0$, a neutron star, above a certain $\beta_0$-dependent critical 
mass, can have an effective scalar coupling $\alpha_A$ of order unity. 
This non-perturbative strong-field behavior, the so-called ``spontaneous 
scalarization'' \index{spontaneous scalarization}
of a neutron star, was discovered 20 years ago by Damour and 
Esposito-Far{\`e}se \cite{de93}.

Finally, there is the post-Keplerian parameter $\dot{P}_b$, related to the
damping of the orbit due to the emission of gravitational waves. We have seen
above that in alternative gravity theories the back reaction from the 
gravitational wave emission might enter the equations of motion already at the 
1.5 post-Newtonian level, giving rise to a $\dot{P}_b \propto {\cal V}_b^3/c^3$. 
To leading order one finds in mono-scalar-tensor gravity the dipolar 
\index{dipolar radiation damping}
contribution from the scalar field \cite{ear75,will77,de96a}:
\begin{equation}\label{eq:PbdotD}
   \dot{P}_b = -2\pi \, \frac{m_p m_c}{M^2} \, 
                \frac{1+e^2/2}{(1-e^2)^{5/2}} \,
                \frac{{\cal V}_b^3}{c^3} \,
                \frac{(\alpha_p - \alpha_c)^2}{1 + \alpha_p\alpha_c} + {\cal O}({\cal V}_b^5/c^5) \;.
\end{equation}
As one can see, the change in the orbital period due to dipolar radiation 
depends strongly on the difference in the effective scalar coupling $\alpha_A$.
Binary pulsar systems with a high degree of asymmetry in the compactness of 
their components are therefore ideal to test for dipolar radiation. An order 
unity difference in the effective scalar coupling would lead to a change in
the binary orbit, which is several orders of magnitude ($\sim c^2/{\cal V}_b^2$) 
stronger than the quadrupolar damping predicted by GR.

At the 2.5 post-Newtonian level ($\propto {\cal V}_b^5/c^5$), in general, there 
are several contributions entering the $\dot{P}_b$ calculation:
\begin{itemize}
\item Monopolar waves for eccentric orbits.
\item Higher order contributions to the dipolar wave damping.
\item Quadrupolar waves from the tensor field, and the fields that are also
      responsible for the monopolar and/or dipolar waves. 
\end{itemize}
For scalar-tensor gravity these expressions can be found in \cite{de92}. For
GR one finds from the well-known {\it quadrupole formula} \cite{ein18,pet64}:
\index{quadrupole formula}
\begin{equation}\label{eq:PbdotGR}
  \dot{P}_b^{\rm GR} 
    = -\frac{192\pi}{5} \, \frac{m_p m_c}{M^2} \,
       \frac{1 + 73 e^2/24 + 37 e^4/96}{(1 - e^2)^{7/2}}\,\frac{V_b^5}{c^5} \;.
\end{equation}

Apart from a change in the orbital period, gravitational wave damping will
also affect other post-Keplerian parameters. While gravitational waves carry
away orbital energy and angular momentum, Keplerian parameters like
the eccentricity and the semi-major axis of the pulsar orbit change as well.
The corresponding post-Keplerian parameters are $\dot{e}$ and $\dot{x}$
respectively. However, these changes affect the arrival times of the pulsar
signals much less than the $\dot{P}_b$, and therefore do (so far) not play a
role in the radiative tests with binary pulsars.

As already mentioned in Section~\ref{sec:EOM}, there is no generic connection
between the higher-order gravitational wave damping effects and the parameters
${\cal G}$, $\varepsilon$, and $\xi$ of the modified Einstein-Infeld-Hoffmann 
formalism. Such higher order, mixed radiative and strong-field effects depend in 
a complicated way on the structure of the gravity theory \cite{dt92}.

The post-Keplerian parameters are at the foundation of many of the gravity tests
conducted with binary pulsars. As shown above, the exact functional dependence 
differs for given theories of gravity. A priori, the masses of the pulsar and
the companion are undetermined, but they represent the only unknowns in this set 
of equations. Hence, once two post-Keplerian parameters are measured, the 
corresponding equations can be 
solved for the two masses, and the values for other post-Keplerian 
parameters can be predicted for an assumed theory of gravity. Any further 
post-Keplerian measurement must therefore be consistent with that prediction, 
otherwise the assumed theory has to be rejected. In other words, if $N \ge 3$ 
post-Keplerian parameters can be measured, a total of $N - 2$ independent 
tests can be performed. The method is very powerful, as any additionally 
measured post-Keplerian parameter is potentially able to fail the prediction 
and hence to falsify the tested theory of gravity. The standard graphical 
representation of such tests, as will become clear below, is the mass-mass 
\index{mass-mass diagram}
diagram. Every measured post-Keplerian parameter defines a curve of certain 
width (given by the measurement uncertainty of the post-Keplerian parameter) in 
a $m_p$-$m_c$ diagram. A theory has passed a binary pulsar test, if there is a 
region in the mass-mass diagram that agrees with all post-Keplerian parameter 
curves.


\section{Gravitational wave damping}


\subsection{The Hulse-Taylor pulsar}
\label{sec:htpsr}

\index{The Hulse-Taylor pulsar}

The first binary pulsar to ever be observed happened to be a rare double 
neutron star system. It was discovered by Russell Hulse and Joseph Taylor in 
summer 1974 \cite{ht75}. The pulsar, PSR~B1913+16, has a rotational period of 
59 ms and is in a highly eccentric ($e = 0.62$) 7.75-hour orbit around an unseen 
companion. Shortly after the 
discovery of PSR~B1913+16, it has been realized that this system may allow the 
observation of gravitational wave damping within a time span of a few years 
\cite{bzns75,wag75}.

The first relativistic effect seen in the timing observations of the
Hulse-Taylor pulsar was the secular advance of periastron $\dot{\omega}$. 
Thanks to its large value of 4.2\,deg/yr, this effect was well measured already 
one year after the discovery \cite{thf+76}. Due to the, a priori, unknown masses 
of the system, this measurement could not be converted into a quantitative 
gravity test. However, assuming GR is correct, equation~(\ref{eq:omdotGR}) 
gives the total mass $M$ of the system. From the modern 
value given in table~\ref{tab:pars1913} one finds $M = m_p + m_c = 2.828378 \pm 
0.000007 \,M_\odot$ \cite{wnt10}.\footnote{Strictly speaking, this is the total 
mass of the system scaled with 
an unknown Doppler factor $D$, i.e.\ $M^{\rm observed} = D^{-1} 
M^{\rm intrinsic}$ \cite{dt92}. For typical velocities, $D - 1$ is expected
to be of order $10^{-4}$, see for instance \cite{wkk00}. In gravity tests based 
on post-Keplerian parameters, the factor $D$ drops out and is therefore 
irrelevant \cite{dd86}.}

It took a few more years to measure the Einstein delay (\ref{eq:D_Einstein}) with good precision.
In a single orbit this effect is exactly degenerate with the Roemer delay,
and only due to the relativistic precession of the orbit these two delays
become separable \cite{bzns75,bt75}. By the end of 1978, the 
timing of PSR~B1913+16 yielded a measurement of the post-Keplerian parameter
$\gamma$, which is the amplitude of the Einstein delay \cite{tfm79}. Together
with the total mass from $\dot{\omega}^{\rm GR}$, equation~(\ref{eq:gammaGR})
can now be used to calculate the individual masses. With the modern
value for $\gamma$ from table~\ref{tab:pars1913}, and the total mass given 
above, one finds the individual masses $m_p = 1.4398 \pm 0.0002 \, M_\odot$ and 
$m_c = 1.3886 \pm 0.0002 \, M_\odot$ for pulsar and companion respectively 
\cite{wnt10}.

\begin{figure}[H]
\centering
\includegraphics[height=90mm]{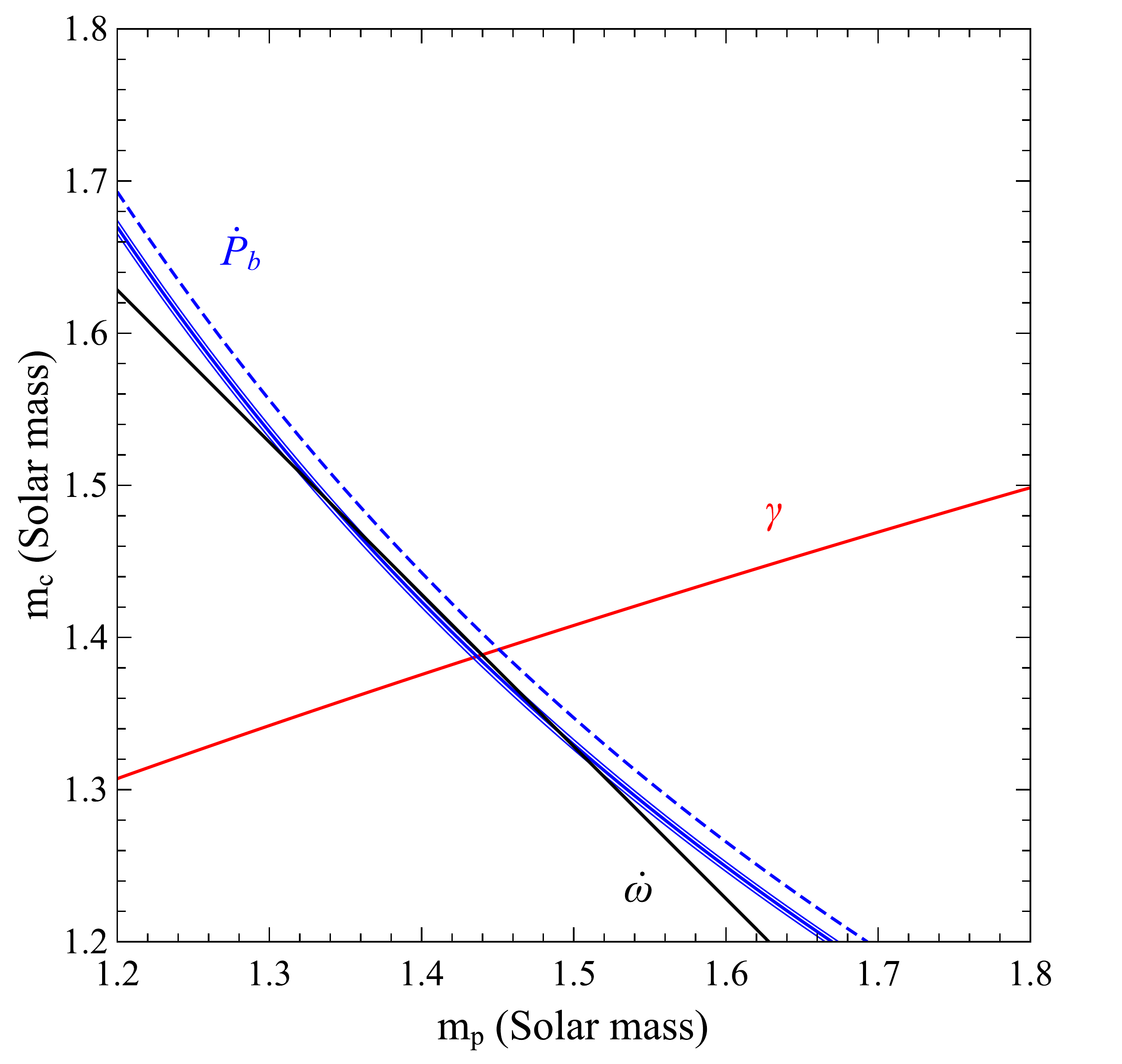}
\caption{Mass-mass diagram for PSR~B1913+16 based on GR and the three 
observed post-Keplerian 
parameters $\dot{\omega}$ (black), $\gamma$ (red) and $\dot{P}_b$ (blue).
The dashed $\dot{P}_b$ curve is based on the observed $\dot{P}_b$, without
corrections for Galactic and Shklovskii effects. The solid $\dot{P}_b$ curve
is based on the corrected (intrinsic) $\dot{P}_b$, where the thin lines indicate
the one-sigma boundaries. Values are taken from table~\ref{tab:pars1913}.
\label{fig:mm1913}}
\end{figure}

With the knowledge of the two masses, $m_p$ and $m_c$, the binary 
system is fully determined, and 
further GR effects can be calculated and
compared with the observed values, providing an intrinsic consistency check
of the theory. In fact, Taylor {\it et al.} \cite{tfm79} reported the 
measurement of a decrease in the orbital period $\dot{P}_b$, consistent with
the quadrupole formula (\ref{eq:PbdotGR}). This was the first proof for the
existence of gravitational waves as predicted by GR. In the meantime the 
$\dot{P}_b$ is measured with a precision of 0.04\% (see 
table~\ref{tab:pars1913}). However, this is not the precision with which 
the validity of the quadrupole formula is verified in the PSR~B1913+16
system. The observed $\dot{P}_b$ needs to be corrected for extrinsic effects,
most notably the differential Galactic acceleration and the Shklovskii effect,
\index{differential Galactic acceleration}
\index{Shklovskii effect}
to obtain the intrinsic value caused by gravitational wave damping 
\cite{shk70,dt91}. The extrinsic contribution due to the Galactic 
gravitational field (acceleration {\bf g}) and the proper motion (transverse
angular velocity in the sky $\mu$) are given by
\begin{equation}\label{eq:dPbdotGal}
  \delta\dot{P}_b^{\rm ext} = \frac{P_b}{c} \left[ \hat{\bf K}_0 \cdot
    ({\bf g}_{\rm PSR} - {\bf g}_\odot) + \mu^2 d \right] \;,
\end{equation}
where $\hat{\bf K}_0$ is the unit vector pointing towards the pulsar,
which is at a distance $d$ from the Solar system. For PSR~B1913+16, $P_b$ and
$\hat{\bf K}_0$ are measured with very high precision, and also $\mu$
is known with good precision ($\sim 8$\%). However, there is a large 
uncertainty in the distance $d$, which is also needed to calculate
the Galactic acceleration of the PSR~B1913+16 system, ${\bf g}_{\rm PSR}$,
in equation~(\ref{eq:dPbdotGal}).
Due to its large distance, there is no direct parallax measurement for $d$,
and estimates for $d$ are based on model dependent methods, like the measured
column density of free electrons between PSR~B1913+16 and the Earth. Such 
methods are known to have large systematic uncertainties, and for this reason
the distance to PSR~B1913+16 is not well known: $d = 9.9 \pm 3.1$\,kpc 
\cite{wsx+08,wnt10}. In addition, there are further uncertainties, e.g.~in the 
Galactic gravitational potential and the distance of the Earth to the Galactic
center. Accounting for all these uncertainties leads to an agreement between
$\dot{P}_b - \delta\dot{P}_b^{\rm ext}$ and $\dot{P}_b^{\rm GR}$ at the level
of about $0.3\%$ \cite{wnt10}. The corresponding mass-mass diagram is given
in figure~\ref{fig:mm1913}. As the precision of the radiative test with 
PSR~B1913+16 is limited by the model-dependent uncertainties in 
equation~(\ref{eq:dPbdotGal}), it is not expected that this test can be
significantly improved in the near future.

\begin{figure}[H]
\centering
\includegraphics[height=90mm]{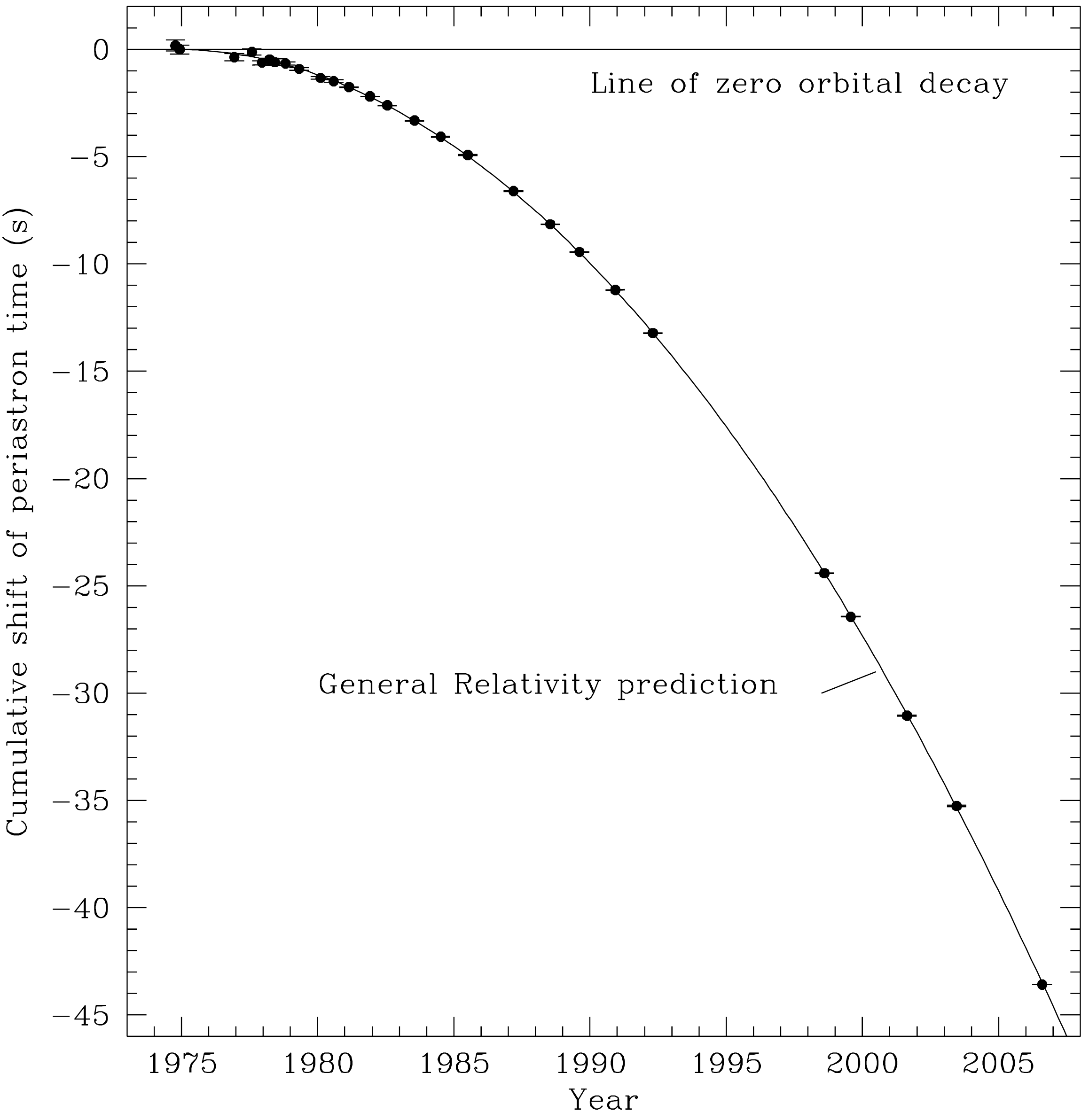}
\caption{Shift in the time of periastron passage of PSR~B1913+16 due to 
gravitational wave damping. The parabola represents the GR prediction and
the data points the timing measurements, with (vertical) error bars mostly too 
small to be resolved. The observed shift in periastron time is a direct 
measurement of the change in the world-line of the pulsar due to the 
back-reaction of the emitted gravitational waves 
(cf.~figure~\ref{fig:spacetime}). The corresponding spatial shift amounts to 
about 20\,000\,km. Figure is taken from \cite{wnt10}.
\label{fig:perishift}}
\end{figure}

Finally, besides the mass-mass diagram, there is a different way to illustrate 
the test of gravitational wave damping with PSR~B1913+16. According to
equation~(\ref{eq:Kepler}), the change in the orbital period, i.e.~the 
post-Keplerian parameter $\dot{P}_b$, is measured from a shift in the time
of periastron passage, where $U$ is a multiple of $2\pi$. One finds 
for the shift in periastron time, as compared to an orbit with zero decay
\index{shift in periastron time}
\begin{equation}\label{eq:perishift}
  \Delta T = \frac{1}{2} P_b \dot{P}_b n^2 + {\cal O}(P_b\dot{P}_b^2 n^3) \;,
\end{equation}
where $n = 0,1,2,\dots$ denotes the number of the periastron passage, and is 
given by $n \simeq (T - T_0)/P_b$.
Equation~(\ref{eq:perishift}) represents a parabola in time, which can be 
calculated with high precision using the masses that come from 
$\dot\omega^{\rm GR}$ and $\gamma^{\rm GR}$ (see above). On the other hand, the 
observed cumulative shift in periastron can be extracted from the timing 
observations with high precision. A comparison of observed and predicted 
cumulative shift in the time of the periastron passage is given in 
figure~\ref{fig:perishift}.

\begin{table}
\caption{Observed orbital timing parameters of PSR~B1913+16, based on the 
Damour-Deruelle timing model (taken from \cite{wnt10}). Figures in parentheses 
represent estimated uncertainties in the last quoted digit. 
\label{tab:pars1913} }
\vspace{1mm}
\centerline{
{\small
\begin{tabular}{llr}
\hline
$T_0$        & time of periastron passage (MJD) & 52144.90097841(4) \\
$x$          & projected semi-major axis of the pulsar orbit (s) 
             & 2.341782(3) \\
$e$          & orbital eccentricity & 0.6171334(5) \\
$P_b$        & orbital period at $T_0$ (d) & 0.322997448911(4) \\
$\omega_0$   & longitude of periastron at $T_0$ (deg) & 292.54472(6) \\
$\dot\omega$ & secular advance of periastron (deg/yr) & 4.226598(5) \\
$\gamma$     & amplitude of Einstein delay (ms) & 4.2992(8) \\
$\dot{P}_b$  & secular change of orbital period & $-2.423(1) \times 10^{-12}$ \\
\hline
\end{tabular}
}}
\end{table}

\subsection{The Double Pulsar --- The best test for Einstein's quadrupole 
formula, and more}
\label{sec:dp}

\index{Double Pulsar}

\begin{figure}[H]
\centering
\includegraphics[height=100mm]{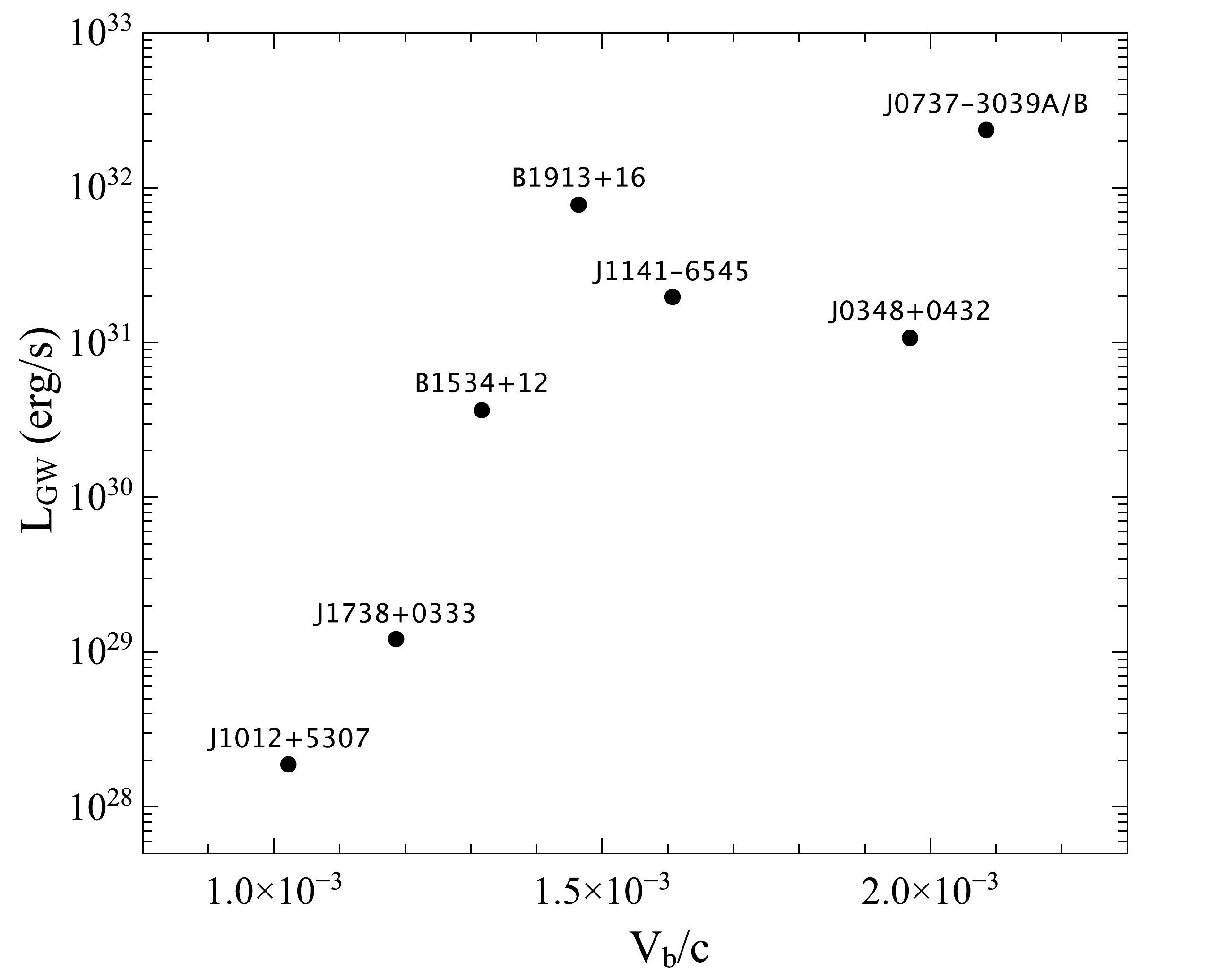}
\caption{Short-orbital-period ($P_b < 1\,{\rm day}$) binary pulsars used for gravity tests. The velocity
$V_b$ (divided by the speed of light $c$) is a direct measure for the strength 
of post-Newtonian effects in the 
orbital dynamics. The gravitational wave luminosity $L_{\rm GW}$ is an indicator 
for the strength of radiative effects that cause secular changes to the orbital 
elements due to gravitational wave damping.
\label{fig:VbLGW}}
\end{figure}

In 2003 a binary system was discovered where, at first, one member was 
identified 
as a pulsar with a 23\,ms period \cite{bdp+03}. About half a year later, 
the companion was also recognized as a radio pulsar with a period of 2.8\,s 
\cite{lbk+04}. Both pulsars, known as PSRs~J0737$-$3039A and 
J0737$-$3039B, respectively, (or $A$ and $B$ hereafter),
orbit each other in less than 2.5 hours in a mildly eccentric ($e = 0.088$)
orbit. As a result, the system is not only the first and only double neutron 
star system where both neutron stars are visible as active radio pulsars, but it 
is also the most relativistic binary pulsar laboratory for gravity known to 
date (see figure~\ref{fig:VbLGW}). Just to give an example for the strength of
relativistic effects, the advance of periastron, $\dot{\omega}$, is 17 degrees 
per year, meaning that the eccentric orbit does a full rotation in just 21 
years. In this subsection, we briefly discuss the properties of this unique 
system, commonly referred to as the {\it Double Pulsar}, 
and highlight some of the gravity tests that are based on the radio observations 
of this system. For detailed reviews of the Double Pulsar see \cite{ks08,kw09}.

In the Double Pulsar system a total of six post-Keplerian parameters have 
been measured by now. Five arise from four different relativistic effects 
visible in pulsar timing \cite{ksm+06}, while a sixth one can be determined from 
the effects of geodetic precession, which will be discussed in detail in
Section~\ref{sec:dpSO} below. The relativistic precession of the orbit,
$\dot{\omega}$, was measured within a few days after timing of the system 
commenced, and by 2006 it was already known with a precision of 0.004\% (see
table~\ref{tab:0737pars}).  At the same time the measurement of the amplitude of 
Einstein delay, $\gamma$, reached 0.7\% (see table~\ref{tab:0737pars}). Due to 
the periastron precession of 17 degrees per year, the Einstein delay was soon
well separable from the Roemer delay. Two further post-Keplerian parameters 
came from the detection of the Shapiro delay: 
the {\em shape} and {\em range} parameters $s$ and $r$. They were measured with 
a precision of 0.04\% and 5\%, respectively (see table~\ref{tab:0737pars}). From
the measured value $s = \sin i = 0.99974_{-0.00039}^{+0.00016}$ 
($i = 88.7^\circ{}_{-0.8^\circ}^{+0.5^\circ}$) 
one can already see how exceptionally edge-on this system is.\footnote{The only 
binary pulsar known to be (most likely) even more edge-on is PSR~J1614$-$2230 
with $s = \sin i = 0.999894 \pm 0.000005$ ($i = 89.17^\circ \pm 0.02^\circ$) 
\cite{dpr+10}. For this wide-orbit system ($P_b \approx 8.7\,{\rm d}$), however, 
no further post-Keplerian parameter is known that could be used in a gravity 
test.}
Finally, the decrease of the orbital period due to gravitational wave damping
was measured with a precision of 1.4\% just three years after the discovery of 
the system (see table~\ref{tab:0737pars}). 

\begin{table}[ht]
\caption{A selection of observed orbital timing parameters of the Double Pulsar, 
based on the Damour-Deruelle timing model (taken from \cite{ksm+06}). All 
post-Keplerian parameters below are obtained from the timing of pulsar $A$.
The timing precision for pulsar $B$ is considerably lower, and allows only
for a, in comparison, low precision measurement ($\sim 0.3$\%) of $\dot{\omega}$ 
\cite{ksm+06}. Figures in parentheses represent estimated uncertainties in the 
last quoted digit. 
\label{tab:0737pars} }
\vspace{1mm}
\centerline{\small
\begin{tabular}{llr}
\hline
$x_A \equiv a_A\sin i/c$ & projected semi-major axis of pulsar $A$ (s) 
             & 1.415032(1) \\
$x_B \equiv a_B\sin i/c$ & projected semi-major axis of pulsar $B$ (s) 
             & 1.5161(16) \\
$e$          & orbital eccentricity & 0.0877775(9) \\
$P_b$        & orbital period (d) & 0.10225156248(5) \\
$\dot\omega$ & secular advance of periastron (deg/yr) & 16.89947(68) \\
$\gamma$     & amplitude of Einstein delay for $A$ (ms) & 0.3856(26) \\
$\dot{P}_b$  & secular change of orbital period & $-1.252(17)\times 10^{-12}$\\
$s$          & {\it shape} of Shapiro delay for $A$ & $0.99974(-39,+16)$ \\
$r$          & {\it range} of Shapiro delay for $A$ ($\mu$s) & 6.21(33) \\
\hline
\end{tabular}
}
\end{table}

A unique feature of the Double Pulsar is its nature as a ``dual-line source'',
i.e.~we measure the orbits of both neutron stars
at the same time. Obviously, the sizes of the two orbits are not independent
from each other as they orbit a common center of mass.  
In GR, up to first post-Newtonian order the relative size 
of the orbits is identical to the inverse ratio of masses. 
\index{ratio of masses in binary pulsars}
Hence, by measuring
the orbits of the two pulsars (relative to the centre of mass), we obtain a
precise measurement of the mass ratio. This ratio is directly observable, as
the orbital inclination angle is obviously identical for both pulsars, i.e.
\begin{equation}
R \equiv \frac{m_{\rm{A}}}{m_{\rm{B}}} 
  =      \frac{a_{\rm{B}}}{a_{\rm{A}}} 
  =      \frac{a_{\rm{B}}\sin i/c}{a_{\rm{A}}\sin i/c} 
  \equiv \frac{x_{\rm{B}}}{x_{\rm{A}}} \:.
\end{equation}
This expression is not just limited to GR. In fact, it is valid up to first
post-Newtonian order and free of any explicit strong-field effects in any 
Lorentz-invariant theory of gravity (see 
\cite{dam09} for a detailed discussion). Using the parameter values of 
table~\ref{tab:0737pars}, one finds that in the Double Pulsar the masses are 
nearly equal with $R=1.0714 \pm 0.0011$.

\begin{figure}[H]
\centering
\includegraphics[height=90mm]{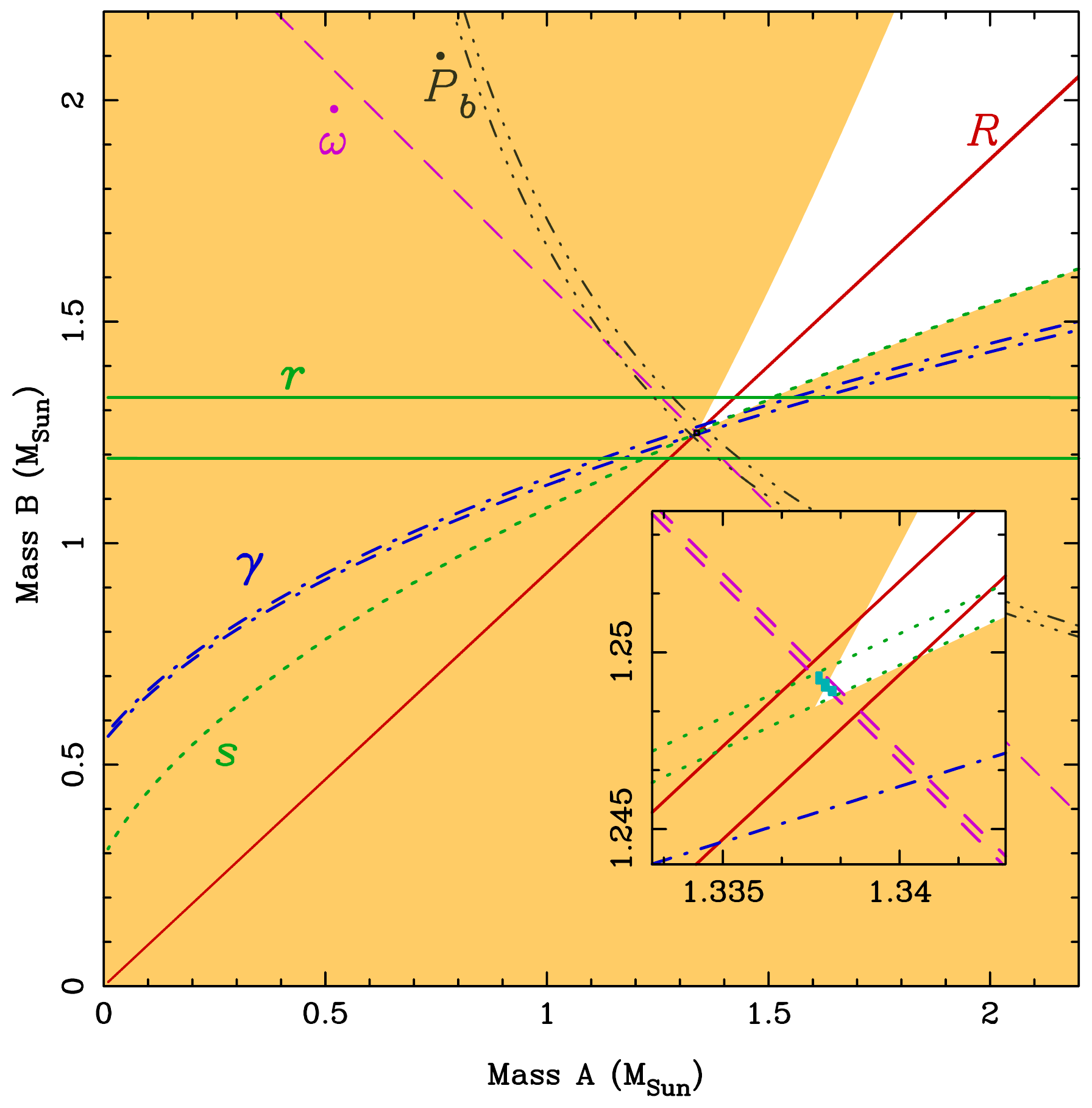}
\caption{GR mass-mass diagram based on timing observations of the Double Pulsar.
The orange areas are excluded simply by the fact that $\sin i \le 1$. 
The figure is taken from \cite{kw09} ($\Omega_{\rm SO}$ lines removed)
and based on the timing solution published in \cite{ksm+06}.
\label{fig:mm_dp}}
\end{figure}

As it turns out, all the post-Keplerian parameters measured from timing are 
consistent with GR. In addition, the region of allowed masses agrees well
with the measured mass ratio $R$ (see figure~\ref{fig:mm_dp}). One has to keep
in mind, that the test presented here is based on data published in 2006 
\cite{ksm+06}. In the meantime continued timing lead to a significant 
decrease in the uncertainties of the post-Keplerian parameters of the Double 
pulsar. This is especially the case for $\dot{P}_b$, for which the uncertainty 
typically decreases with $T_{\rm obs}^{-2.5}$ \cite{bt76}, $T_{\rm obs}$ being 
the total time span of timing observations. The new results
will be published in an upcoming publication (Kramer {\it et al.}, in prep.).
As reported in \cite{kra12}, presently the Double Pulsar provides the best test 
for the GR quadrupole formalism for gravitational wave generation, with an 
uncertainty well below the 0.1\% level. As discussed above, the Hulse-Taylor
pulsar is presently limited by uncertainties in its distance. This raises the 
valid question, at which level such uncertainties will start to limit the
radiative test with the Double Pulsar as well. Compared to the Hulse-Taylor 
pulsar, the Double Pulsar is much closer to Earth. Because of this, a direct 
distance
estimate of $1.15_{-0.16}^{+0.22}$\,kpc based on a parallax measurement with
long-baseline interferometry was obtained \cite{dbt09}. Thus, with the
current accuracy in the measurement of distance and transverse velocity, 
GR tests based on $\dot{P}_b$ can be taken to the 0.01\% level. We will come
back to this in Section~\ref{sec:summary}, where we discuss some future tests 
with the Double Pulsar.

With the large number of post-Keplerian parameters and the known mass ratio,
the Double Pulsar is the most over-constrained binary pulsar system. For
this reason, one can do more than just testing specific gravity theories.
The Double Pulsar allows for certain generic tests on the orbital dynamics, time 
dilation, and photon propagation of a spacetime with two strongly 
self-gravitating bodies
\cite{kw09}. First, the fact that the Double Pulsar gives access to the mass 
ratio, $R$, in any Lorentz-invariant theory of gravity, allows us to determine 
$m_{\rm A}/M = R/(1+R) = 0.51724 \pm 0.00026$ and $m_{\rm B}/M = 1/(1+R) = 
0.48276 \pm 0.00026$. With this information at hand, the measurement of the 
shape of the Shapiro delay $s$ can be used to determine ${\cal V}_b$ via 
equation~(\ref{eq:PKagSINI}):  ${\cal V}_b/c = (2.0854 \pm 0.0014) \times 
10^{-3}$. At this point, the measurement of the post-Keplerian 
parameters $\dot\omega$, $\gamma$, and $r$ (equations~(\ref{eq:PKagOMDOT}),
(\ref{eq:PKagGAMMA}), (\ref{eq:PKagM2})) can be used to impose restrictions on 
the ``strong-field'' parameters of Lagrangian (\ref{eq:LO}) \cite{kw09}:
\index{strong-field parameters}
\begin{eqnarray}
  \frac{2\varepsilon - \xi}{5} &=& 0.9995 \pm 0.0016 \;,\label{eq:epsxi}\\
  \frac{G_{0{\rm B}}}{\cal G} + {\cal K}_{\rm A}^{\rm B}  &=& 1.005  \pm 0.010  \;,\\
  \frac{\varepsilon_{0{\rm B}} + 1}{4}\,
  \frac{G_{0{\rm B}}}{{\cal G}} &=& 
  1.009 \pm 0.054 \;.
\end{eqnarray}
This is in full agreement with GR, which predicts one for all three of 
these expressions. Consequently, nature cannot deviate much from 
GR in the quasi-stationary strong-field regime of gravity (G2 in 
figure~\ref{fig:gravregimes}).


\subsection{PSR J1738+0333 --- The best test for scalar-tensor gravity}
\label{sec:1738}

\index{PSR J1738+0333}

The best ``pulsar clocks'' are found amongst the fully recycled millisecond 
pulsars, which have rotational periods less than about 10\,ms (see e.g.\
\cite{vbc+09}). A result of the stable mass transfer between companion
and pulsar in the past --- responsible for the recycling of the pulsar --- is a 
very efficient 
circularization of the binary orbit, that leads to a pulsar-white dwarf system 
with very small residual eccentricity \cite{phi92}. For such systems, the 
post-Keplerian parameters $\dot{\omega}$ and $\gamma$ are generally not
observable. There are a few cases where the orbit is seen sufficiently edge-on,
so that a measurement of the Shapiro delay gives access to the two 
post-Keplerian parameters $r$ and $s$ with good precision 
(see e.g.~\cite{rt91}, which was the first detection of a Shapiro delay in a
binary pulsar). With these two parameters the system is then fully
determined, and in principle can be used for a gravity test in combination with
a third measured (or constrained) post-Keplerian parameter (e.g.~$\dot{P}_b$).
Besides the Shapiro delay parameters, some of the circular binary pulsar
systems offer a completely different access to their masses, which is not solely 
based on the timing observations in the radio frequencies. If the companion 
star is bright enough for optical spectroscopy, then we have a dual-line system,
where the Doppler shifts in the spectral lines can be used, together 
with the timing observations of the pulsar, to determine the mass ratio $R$.
Furthermore, if the companion is a white dwarf, the spectroscopic information in 
combination with models of the white dwarf and its atmosphere can be used to 
determine the mass of the white dwarf $m_c$, ultimately giving the mass of the 
pulsar via $m_p = R \, m_c$. As we will see in this and the following 
subsection, two of the best binary pulsar systems for gravity tests have their 
masses determined through such a combination of radio and optical astronomy.

\begin{figure}[H]
\centering
\includegraphics[height=90mm]{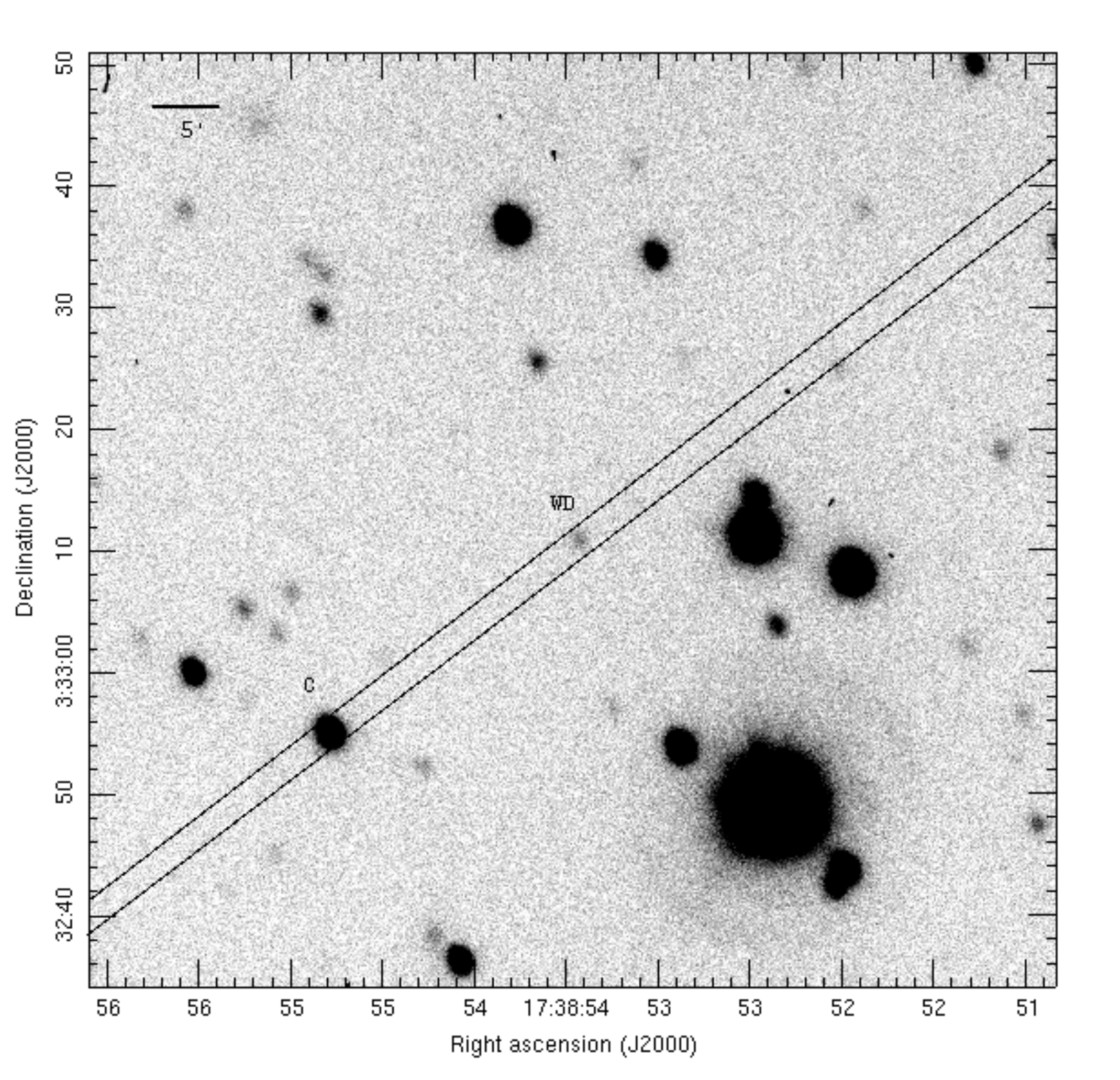}
\caption{Optical finding chart for the PSR~J1738+0333 companion. Indicated 
are the white dwarf companion (WD), the slit orientation used during the
observation and the comparison star (C) that was included in the slit. The white 
dwarf is sufficiently bright to allow for high signal-to-noise spectroscopy (see 
\cite{avk+12} for details, where this figure is taken from).
\label{fig:1738comp}}
\end{figure}

PSR~J1738+0333 was discovered in 2001 \cite{jac05}. It has a spin period $P$ of 
5.85\,ms and is a member of a low-eccentricity ($e < 4 \times 10^{-7}$) binary 
system with an orbital period $P_b$ of just 8.5 hours. The companion is an
optically bright low-mass white dwarf (see figure~\ref{fig:1738comp}).
Extensive timing observation over a period of 10 years allowed a 
determination of astrometric, spin and orbital parameters with high precision
\cite{fwe+12}, most notably
\begin{itemize}
\item A change in the orbital period of 
      $(-17.0 \pm 3.1) \times 10^{-15}$.
\item A timing parallax, which gives a model independent distance estimate of
      $d = 1.47 \pm 0.10$\,kpc. 
\end{itemize}
The latter is important to correct for the Shklovskii effect and the 
differential Galactic acceleration to obtain the intrinsic $\dot{P}_b$ 
(cf.~equation~(\ref{eq:dPbdotGal})). Additional spectroscopic observations of 
the white dwarf gave the mass ratio $R = 8.1 \pm 0.2$ and the companion mass 
$m_c = 0.181_{-0.005}^{+0.007}\,M_\odot$, and consequently the pulsar mass
$m_p = 1.47_{-0.06}^{+0.07}\,M_\odot$ \cite{avk+12}. It is important to 
note, that the mass determination for PSR~B1738+0333 is free of any 
explicit strong-field contributions, since this is the case for the mass ratio 
\cite{dam09}, and certainly for the mass of the white dwarf, which is a weakly 
self-gravitating body, i.e.~a gravity regime that has been well tested in the 
Solar system (G1 in figure~\ref{fig:gravregimes}). 

\begin{figure}[H]
\centering
\includegraphics[height=90mm]{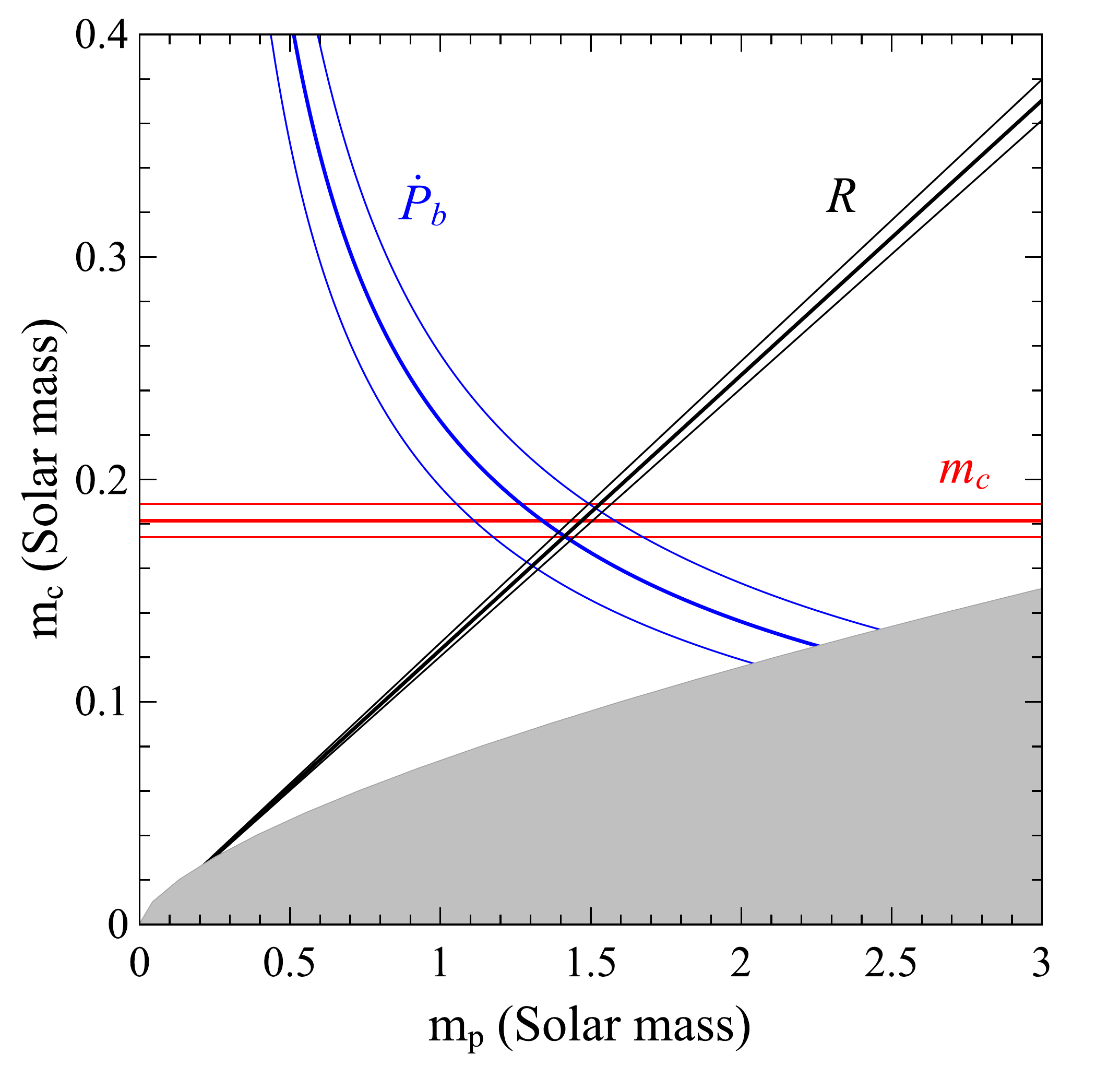}
\caption{GR mass-mass diagram based on the timing observations of PSR~J1738+0333 
and the optical observations of its white-dwarf companion respectively. The thin 
lines indicate the one-sigma errors of the measured parameters. The grey area is 
excluded by the condition $\sin i \le 1$.
\label{fig:mm_1738}}
\end{figure}

After using equation~(\ref{eq:dPbdotGal}) to correct for the
Shklovskii contribution, $\delta\dot{P}_b = P_b\mu^2d/c = (8.3_{-0.5}^{+0.6})
\times 10^{-15}$, and the contribution from the Galactic differential 
acceleration, $\delta\dot{P}_b = (0.58_{-0.14}^{+0.16})\times 10^{-15}$, one 
finds an intrinsic orbital period change due to gravitational wave damping of
$\dot{P}_b^{\rm intr} = (-25.9 \pm 3.2) \times 10^{-15}$. This value agrees well 
with the prediction of GR, as can be seen in figure~\ref{fig:mm_1738}. 

\begin{figure}[H]
\centering
\includegraphics[height=110mm]{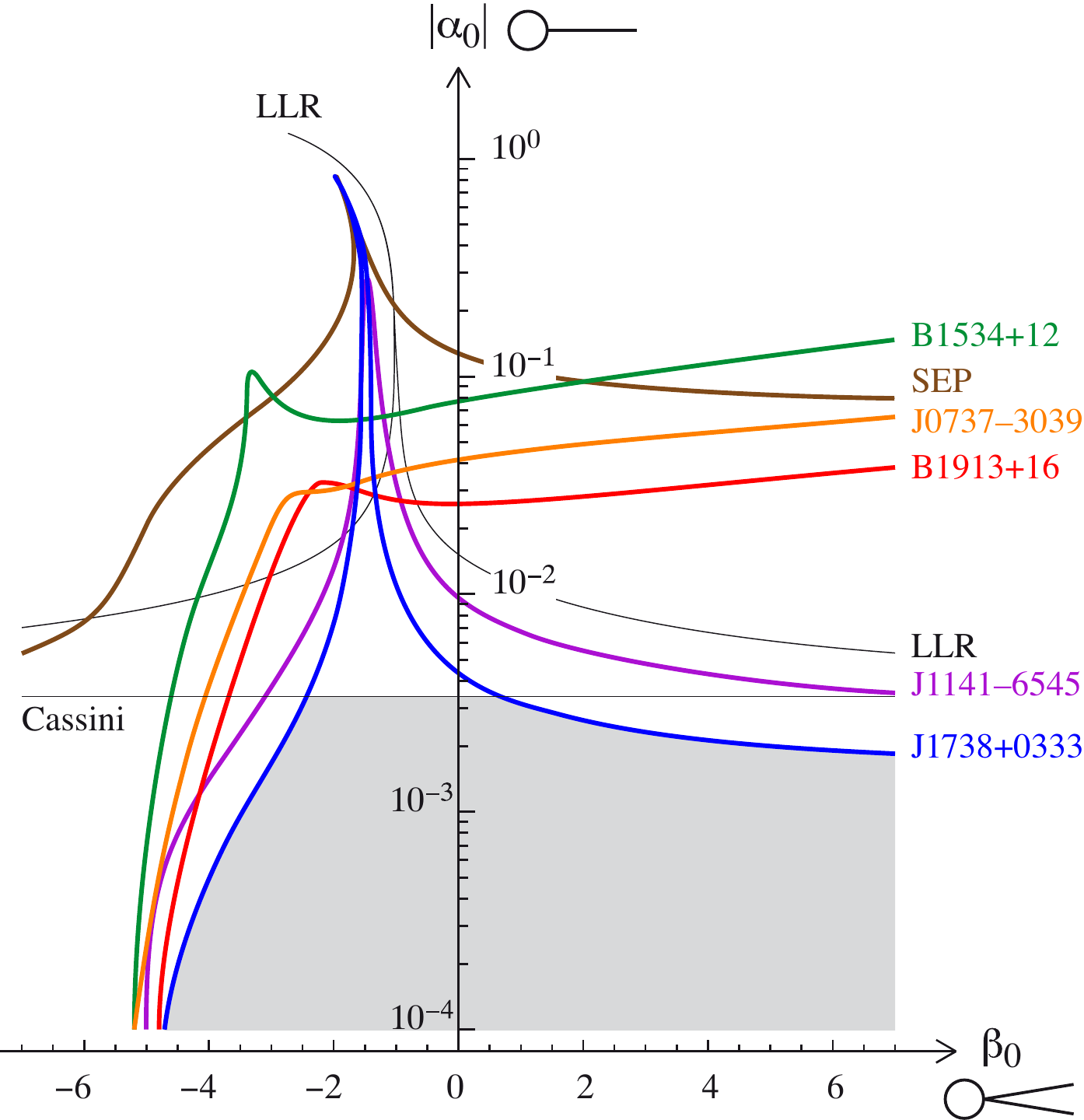}
\caption{Constraints on the class of $T_1(\alpha_0,\beta_0)$ scalar-tensor 
theories of \cite{de93,de96a}, from different binary pulsar and Solar system
(Cassini and Lunar Laser Ranging) experiments. The grey
area indicates the still allowed $T_1$ theories, and includes GR ($\alpha_0 = 
\beta_0 = 0$). It is obvious that PSR~J1738+0333 is the most constraining  
experiment for most of the $\beta_0$ range, and is even competitive with Cassini 
in testing the Jordan-Fierz-Brans-Dicke theory ($\beta_0 = 0$). As can be 
clearly seen, the double neutron-star systems PSR~B1534+12 \cite{sttw02}, 
PSR~B1913+16 (Hulse-Taylor pulsar) and PSR~J0737$-$3039A/B (Double Pulsar) are 
considerably less 
constraining, as explained in the text. PSR~J1141$-$6545 is also well suited for 
a dipolar radiation test \cite{bbv08}, since it also has a white dwarf companion 
\cite{abw+11}. Figure is taken from \cite{fwe+12}.
\label{fig:stg_1738}}
\end{figure}

The radiative test with PSR~J1738+0333 represents a $\sim 15$\% verification of  
GR's quadrupole formula. A comparison with the $< 0.1$\% test from the 
Double Pulsar (see Section~\ref{sec:dp}) raises the valid question of whether 
the PSR~J1738+0333 experiment is teaching us something new about the nature of 
gravity and the validity of GR. To address this question, let's have a look at
equation~(\ref{eq:PbdotD}). Dipolar radiation can be a strong source of 
gravitational wave damping, if there is a sufficient difference between the
effective coupling parameters $\alpha_p$ and $\alpha_c$ of pulsar and 
companion respectively. For the Double Pulsar, where we have two neutron stars 
with $m_p \approx m_c$, one generally expects that $\alpha_p \approx \alpha_c$, 
and therefore the effect of dipolar radiation would be strongly suppressed. On 
the other hand, in the PSR~J1738+0333 system there is a large difference in the 
compactness of the two bodies. For the weakly self-gravitating white-dwarf 
companion $\alpha_c \simeq \alpha_0$, i.e.~it assumes the weak-field 
value\footnote{From the Cassini experiment \cite{bit03} one obtains 
$|\alpha_0| < 3 \times 10^{-3}$ (95\% confidence).}, 
while the strongly self-gravitating pulsar can have an $\alpha_p$ that 
significantly deviates from $\alpha_0$. In fact, as discussed in 
Section~\ref{sec:PPK}, $\alpha_p$ can even be of oder 
unity in the presence of effects like strong-field scalarization. In the absence
of non-perturbative strong-field effects one can do a first order estimation
$(\alpha_p - \alpha_c) \propto (\epsilon_p - \epsilon_c) + 
{\cal O}(\epsilon^2)$. For the Double Pulsar one finds $(\epsilon_p - 
\epsilon_c)^2 \approx 6 \times 10^{-5}$, which is significantly smaller
than for the PSR~J1738+0333 system, which has 
$(\epsilon_p - \epsilon_c)^2 \approx 0.012$.\footnote{These numbers are
based on the equation of state MPA1 in \cite{mpa87}. Within GR, MPA1 
has a maximum neutron-star mass of $2.46\,M_\odot$, which can also account
for the high-mass candidates of \cite{frb+08,vbk11,rfs+12}.}
 As a consequence, the orbital decay
of asymmetric systems like PSR~J1738+0333 could still be dominated by dipolar 
radiation, even if the Double Pulsar agrees with GR. For this reason, 
PSR~J1738+0333 is particularly useful to test
gravity theories that violate the strong equivalence principle and 
therefore predict the emission of dipolar radiation. A well known class of
gravity theories, where this is the case, are scalar-tensor theories. As it
turns out, PSR~J1738+0333 is currently the best test system for these 
alternatives to GR (see figure~\ref{fig:stg_1738}). In terms of 
equation~(\ref{eq:PbdotD}), one finds
\begin{equation}\label{eq:1738Dtest}
  |\alpha_p - \alpha_c| < 2 \times 10^{-3} 
  \quad \mbox{(95\% confidence)} \;,
\end{equation}
where for the weakly self-gravitating white dwarf companion $\alpha_c \simeq 
\alpha_0$. This limit can be interpreted as a generic limit on dipolar 
radiation, where $\alpha_p - \alpha_c$ is the difference of some hypothetical
(scalar- or vector-like) ``gravitational charges'' \cite{gw02}.

\subsection{PSR J0348+0432 --- A massive pulsar in a relativistic orbit}

\index{PSR J0348+0432}

\begin{figure}[H]
\centering
\includegraphics[height=100mm]{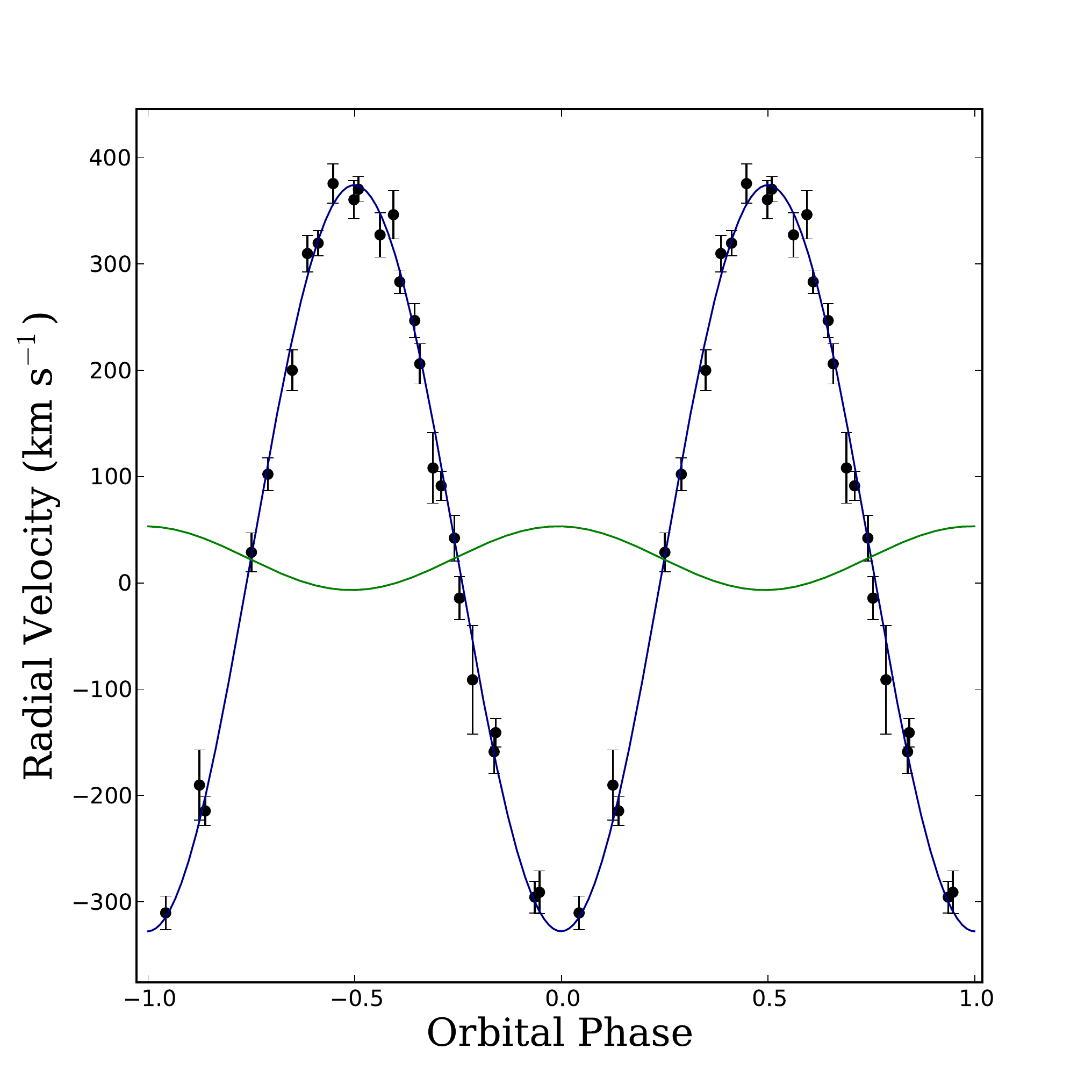}
\caption{Spectroscopically measured radial velocities for the white-dwarf 
companion of PSR~J0348+0432. For illustration purposes the data are plotted 
twice. The fitted sinusoidal curve (blue) has an amplitude of $351 \pm 4 \,
{\rm km/s}$. As a comparison, the sinusoidal green line shows the radial 
velocity of the pulsar as derived from the timing solution. The amplitude of 
the green line is known with very high precision: $30.008235 \pm 0.000016 \,
{\rm km/s}$. The ratio of the amplitudes gives the mass ratio $R$.
Figure is taken from \cite{afw+13}.
\label{fig:0348_vr}}
\end{figure}

PSR~J0348+0432 was discovered in 2007 in a drift scan survey using the Green 
Bank radio telescope (GBT) \cite{blr+13,lbr+13}. PSR~J0348+0432 is a mildly 
recycled radio-pulsar with a spin period of 39 ms. Soon it was found to be 
in a 2.46-hour 
orbit with a low-mass white-dwarf companion. In fact, the orbital period is only
15 seconds longer than that of the Double Pulsar, which by itself makes this 
already an interesting system for gravity. Initial timing observations of 
the binary yielded an accurate astrometric position, which allowed for an 
optical identification of its companion \cite{afw+13}. As it turned out, the 
companion is a relatively bright white dwarf with a spectrum that shows
deep Balmer lines. Like in the case of PSR~J1738+0333, one could use
high-resolution
optical spectroscopy to determine the mass ratio $R = 11.70 \pm 0.13$ (see 
figure~\ref{fig:0348_vr}) and the companion mass $m_c = 0.172 \pm 0.003 \, 
M_\odot$. For the mass of the pulsar one then finds $m_p = R\,m_c = 2.01 \pm 
0.04\,M_\odot$, which is presently the highest, well determined neutron star 
mass, and only the second neutron star with a well determined mass close to 
2\,$M_\odot$.\footnote{The first well determined two Solar mass neutron star is 
PSR~J1614$-$2230 \cite{dpr+10}, which is in a wide orbit and therefore does not 
provide any gravity test.}

\begin{figure}[H]
\centering
\includegraphics[height=90mm]{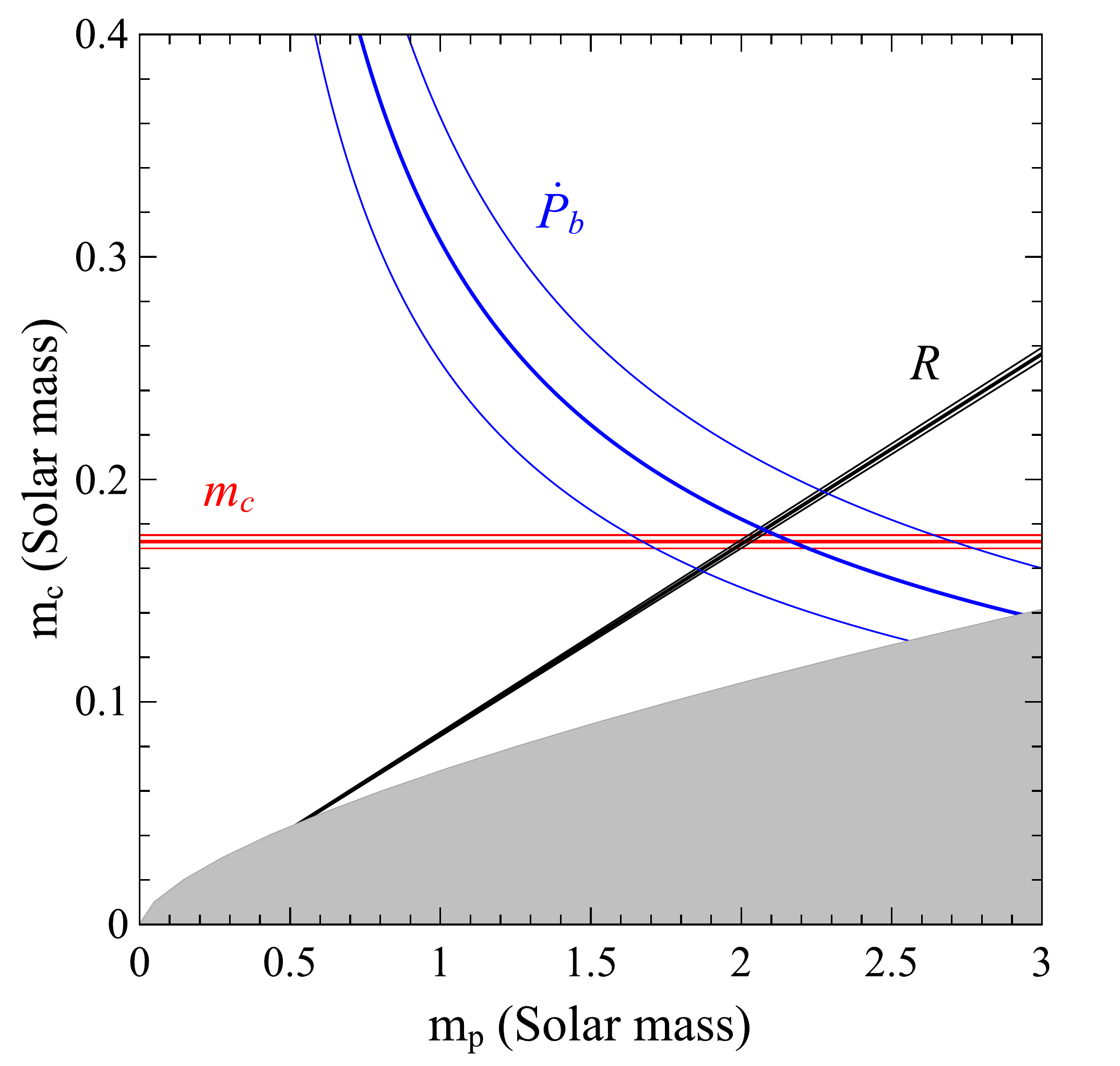}
\caption{GR mass-mass diagram based on timing and optical observations of the
PSR~J0348+0432 system. The thin lines indicate the one-sigma errors of the
measured parameters. The grey area is excluded by the condition $\sin i \le 1$.
\label{fig:mm_0348}}
\end{figure}

Since its discovery there have been regular timing observations of 
PSR~J0348+0432 with three of the major radio telescopes in the world,
the 100-m Green Bank Telescope, the 305-m radio telescope at the Arecibo 
Observatory, and the 100-m Effelsberg radio telescope. Based on the timing data,
in 2013 Antoniadis {\it et al.}\ \cite{afw+13} reported the detection of a 
decrease in the orbital period of $\dot{P}_b = (-2.73 \pm 0.45) \pm 10^{-13}$ 
that is in full agreement with GR (see figure~\ref{fig:mm_0348}). In numbers:
\begin{equation}\label{eq:0348GRtest}
  \dot{P}_b/\dot{P}_b^{\rm GR} = 1.05 \pm 0.18 \;.
\end{equation}
As it turns out, using the distance inferred from the photometry of the white 
dwarf ($d \sim 2.1 \, {\rm kpc}$) corrections due to the Shklovskii effect and 
differential acceleration in the Galactic potential (see 
equation~(\ref{eq:dPbdotGal})) are negligible compared to the measurement 
uncertainty in $\dot{P}_b$.

\begin{figure}[H]
\centering
\includegraphics[height=90mm]{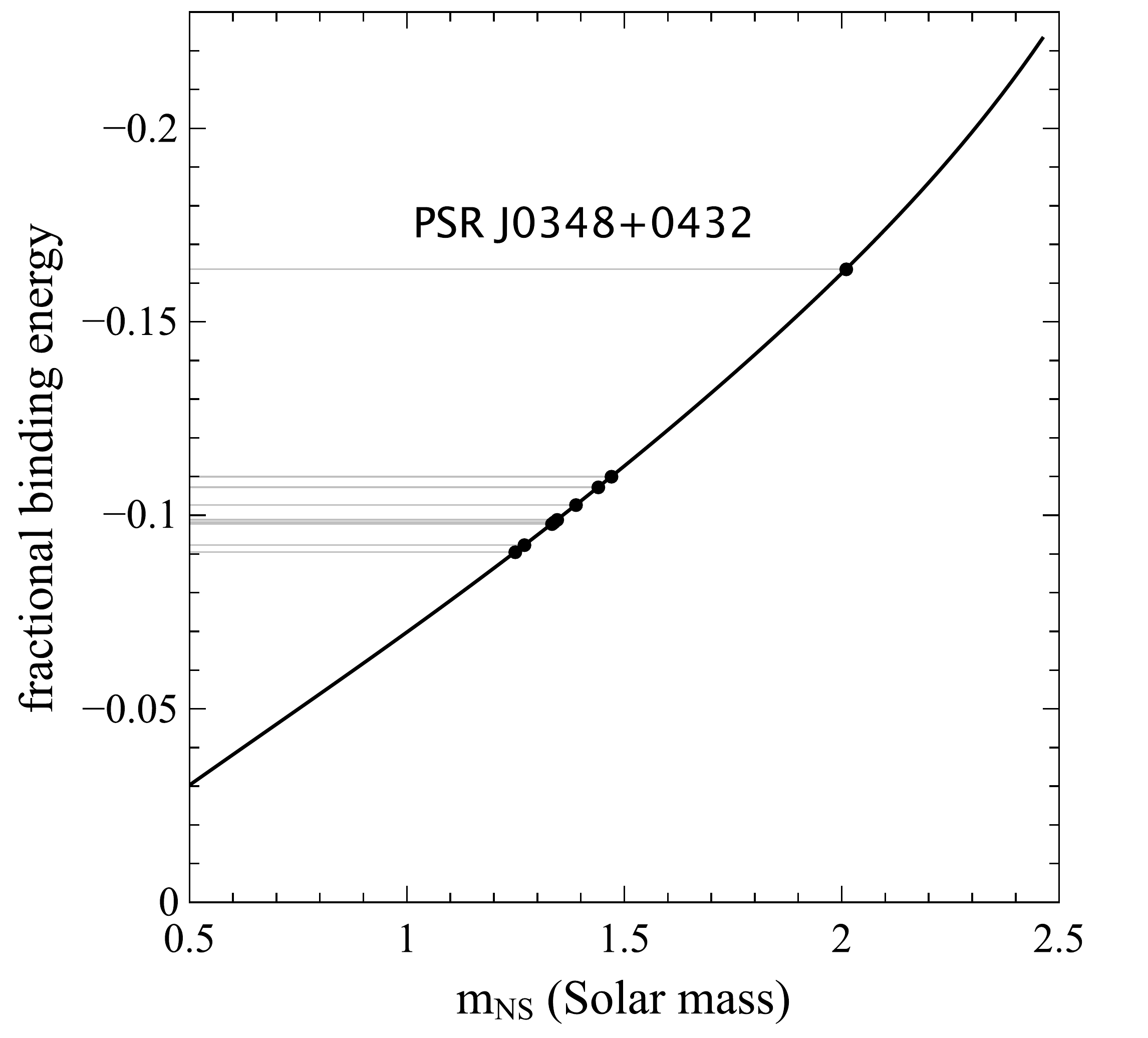}
\caption{Fractional gravitational binding energy 
\index{gravitational binding energy}
of a neutron star as a 
function of its (inertial) mass, based on equation of state MPA1 \cite{mpa87}. 
The plot clearly shows the prominent position of PSR~J0348+0432. The other 
dots indicate the neutron star masses of the individual test systems in 
figure~\ref{fig:stg_1738}.
\label{fig:eps}}
\end{figure}

Like PSR~1738+0333, PSR~J0348+0432 is a system with a large asymmetry
in the compactness of the components, and therefore well suited for a 
dipolar radiation test. Using equation~(\ref{eq:PbdotD}), the limit 
(\ref{eq:0348GRtest}) can be converted into a limit on additional
gravitational scalar or vector charges:
\begin{equation}\label{eq:0348Dtest}
  |\alpha_p - \alpha_0| < 5 \times 10^{-3} 
  \quad \mbox{(95\% confidence)} \;.
\end{equation}
This limit is certainly weaker than the limit (\ref{eq:0348GRtest}), but
it has a new quality as it tests a gravity regime in neutron stars that has not
been tested before. Gravity tests before \cite{afw+13} were confined to 
``canonical'' neutron star masses of $\sim 1.4\,M_\odot$. PSR~J0348+0432 for the 
first time allows a test of the relativistic motion of a massive neutron star, 
which in terms of gravitational self-energy lies clearly outside the tested 
region (see figure~\ref{fig:eps}).

\begin{figure}[H]
\centering
\includegraphics[height=90mm]{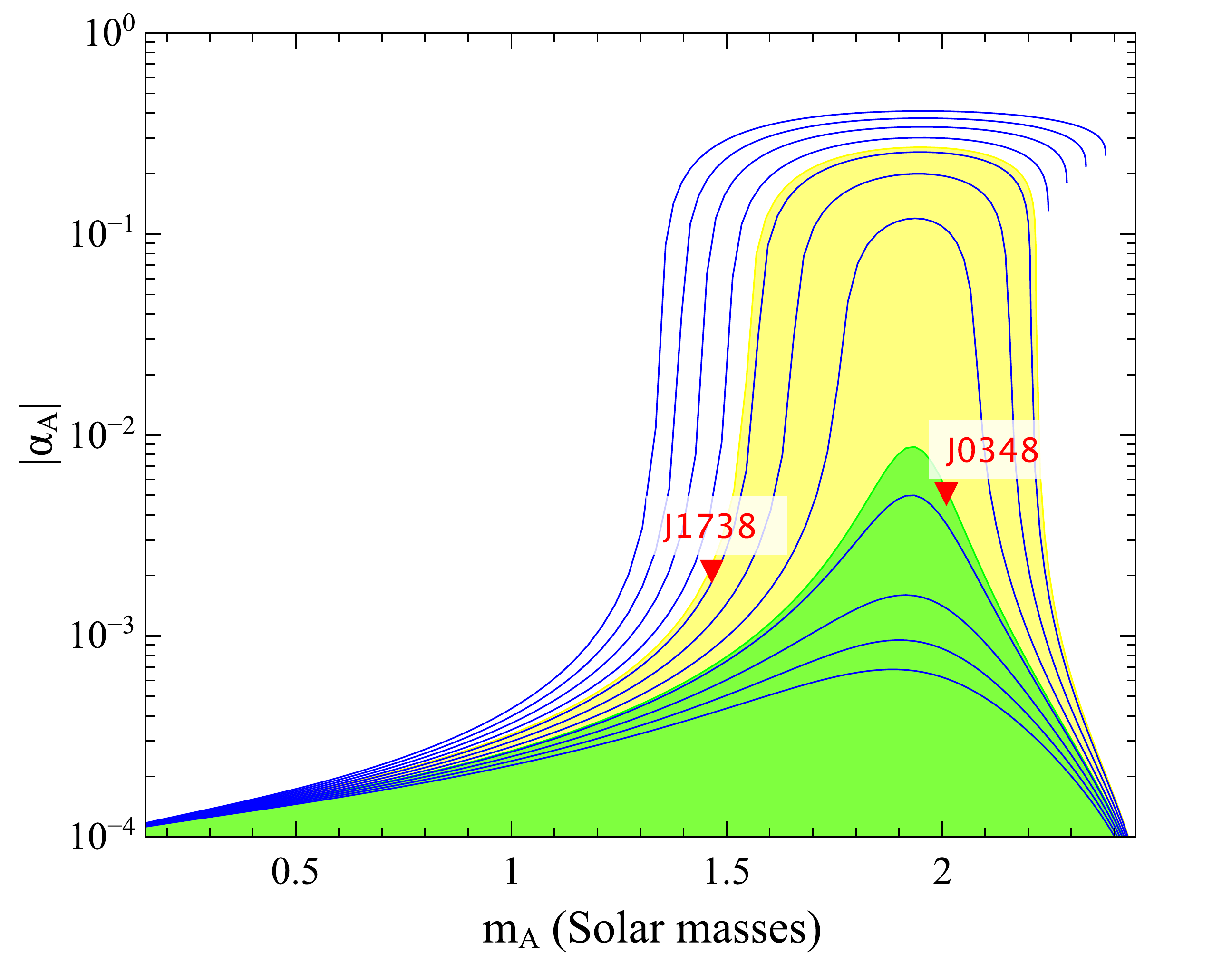}
\caption{Effective scalar coupling as a function of the neutron-star mass, in 
the $T_1(\alpha_0,\beta_0)$ mono-scalar-tensor gravity theory of 
\cite{de93,de96a}. For the linear coupling of matter to the
scalar field we have chosen $\alpha_0 = 10^{-4}$, a value well below the 
sensitivity of any near-future Solar system experiment, like GAIA \cite{hhl+09}. 
The blue curves correspond to stable neutron-star configurations for different 
values of the quadratic coupling $\beta_0$: $-5$ to $-4$ (top to bottom) in 
steps of 0.1. The yellow area indicates the parameter space still allowed by the  
limit~(\ref{eq:1738Dtest}) [label `J1738'], whereas only the green area is in 
agreement with the limit~(\ref{eq:0348Dtest}) [label `J0348']. The plot shows 
clearly how the massive pulsar PSR~J0348+0432 probes deep into a new gravity
regime. Neutron-star calculations are based on equation of state MPA1 
\cite{mpa87} (see \cite{afw+13} for a different equation-of-state). 
\label{fig:0348regime}}
\end{figure}

\begin{figure}[H]
\centering
\includegraphics[height=90mm]{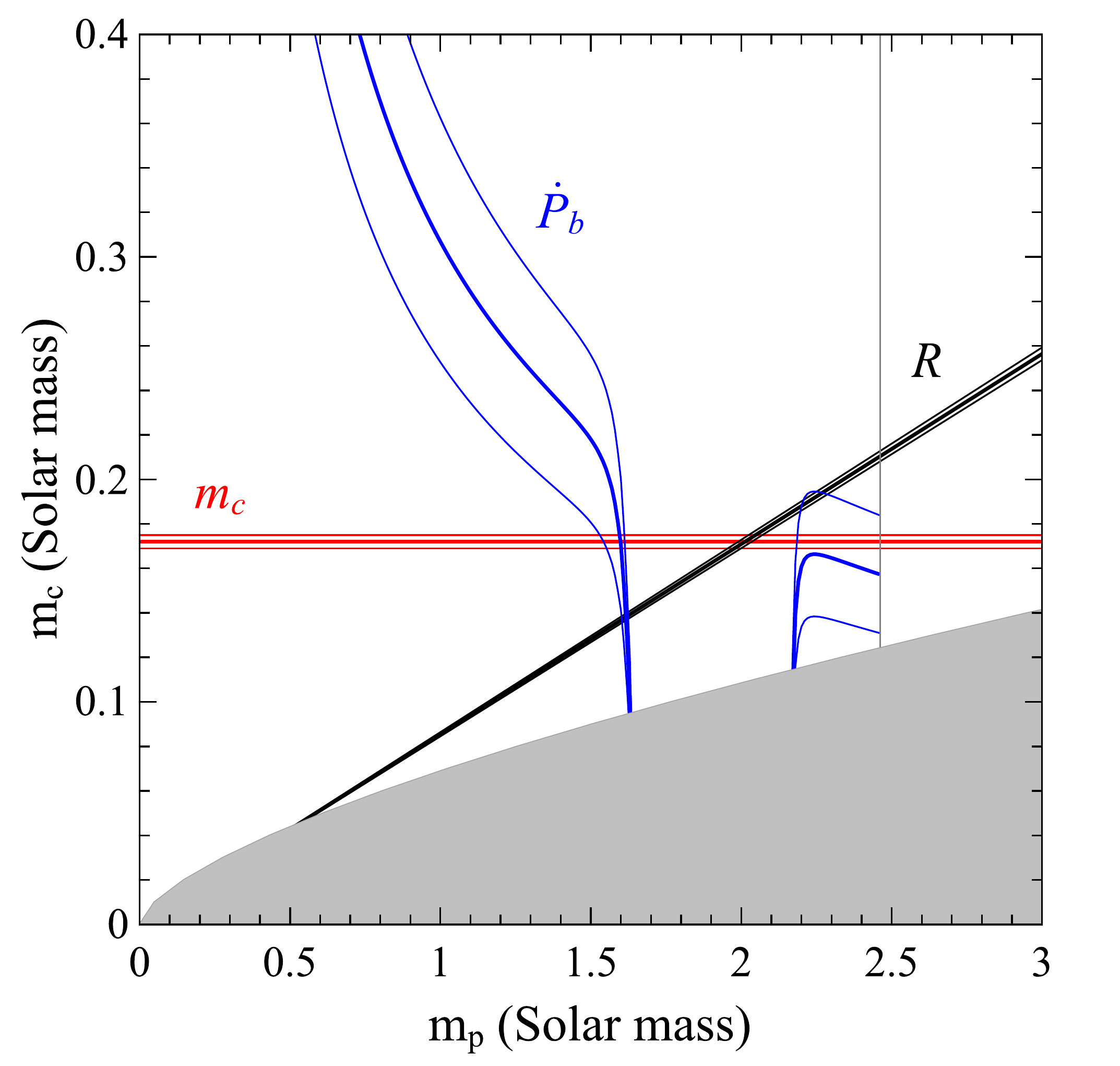}
\caption{Mass-mass diagram based on timing and optical observations of 
the PSR~J0348+0432 system, for the mono-scalar-tensor gravity 
$T_1(10^{-4},-4.5)$. The thin lines indicate the one-sigma errors of the
measured parameters. The vertical grey line is at the maximum mass of a 
neutron star for the given theory and equation-of-state (MPA1). 
The grey area is excluded by the condition $\sin i \le 1$. Obviously \
$T_1(10^{-4},-4.5)$ is clearly falsified by this test, as there is no common 
region for the curves of the three parameters $m_c$, $R$ and $\dot{P}_b$.
\label{fig:mm_0348_stg}}
\end{figure}

Although an increase in fractional binding energy of about 50\% does not 
seem much, in the highly non-linear gravity regime of neutron stars it could
make a significant difference. To demonstrate this, \cite{afw+13} used the 
scalar-tensor gravity $T_1(\alpha_0,\beta_0)$ of \cite{de93,de96a}, which is 
known to behave strongly
non-linear in the gravitational fields of neutron stars, in particular for
$\beta_0 < -4.0$. As shown in figure~\ref{fig:0348regime}, PSR~J0348+0432
excludes a family of scalar-tensor theories that predict significant deviations
from GR in massive neutron stars and were not excluded by previous experiments,
most notably the test done with PSR~J1738+0333. To further illustrate this
in a mass-mass diagram, figure~\ref{fig:mm_0348_stg} shows a gravity theory
with strong-field scalarization in massive neutron stars that passes the
PSR~J1738+0333 experiment, but is falsified by PSR~J0348+0432. 

With PSR~J0348+0432, gravity tests now cover a range of neutron star masses from
1.25\,$M_\odot$ (PSR~J0737$-$3039B) to 2\,$M_\odot$. No significant deviation 
from GR in the orbital motion of these neutron stars was found. These findings 
have interesting implications for the upcoming ground-based gravitational wave 
experiments, as we will briefly discuss in the next subsection.


\subsection{Implications for gravitational wave astronomy}

The first detection of gravitational waves from astrophysical sources by 
ground-based laser interferometers, like LIGO\footnote{www.ligo.org} and  
VIRGO\footnote{www.cascina.virgo.infn.it}, will mark the beginning of a 
new era of gravitational wave astronomy \cite{ss09}. One of the most promising 
sources for these detectors are merging compact binaries, 
\index{merging compact binaries}
consisting of neutron 
stars and black holes, whose orbits are decaying towards a final coalescence due 
to gravitational wave damping. While the signal sweeps in frequency $f$ through 
the detectors' typical sensitive bandwidth $[f_{\rm in}, f_{\rm out}]$ 
from about 20 Hz to a few kHz, the 
gravitational wave signal will be deeply buried in the broadband noise of the 
detectors \cite{ss09}. To detect the signal, one will have to 
apply a matched filtering technique, i.e.~correlate the output of the detector 
with a template wave form. Consequently, it is crucial to know the binary's 
orbital phase with high accuracy for searching and analyzing the signals from 
in-spiraling compact binaries. Typically, one aims to lose less than one 
gravitational wave cycle in a signal with $\sim$ $10^4$ cycles. For this reason, 
within GR such calculations for the phase evolution of compact binaries have 
been conducted with great effort to cover many post-Newtonian orders including 
spin-orbit and spin-spin contributions (see \cite{bla06,bla11} for reviews). 
Table~\ref{tab:pN} illustrates the importance of the individual corrections
to the number of cycles spent in the LIGO/VIRGO band\footnote{The advanced 
LIGO/VIRGO gravitational wave detectors are expected to have a lower end seismic 
noise cut-off at about 10\,Hz \cite{aaa+13}. For a low signal-to-noise ratio the 
low-frequency cut-off is considerably higher. In this review, we adapt a
value of 20\,Hz as the minimum frequency. The maximum frequency of a few kHz
is not important here, since the frequency of the innermost circular orbit
is well below the upper limit of the LIGO/VIRGO band.} 
for two merging non-spinning neutron stars. For a later comparison, the two 
neutron-star masses are chosen to be 2\,$M_\odot$ and 1.25\,$M_\odot$, the 
highest and lowest neutron-star masses observed.

\begin{table}[ht]
\caption{Contributions to the accumulated number of gravitational wave cycles in 
the frequency band of 20\,Hz to 1350\,Hz for a 2\,$M_\odot$/1.25\,$M_\odot$ 
neutron-star pair (cf.~equations~(235),(236) in \cite{bla06}). The frequency of 
1350\,Hz corresponds to the innermost circular orbit of the merging binary 
system \cite{bla06}.  
\label{tab:pN} }
\vspace{1mm}
\centerline{
\begin{tabular}{lcr}
\hline
&  correction to LO & number of cycles \\
\hline
 LO (leading order)      &  --- 	        & 4158.6 \\
 1pN    &  $(v/c)^2$		& 196.4\\
 1.5pN  &  $(v/c)^3$ 	& $-$123.6\\
 2pN    &  $(v/c)^4$		& 7.2\\
 2.5pN  &  $(v/c)^5$ 	& $-$10.3\\
 3pN    &  $(v/c)^6$		& 2.4\\
 3.5pN  &  $(v/c)^7$ 	& $-$0.9\\
\hline
\end{tabular}
}
\end{table}

If the gravitational interaction between two compact masses is different from 
GR, the phase evolution over the last few thousand cycles, which fall into the 
bandwidth of the detectors, deviates from the (GR) template. This will degrade
the ability to accurately determine the parameters of the merging binary,
or in the worst case even prevent the detection of the signal.
In scalar-tensor gravity, for instance, the evolution of the phase is modified 
because the system can now lose additional energy to dipolar waves 
\cite{will94,de98}. Depending on the difference between the effective scalar 
couplings of the two bodies, $\alpha_A$ and $\alpha_B$, the 1.5 post-Newtonian 
dipolar contribution to the equations of motion could drive the gravitational 
wave signal many cycles away from the GR template. For this reason, it is 
desirable that potential 
deviations from GR in the interaction of two compact objects can be tested and 
constrained prior to the start of the advanced gravitational wave detectors. 
With its location at the high end of the measured neutron-star masses, 
PSR~J0348+0432 with its limit~(\ref{eq:0348Dtest}) plays a particularly 
important role in such constraints.

The change in the number of cycles that fall into the frequency band of a 
gravitational wave detector due to a dipolar contribution is given, to leading
order, by \cite{will94,de98} 
\begin{equation}\label{eq:dN}
  \Delta N \approx -\frac{25}{21504 \,\pi}
                   \left(\frac{m_A m_B}{M^2}\right)^{2/5} 
                   (\alpha_A - \alpha_B)^2
                   \left(u_{\rm in}^{-7/3} - u_{\rm out}^{-7/3}\right) \;,
\end{equation}
where $u \equiv \pi (G{\cal M}c^{-3}) f$, and ${\cal M} \equiv (m_A 
m_B)^{3/5}
M^{-1/5}$ is the chirp mass. Equation~(\ref{eq:dN}) is based on the assumption
that $\alpha_0$, $\alpha_A$ and $\alpha_B$ are considerably smaller than unity,
which is supported by binary pulsar experiments. For a 2/1.25\,$M_\odot$
double neutron-star merger, one finds from equation~(\ref{eq:dN}) and
the limit~(\ref{eq:0348Dtest})
\begin{equation}\label{eq:dNlimit}
  |\Delta N (f_{\rm in} = 20\,{\rm Hz}, f_{\rm out} = f_{\rm ICO})|
  < 0.4 \;, 
\end{equation}
where $f_{\rm ICO} \approx 1350\,{\rm Hz}$ is the gravitational wave frequency 
of the innermost circular orbit (cf.~\cite{bla06}). The exact value of 
$f_{\rm ISCO}$ does not play an important role in equation~(\ref{eq:dN}),
since $f_{\rm in} \ll f_{\rm ICO}$.
This result is based on the extreme assumption, that the light neutron star
has an effective scalar coupling which corresponds to the well constrained
weak-field limit, i.e.~$\alpha_B = \alpha_0$. If the companion of the 
2\,$M_\odot$ neutron star is a
10\,$M_\odot$ black hole, then the constraints on $\Delta N$ that can be derived 
from binary pulsar experiments are even tighter (see \cite{afw+13}).
A comparison with table~\ref{tab:pN} shows that the limit~(\ref{eq:dNlimit})
is already below the contribution of the highest order correction calculated.

As explained in \cite{afw+13}, binary pulsar experiment cannot exclude 
significant deviations associated with short-range fields (e.g.~massive
scalar fields), which could still impact the mergers for ground-based 
gravitational wave detectors. Also, there is the possibility of the occurrence 
of effects like dynamical scalarization that, depending on the specifics of the 
theory and the masses, could start to influence the merger at $f < f_{\rm ICO}$ 
\cite{bppl13}, and consequently limit the validity of (\ref{eq:dNlimit}) to a  
smaller frequency band. Nevertheless, the constraints on dipolar
radiation obtained from binary pulsars provide added confidence in the use of
elaborate GR templates to search for the signals of compact merging binaries
in the LIGO/VIRGO data sets.


\section{Geodetic precession}
\label{sec:gp}

\index{geodetic precession}

A few months after the discovery of the Hulse-Taylor pulsar, Damour and Ruffini 
\cite{dr74} proposed a test for geodetic precession in that system. If the pulsar spin is sufficiently tilted with respect to the orbital angular momentum,
the spin direction should gradually change over time (see 
Section~\ref{sec:geodprec}). A change in
the orientation of the spin-axis of the pulsar with respect to the line-of-sight 
should lead to changes in the observed pulse profile. These pulse-profile 
changes 
\index{pulse-profile changes}
manifest themselves in various forms \cite{dr75}, such as changes in the 
amplitude ratio or separation of pulse components \cite{wrt89,kra98}, the 
shape of the characteristic swing of the 
linear polarization \cite{sta04}, or the absolute value of the position angle of 
the polarization in the sky \cite{kw09}. In principle, such changes could allow 
for a measurement of the precession rate and by this yield a test of GR. In 
practice, it turned out to be rather difficult to convert changes in the 
pulse profile into a quantitative test for the precession rate.
Indeed, the Hulse-Taylor pulsar, in spite of prominent profile changes due to 
geodetic precession \cite{wrt89,kra98}, does not (yet) allow for a quantitative 
test of geodetic precession. This is mostly due to uncertainties in the 
orientation of the magnetic axis and the intrinsic beam shape \cite{wt02}.

Profile and polarization changes due to geodetic precession have been observed 
in other binary pulsars as well \cite{mks+10,dkc+12}, but again did not lead to 
a quantitative gravity test.
A complete list of binary pulsars that up to date show signs of geodetic 
precession can be found in \cite{kra12}. Out of the six pulsars listed in
\cite{kra12}, so far only two allowed for quantitative constraints on their 
rate of geodetic precession. These two binary pulsars will be discussed in more 
details in the following.


\subsection{PSR B1534+12}

\index{PSR B1534+12}

PSR~B1534+12 is a 38\,ms pulsar, which was discovered in 1991 \cite{wol91}. It
is a member of an eccentric ($e = 0.27$) double neutron-star system 
with an orbital period of about 10 hours. 
Subsequent timing observations lead to the determination of five post-Keplerian 
parameters: $\dot\omega$, $\gamma$, $\dot{P}_b$, and $r$, $s$ from the 
Shapiro delay \cite{sttw02}. The large uncertainty in the distance to this 
system still prevents its usage in a gravitational wave test, since the observed 
$\dot{P}_b$ has a large Shklovskii contribution, which one cannot 
properly correct for. The other four post-Keplerian parameters are nevertheless 
useful to test quasi-stationary strong-field effects. However, these tests are 
generally less constraining than tests from other pulsars (see e.g.~figure~\ref{fig:stg_1738}).

Continued observations of PSR~B1534+12 with the 305-m Arecibo radio telescope
revealed systematic changes in the the observed pulsar profile by about 1\% per 
year, as well as changes in the polarization properties of the pulsar 
\cite{stta00}. As outlined above, such changes are expected from geodetic 
precession. Using equation~(\ref{eq:OmGP}) and the parameters 
from \cite{sttw02}, one finds that GR predicts a precession rate of
\begin{equation}\label{eq:OmGP1534}
  \Omega^{\rm SO} = 0.51\,{\rm deg/yr}
\end{equation}
for PSR B1534+12.

Besides the secular changes visible in the high signal-to-noise ratio pulse 
profile and polarization data
of PSR~B1534+12, Stairs~{\it et al.}~\cite{sta04} reported the detection of 
special-relativistic aberration of the revolving pulsar beam due to orbital 
motion. Aberration periodically shifts the observed angle between the line of 
sight and spin axis of PSR~B1534+12 by an amount that depends on the orientation 
of the pulsar spin, and therefore contains additional geometrical information.
Combining these observations, Stairs~{\it et al.}~\cite{sta04} were able to 
determine the system geometry, including the misalignment between the spin of 
PSR~B1534+12 and the angular momentum of the binary motion, and constrain the 
rate of geodetic precession to
\begin{equation}
\begin{array}{lcll}
  \Omega^{\rm SO} &=& 0.44_{-0.16}^{+0.48}\,{\rm deg/yr} & 
  \quad \mbox{(68\% confidence)} \;, \\[2mm]
  \Omega^{\rm SO} &=& 0.44_{-0.24}^{+4.6}\,{\rm deg/yr} & 
  \quad \mbox{(95\% confidence)} \;.
\end{array}
\end{equation}
Although the uncertainties are comparably large, these were the first 
beam-model-independent constraints on the geodetic precession rate of a 
binary pulsar. As can be seen, these model-independent constraints on the 
precession rate are consistent with the prediction by GR, as given in equation~(\ref{eq:OmGP1534}).


\subsection{The Double Pulsar}
\label{sec:dpSO}

\begin{figure}[H]
\centering
\includegraphics[height=90mm]{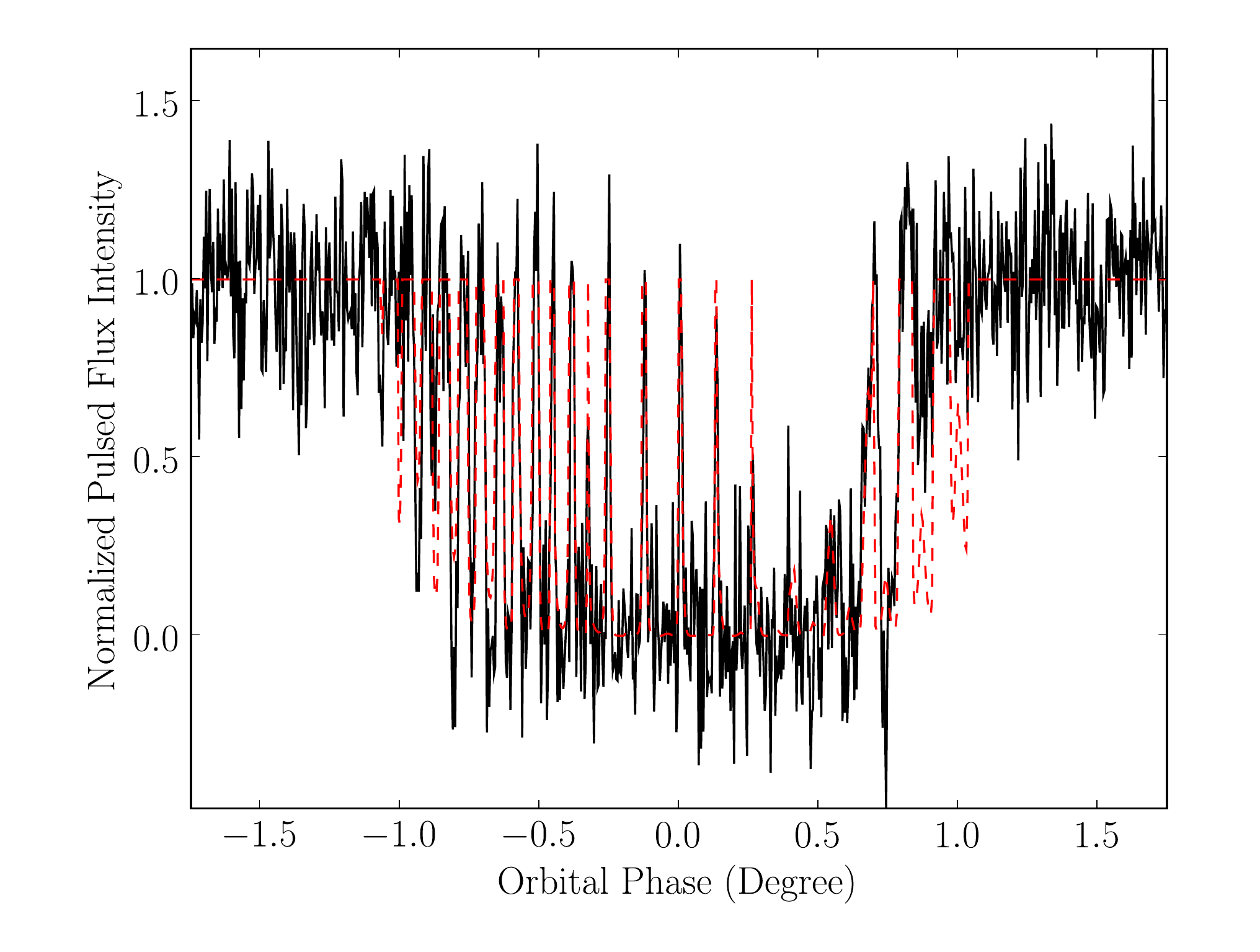}
\caption{
Average eclipse profile of pulsar~$A$ observed at 820 MHz over a 5-day period 
around 11 April 2007 (black line). The model based on a co-rotating 
magnetosphere gives a good explanation of the eclipse profile (red dashed line).
Figure is taken from \cite{bkk+08}. 
\label{fig:eclipseA}}
\end{figure}

\begin{figure}[H]
\centering
\includegraphics[height=90mm]{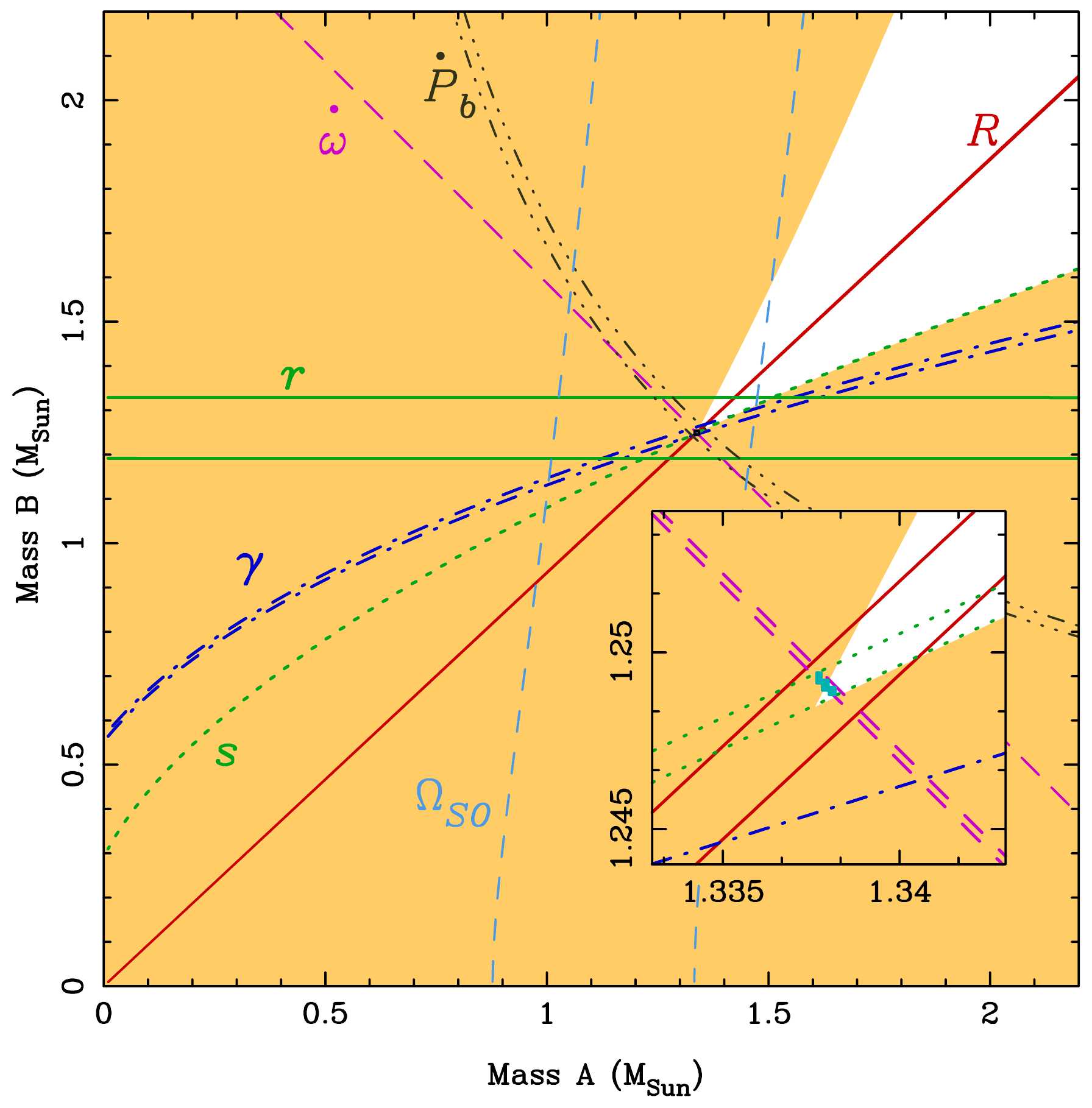}
\caption{GR mass-mass diagram for the Double Pulsar.
Same as in Section~\ref{sec:dp} (figure~\ref{fig:mm_dp}), plus the 
inclusion of the constraints from the geodetic precession of pulsar~$B$ ($
\Omega_{\rm SO}$). Figure is taken from \cite{kw09}.
\label{fig:mm_dp_2}}
\end{figure}

In Section~\ref{sec:dp}, we have seen the Double Pulsar as one of the most 
exciting ``laboratories'' for relativistic gravity, with a wealth of 
relativistic 
effects measured, allowing the determination of 5 post-Keplerian parameters
from timing observations: 
$\dot\omega$, $\gamma$, $\dot{P}_b$, $r$, $s$. Calculating the inclination 
angle of the orbit $i$ from $s = \sin i$, one finds that the line-of-sight is
inclined with respect to the plane of the binary orbit by just about $1.3^\circ$
\cite{ksm+06}. As a consequence, during the superior conjunction the signals of 
pulsar~$A$ pass pulsar~$B$ at a distance of only 20\,000\,km. This is small compared
to the extension of pulsar~$B$'s magnetosphere, which is roughly given by the 
radius of the light-cylinder\footnote{The light-cylinder is defined as the 
surface where the co-rotating frame reaches the speed of light.} 
$r_{\rm lc} \equiv cP/2\pi \sim$ 
130\,000\,km. And indeed, at every superior conjunction pulsar~$A$ gets eclipsed
for about 30 seconds due to absorption by the plasma in the magnetosphere of 
pulsar~$B$ \cite{lbk+04}. A detailed analysis revealed that during every eclipse
the light curve of pulsar~$A$ shows flux modulations that are spaced by half or 
integer numbers of pulsar~$B$'s rotational period \cite{mllp+04} 
(see figure~\ref{fig:eclipseA}). This pattern 
can be understood by absorbing plasma that co-rotates with pulsar~$B$ and is 
confined within the closed field lines of the magnetic dipole of pulsar~$B$. As
such, the orientation of pulsar~$B$'s spin is encoded in the observed light curve 
of pulsar~$A$ \cite{bkk+08}. Over the course of several years, Breton 
{\it et al.}~\cite{bkk+08} observed characteristic shifts in the eclipse 
pattern, that can be directly related to a precession of the spin of pulsar~$B$. 
From this analysis, Breton {\it et al.} were able to derive a precession rate of
\begin{equation}
  \Omega^{\rm SO} = 4.77_{-0.65}^{+0.66}\,{\rm deg/yr} \;.
\end{equation}
The measured rate of precession is consistent with that predicted by GR 
($\Omega_{\rm GR}^{\rm SO} = 5.07\,{\rm deg/yr}$) within its one-sigma uncertainty. This is the sixth(!) post-Keplerian parameter measured in the Double-Pulsar system (see figure~\ref{fig:mm_dp_2}). Furthermore,
for the coupling function $\Gamma_B^A$, which parametrizes strong-field
deviation in alternative gravity theories (see equation~(\ref{eq:OmGPag})), one
finds
\begin{equation}
  \Gamma_B^A/{\cal G} = 1.90 \pm 0.22 \;,
\end{equation}
which agrees with the GR value $\Gamma_B^A/G = 2$.
Although the geodetic precession of a gyroscope was 
confirmed to better than 0.3\% by the Gravity Probe B experiment \cite{edp+11}, 
the clearly less precise test with Double Pulsar~$B$ (13\%) for the first time 
gives a good measurement of this effect for a strongly self-gravitating 
``gyroscope'', and by this represents a qualitatively different test.

The geodetic precession of pulsar~$B$ not only changes the pattern of the
flux modulations observed during the eclipse of pulsar~$A$, it also changes
the orientation of pulsar~$B$'s emission beam with respect to our 
line-of-sight. As a result of this, geodetic precession has by now turned 
pulsar~$B$ in such a way, that since 2009 it is no longer seen by radio 
telescopes on Earth \cite{pmk+10}. From their model, 
Perera {\it et al.}~\cite{pmk+10} predicted 
that the reappearance of pulsar~$A$ is expected to happen around 2035 with the 
same part of the beam, but could be as early as 2014 if one assumes a symmetric 
beam shape.

Finally, for pulsar~$A$ GR predicts a precession rate of 4.78\,deg/yr, which is 
comparable to that of pulsar~$B$. However, since the light-cylinder radius of 
pulsar~$A$ ($\sim 1000\,{\rm km}$) is considerably smaller than that of 
pulsar~$B$,
there are no eclipses that could give insight into the orientation of its spin.
Moreover, long-term pulse profile observations 
indicate that the misalignment between the spin of pulsar~$A$ and the orbital 
angular momentum is less than $3.2^\circ$ (95\% confidence) \cite{fsk+13}. 
For such a close 
alignment, geodetic precession is not expected to cause any significant 
changes in the spin direction (cf.~equations (\ref{eq:OmGP}) and 
(\ref{eq:OmGPag})). This, on the other hand, is good news for tests based 
on timing observations. One does not expect a complication in the analysis of 
the pulse arrival times due to additional modeling of a changing pulse 
profile, like this is, for instance, the case in PSR~J1141$-$6545 \cite{bbv08}.


\section{The strong equivalence principle}
\label{sec:SEP}

\index{strong equivalence principle}

The strong equivalence principle (SEP) extends the weak equivalence principle 
(WEP) to the universality of free fall (UFF) of self-gravitating bodies. In GR,
WEP and SEP are fulfilled, i.e.\ in GR the world line of a body is 
independent of its chemical composition and gravitational binding energy. 
Therefore, a detection of a SEP violation would directly falsify GR. On the 
other hand, alternative theories of gravity generally violate SEP. This is 
also the case for most metric theories of gravity \cite{will93}. For a weakly 
self-gravitating body in a weak external gravitational field one can simply 
express a violation of SEP as a difference between inertial and gravitational 
mass that is proportional to the gravitational binding energy $E_{\rm grav}$ of 
the mass:
\begin{equation}\label{eq:mGmI}
  \frac{m_{\rm G}}{m_{\rm I}} \simeq 
    1 + \eta \, \frac{E_{\rm grav}}{m_{\rm I} c^2} \equiv 
    1 + \eta \, \epsilon \;.
\end{equation}
The Nordtvedt parameter 
\index{Nordtvedt parameter}
$\eta$ is a theory dependent constant. In the 
parameterized post-Newtonian (PPN) framework, $\eta$ is given 
as a combination of different PPN parameters (see \cite{will93} for details).
As a consequence of (\ref{eq:mGmI}), the Earth ($\epsilon \approx -5 \times
10^{-10}$) and the Moon ($\epsilon \approx -2 \times
10^{-11}$) would fall differently in the gravitational field of the Sun 
(Nordtvedt effect \cite{nor68}). The parameter $\eta$ is therefore tightly 
constrained by the lunar-laser-ranging (LLR) experiments to $\eta = (3.0 \pm 
3.6) \times 10^{-4}$, which is in perfect agreement with GR where $\eta = 0$ 
\cite{mhb12}.

In view of the smallness of the self-gravity of Solar system bodies, the LLR 
experiment says nothing about strong-field aspects of SEP. SEP could still be
violated in extremely compact objects, like neutron stars, meaning that a 
neutron star would feel a different acceleration in an external gravitational
field than weakly self-gravitating bodies. For such a strong-field SEP 
violation, the best current limits come from millisecond pulsar-white dwarf 
systems with wide orbits. If there is a violation of UFF by neutron 
stars, then the gravitational field of the Milky Way would polarize the binary 
orbit \cite{ds91}. In comparison with the LLR experiment, such tests have two 
disadvantages: i) the much weaker polarizing external field ($|{\bf g}| \sim 
2 \times 10^{-8}\,{\rm cm}/{\rm s}^2$, as compared to the 
$\sim 0.6\,{\rm cm}/{\rm s}^2$ of the Solar gravitational field at the 
location of the Earth-Moon system),
and ii) the significantly lower precision in the ranging, which is of the order 
of a few $10^3$ cm for the best pulsar experiments ($\sim 1$\,cm for LLR). 
This is almost completely 
counterbalanced by the gravitational binding energy of the neutron star, 
which is a large fraction of its total inertial mass energy ($\epsilon \sim 
-0.1$) and more than eight orders of magnitude larger than that of the Earth. 
This results in experiments with comparable limits on a SEP violation, which 
nonetheless are complementary since they probe different regimes of binding 
energy. The recent discovery of a millisecond pulsar in a hierarchical triple 
(see \cite{lyn12} and Ransom {\it et al.}, in prep.) might allow for a significant improvement in testing SEP, as it combines a strong 
external field ${\bf g}$ with a large fractional binding energy $\epsilon$.

Since beyond the first post-Newtonian approximation there is no general PPN 
formalism available, discussions of gravity tests in this regime are done in 
various theory-specific frameworks. A particularly suitable example for a 
framework that allows a detailed investigation of higher order/strong-field 
deviations from GR, is the above mentioned two-parameter class of 
mono-scalar-tensor theories $T_1(\alpha_0,\beta_0)$ of \cite{de93,de96a}, which 
for certain values of $\beta_0$ exhibit significant strong-field deviations 
from GR, and a correspondingly strong violation of SEP for neutron stars. To 
illustrate 
this violation of SEP, it is sufficient to look at the leading ``Newtonian'' 
terms in the equations of motion of a three body system with masses $m_a$ 
($a = 1,2,3$) \cite{de92}:
\begin{equation}\label{eq:3bodyEOM}
  \ddot{\bf x}_a = -\sum_{b \ne a} {\cal G}_{ab} m_b 
                   \frac{{\bf x}_a - {\bf x}_b}{|{\bf x}_a - {\bf x}_b|^3} \;,
\end{equation}
where the body-dependent effective gravitational constant ${\cal G}_{ab}$ is 
related to the bare gravitational constant $G_\ast$ by
\begin{equation}
  {\cal G}_{ab} = G_\ast (1 + \alpha_a\alpha_b).
\end{equation}
As mentioned above, for a neutron star $\alpha_a$ can significantly deviate from
the weak-field value $\alpha_0 \ll 1$. The structure dependence of the effective 
gravitational constant ${\cal G}_{ab}$ has the consequence that the pulsar does 
not fall in the same way as its companion, in the gravitational field of 
our Galaxy. For a binary pulsar with a non-compact
companion, e.g.~a white dwarf, that effect should be most prominent. 
Since both the white dwarf and the Galaxy are weakly self-gravitating bodies, 
their effective scalar coupling can be approximated by $\alpha_0$, and one finds
from equation (\ref{eq:3bodyEOM})
\begin{equation}
  \ddot{\bf x}_{\rm PSR} - \ddot{\bf x}_{\rm WD} \simeq 
     -G (1 + \delta_{\rm P00}) M
     \frac{{\bf x}_{\rm PSR} - {\bf x}_{\rm WD}}
          {|{\bf x}_{\rm PSR} - {\bf x}_{\rm WD}|^3}
     + \delta_{\rm P00} \, {\bf g} + {\bf a}_{\rm PN}\;,
\end{equation}
where $\delta_{\rm P00} \equiv (\alpha_{\rm PSR} - \alpha_0)\alpha_0$, and
where ${\bf g}$ is the gravitational acceleration caused by the Galaxy at the 
location of the binary pulsar.\footnote{Here we used $G \equiv G^\ast
(1 + \alpha_0^2)$, and we dropped terms of order $\alpha_0^3$ and smaller.} 
Also, the contribution from post-Newtonian dynamics, term ${\bf a}_{\rm PN}$, 
has been added, whose most important consequence is the secular precession of 
periastron, $\dot\omega_{\rm PN}$.
The ${\bf g}$-related term reflects the violation of SEP, which modifies the 
orbital dynamics of binary pulsars. This can be confronted with pulsar 
observations to test for a violation of SEP. In the following we briefly 
discuss different tests of SEP with binary pulsars. For a more complete 
review of the topic of this section see \cite{fkw12}. The discussion below
is not specific to scalar-tensor gravity, and the quantity 
$\delta_{\rm P00}$ can be generically seen as the difference between
inertial and gravitational mass.


\subsection{The Damour-Sch\"afer test}
\label{sec:SEP-DS-test}

\index{Damour-Sch\"afer test}

In 1991, when Damour and Sch\"afer first investigated the orbital 
dynamics of a binary pulsar under the influence of a SEP violation \cite{ds91}, 
only four 
binary pulsars were known in the Galactic disk. Two of these (PSR~B1913+16 and 
PSR~B1957+20) were clearly inadequate for that test, not only because of the 
compactness of their orbits, but also because PSR B1913+16 is member of a double 
neutron star system that lacks the required amount of asymmetry in the binding 
energy, necessary for a stringent test of a SEP violation, and PSR~B1957+20 is 
a so called ``black-widow'' pulsar, where the companion suffers  
significant irregular mass losses, due to the irradiation by the pulsar. The 
remaining systems were PSR~B1855+09 \cite{rt91} and PSR~B1953+29 \cite{bbf83}. 
Both of these systems have wide orbits with small eccentricities, 
$e = 2.2 \times 10^{-5}$ and $e = 3.3 \times 10^{-4}$ respectively.

\begin{figure}
  \centerline{\includegraphics[height=50mm]{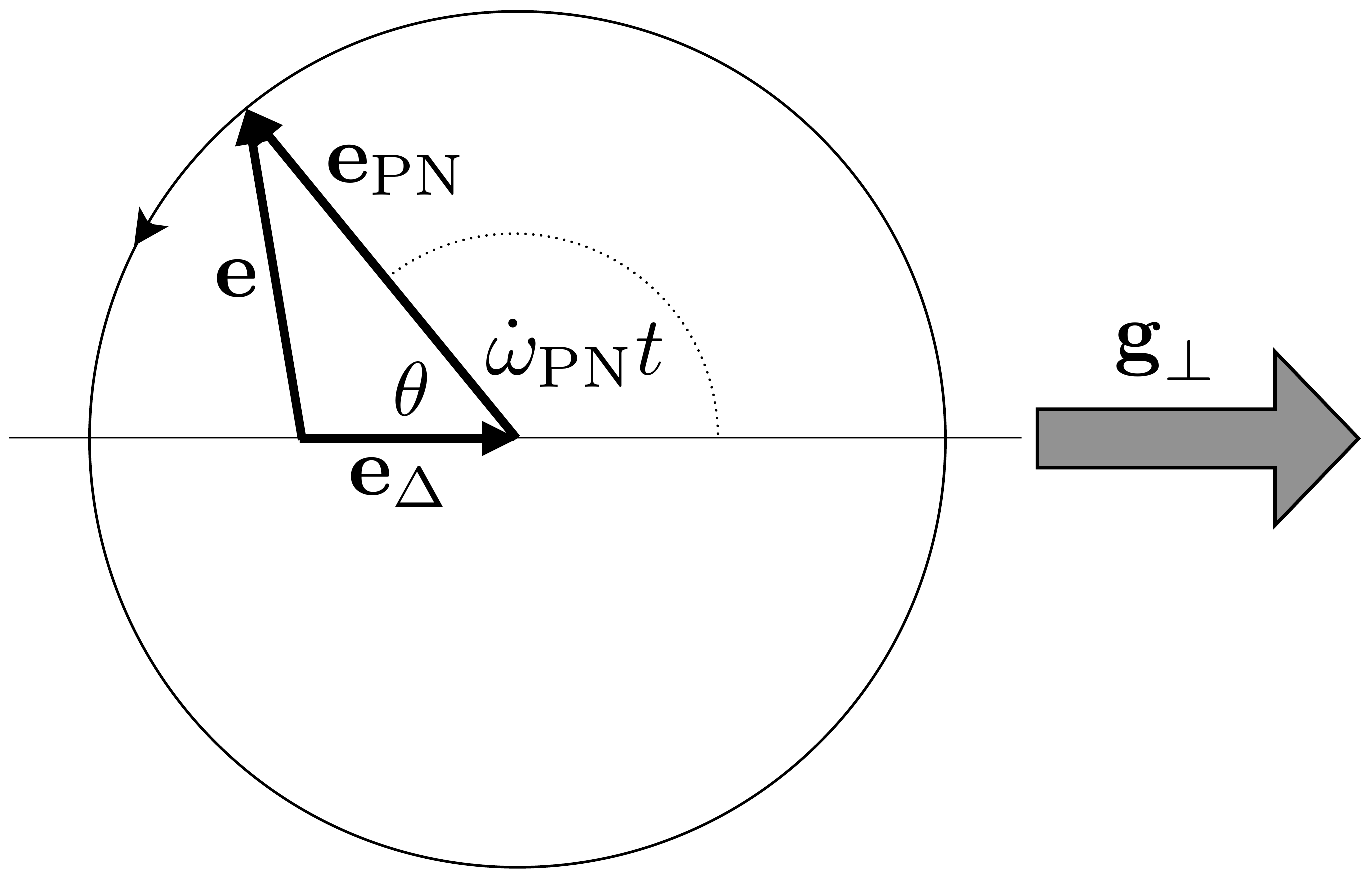}}
  \caption{Time evolution of the orbital eccentricity vector ${\bf{e}} = 
  {\bf{e}}_{\rm PN} + {\bf{e}}_\Delta$ for a small-eccentricity binary,
  in the presence of a SEP violation. The
  vector ${\bf{g}}_{\perp}$ represents the projection of the external
  acceleration in the orbital plane.
  \label{fig:et}}
\end{figure}

Damour and Sch\"afer found for small-eccentricity binary systems
that a violation of SEP leads to a characteristic polarization of the orbit,
which is best represented by a vector addition where the end-point of the 
observed eccentricity vector ${\bf e}(t)$ evolves along a circle in an eccentric 
way (see figure~\ref{fig:et}). The polarizing eccentricity ${\bf e}_\Delta$ is 
proportional to $\delta_{\rm P00}$ and therefore, a limit on $|{\bf e}_\Delta|$
would directly pose a limit on $\delta_{\rm P00}$. Unfortunately neither 
$e_{\rm PN}$ nor $\theta$ in figure~\ref{fig:et} are measurable quantities. 
Also, a direct test for a change in ${\bf e}(t)$ is in many
cases not feasible, as the expected changes are much too small compared to the
available measurement precision from timing (we will discuss exceptions below).
In their test, Damour and Sch\"afer realized if one excludes small $\theta$ 
values with a given probability, one can set an upper limit on 
$|{\bf e}_\Delta|$
without knowing $e_{\rm PN}$. This is the basic idea behind the 
Damour-Sch\"afer test. The angle $\theta$ can be assumed to have a uniform probability 
distribution in the range $[0^\circ, 360^\circ)$ if the following two conditions 
are met:
\begin{description}
\item[DS1)] The system should have a sufficient age, so that one can assume that 
  the relativistic precession of the orbit will have caused the eccentricity 
  vector to have made many turns since the system's birth, thereby effectively 
  randomizing the relative orientation $\theta \equiv \pi - 
  \dot\omega_{\rm PN} t$ (cf.~equation~(\ref{eq:Tomdot}) below).
\item[DS2)] The rate of periastron advance $\dot\omega_{\rm PN}$ should be 
  appreciably larger than the angular velocity of the pulsar's rotation of the 
  Galaxy with 
  which ${\bf g}$ rotates in the reference frame of the binary system. As a 
  result, the projection of the Galactic acceleration vector onto the orbit can 
  be considered constant.
\end{description}
Only if these conditions are met, the Damour-Sch\"afer test can be applied. Both
of the systems considered in \cite{ds91} fulfill these two criteria. Damour 
and Sch\"afer derived 90\% confidence limits of $|\delta_{\rm P00}| < 5.6 
\times 10^{-2}$ and $|\delta_{\rm P00}| < 1.1 \times 10^{-2}$, from 
PSR~B1855+09 and PSR~B1953+29 respectively.

Once the eccentricity of a wide binary pulsar system is measured, there is 
generally little one can do to improve the Damour-Sch\"afer test with that
system. Significant improvement of the Damour-Sch\"afer test has to come from
the discovery of a new system. The larger the orbital period and the smaller the 
orbital eccentricity, the tighter a limit can be derived from a Damour-Sch\"afer
test. In fact, the figure-of-merit for this test is $P_b^2/e$. In the meantime,
quite a few suitable systems have been discovered (see e.g.\ table 4 in 
\cite{gsf+11}). However, one cannot just pick the one with the best 
figure-of-merit from that ensemble, as this introduces a selection 
bias, since it is possible that the small eccentricity of the selected sytem
is actually the result of a SEP violation where by chance $\theta$ is small. In 
fact, if one has a large number of systems, there is a high probability that 
there are systems with small $\theta$ \cite{wex00,fkw12}. In this case, 
one has to properly combine all the systems in a statistical test. The latest 
results based on a proper statistical treatment can be found in 
\cite{sfl+05,gsf+11} which give 95\% confidence limits of order 
$5 \times 10^{-3}$. The slightly better limit in \cite{gsf+11} has a caveat by 
including PSR~J1711$-$4322, which has a large figure-of-merit but neither 
fulfills condition DS1 nor DS2. Concerning DS2, as shown in \cite{kk12}, 
PSR~J1711$-$4322 is at a location in the Galactic plane where 
$\dot\omega_{\rm PN}$ is close, or even equal to the Galactic rotation. This can 
lead to a highly non-uniform evolution of $\theta(t)$.


\subsection{Direct tests}

\index{direct tests of strong equivalence principle}

There is an underlying assumption in the Damour-Sch\"afer test for multiple 
systems, which is related to the mass dependence of a SEP violation. 
Constraining a $\delta_{\rm P00}$ from a set of pulsar-white dwarf systems
in a generic way, requires the assumption that $\delta_{\rm P00}$ is
practically independent of the mass of the neutron star, as these systems have 
different pulsar masses. Even in the absence of a non-perturbative behavior, 
where to first order $\delta_{\rm P00}$ is proportional to $\epsilon$ 
(cf.~equation~(\ref{eq:mGmI})), we can have deviations from that assumption of 
oder 30\% along the range of observed neutron star masses. And in the presence 
of non-perturbative strong-field effects, like the spontaneous scalarization 
mentioned above, this assumption is strongly violated. For this reason it is 
desirable to have direct tests, based on long term timing observations of 
individual systems, used to directly constrain $\dot e$ (see \cite{fkw12}
for details). As it requires a number of conditions to be met, like high timing 
precision and knowledge on the orbital orientation, only few systems turn out to 
be suitable at present. In \cite{fkw12} two binary pulsar systems have been 
identified as 
particularly suitable for a direct test of a SEP violation: PSR~J1713+0747 and 
PSR~J1903+0327. While the work on PSR~J1713+0747 is still in progress, 
preliminary results for PSR~J1903+0327 have been published in \cite{fkw12}. 
PSR~J1903+0327 is a millisecond pulsar with good timing precision in a wide 
($x = 105.593$\,lt-s), highly eccentric ($e = 0.44$) orbit \cite{crl+08,fbw+11}. 
The pulsar is comparably massive ($m_p = 1.67\,M_\odot)$ and the companion is a 
Sun-like main sequence star. At present, the limit from PSR~J1903+0327 
($\sim$ few \%) cannot compete with the results of the Damour-Sch\"afer test 
mentioned above. But these limits are expected to improve with time, just by
continuous timing observations.

In summary, there are several advantages of a direct test \cite{fkw12} compared 
to the Damour-Sch\"afer test:
\begin{itemize}
\item The tests are conducted for specific neutron star masses, and therefore
  are meaningful even in the presence of a non-perturbative strong-field 
  behavior.
\item The test no longer requires probabilistic consideration for unknown 
  angles, and therefore cannot only set an upper limit, but also has the 
  potential to detect a SEP violation.
\item There are hardly any relevant additional effects, that lead to a 
  non-zero $\dot e$. 
  For wide binary orbits, $\dot e$ from gravitational wave damping is 
  absolutely negligible. Therefore the precision of this test is expected to 
  just keep on improving with time, at least for the foreseeable future.
\item We do not need to restrict our sample to systems with small 
  eccentricities. In fact, in an eccentric system the violation of SEP would not 
  only cause a change in the orbital eccentricity ($\dot e$) but also, depending 
  on the orientation, change the inclination of the orbital plane, which leads 
  to a change in the projected semi-major axis ($\dot x$). This allows for a 
  unique cross-check, since the ratio $\dot x/(x\dot e)$ for a SEP violation 
  only depends on the orientation and the eccentricity of the pulsar orbit
  \cite{fkw12}.
\end{itemize}


\section{Local Lorentz invariance of gravity}
\label{sec:LLI}

\index{Local Lorentz invariance}

Some alternative gravity theories allow the Universal matter distribution to 
single out the existence of a preferred frame, which breaks the symmetry of 
local Lorentz invariance (LLI) for the gravitational interaction. In the 
post-Newtonian 
parametrization of semi-conservative gravity theories, LLI violation is 
characterized by two parameters, $\alpha_1$ and $\alpha_2$ \cite{will93}. 
Non-vanishing $\alpha_1$ and $\alpha_2$ modify the dynamics of self-gravitating
systems that move with respect to the preferred frame (preferred-frame effects).
In GR one finds $\alpha_1 = \alpha_2 = 0$.
\index{preferred-frame effects}

As the most natural preferred frame, generally one chooses the frame associated 
with the isotropic cosmic microwave background (CMB), meaning that the preferred 
frame is assumed to be fixed by the global matter distribution of the Universe. 
From the five-year Wilkinson Microwave Anisotropy Probe (WMAP) satellite 
experiment, a CMB dipole measurement with high precision was obtained 
\cite{hwh+09}. The CMB dipole corresponds to a motion of the Solar system with 
respect to the CMB with a velocity of $369.0\pm 0.9\,{\rm km/s}$ in direction of 
Galactic longitude and latitude $(l, b) = 
(263.99^\circ \pm 0.14^\circ, 48.26^\circ \pm 0.03^\circ)$. The numbers quoted 
in the next two sub-sections, will be with respect to the CMB frame. A 
generalization to other frames is straightforward, and was done in some of the 
references cited below.

The most important (weak-field) constraints on preferred-frame effects do come 
form Lunar Laser Ranging (LLR) \cite{mwt08},
\begin{equation}\label{eq:a1LLR}
  \alpha_1 = (-0.7 \pm 1.8) \times 10^{-4} \quad\mbox{(95\% CL)} \;,
\end{equation}
and the alignment of the Sun's spin with the total angular momentum of the
planets in the Solar system \cite{nor87},
\begin{equation}
  |\alpha_2| < 2.4 \times 10^{-7} \;.
\end{equation}


\subsection{Constraints on $\hat{\alpha}_1$ from binary pulsars}

In binary pulsars, the isotropic violation of Lorentz invariance in the 
gravitational sector should lead to characteristic preferred frame effects 
in the binary dynamics, if the barycenter of the binary is moving relative to 
the preferred frame with a velocity ${\bf w}$. For small-eccentricity binaries, 
the effects induced by $\hat\alpha_1$ and $\hat\alpha_2$ (the hat indicates 
possible modifications by strong-field effects) decouple, and can therefore be 
tested independently \cite{de92lli,sw12}. 

In case of a non-vanishing $\hat{\alpha}_1$, the observed eccentricity vector 
${\bf e}$ of a small-eccentricity binary pulsar is a vectorial superposition 
of a `rotating eccentricity' ${\bf e}_R(t)$ and a fixed `forced eccentricity' 
${\bf e}_F$: ${\bf e}(t) = {\bf e}_F + {\bf e}_R(t)$ \cite{de92lli}. 
The rotating eccentricity has a constant length $e_R$, and rotates with the 
relativistic precession of periastron, $\dot\omega$, in the orbital plane. 
This is identical to the dynamics caused by a violation of the strong 
equivalence principle (cf.\ Section~\ref{sec:SEP}), with the forced eccentricity
this time pointing into the direction of $\hat{\bf L} \times {\bf w}$. As a 
consequence, the binary orbit changes from a less to a more eccentric 
configuration and back on a time scale of 
\begin{equation}\label{eq:Tomdot}
  T_{\dot{\omega}} \equiv \frac{2\pi}{\dot\omega} \simeq (1140\,{\rm yr}) 
      \left(\frac{P_b}{1\,{\rm day}}\right)^{5/3} 
      \left(\frac{M}{2 M_\odot}\right)^{-2/3} \;,
\end{equation}
where we have assumed that the true $\dot\omega$ does not deviate significantly 
from the one predicted by GR (equation (\ref{eq:omdotGR})), an assumption that 
is well justified by other binary-pulsar experiments, like the generic tests in 
the Double Pulsar (cf.~Section~\ref{sec:dp}). 

The forced eccentricity ${\bf e}_F$ is determined by the strength of the 
preferred frame effect. Its magnitude is approximately given by
\begin{equation}
  e_F \simeq 0.093 \, \hat{\alpha}_1 \, 
             \frac{m_p - m_c}{M}
             \left(\frac{M}{2 M_\odot}\right)^{-1/3}
             \left(\frac{P_b}{1\,{\rm day}}\right)^{1/3}
             \left(\frac{w \, \sin\psi}{300\,{\rm km/s}}\right) \;,
\end{equation}
where $\psi$ is the angle between ${\bf w}$ and $\hat{\bf L}$
(see \cite{de92lli} for a detailed expression). The observation of 
small eccentricities in binary pulsars, like $e \sim 10^{-7}$ for PSR~J1738+0333 
does not directly constrain $\hat{\alpha}_1$. The orientation of the a priory
unknown intrinsic ${\bf e}_R$ could be such, that it compensates for a large 
${\bf e}_F$. If the system is sufficiently old, one can assume a uniform 
probability distribution in $[0^\circ,360^\circ)$ for $\theta(t)$. Like in the 
Damour-Sch\"afer test for 
SEP, one can now set a probabilistic upper limit on $e_F$, and by this on 
$\hat\alpha_1$, by excluding $\theta$ values close to alignment of ${\bf e}_R$ 
and ${\bf e}_F$. Based on this method, \cite{de92lli} found a limit of 
$|\hat\alpha_1| < 5 \times 10^{-4}$ with 90\% confidence. 

But even if $\theta$ happens to be close to $0^\circ$, due to the relativistic 
precession it will not remain there, and a large $e_F$ cannot remain 
hidden for ever. In fact, if $\dot\omega$ is sufficiently large 
(greater than $\sim 1^\circ$ per year) a significant change in the orbital 
eccentricity should become observable over time scales of a few years, even if 
at the start of the observation there was a complete cancellation between 
${\bf e}_R$ and ${\bf e}_F$. This can be used to constrain $\hat\alpha_1$ 
\cite{sw12}. Hence, in contrast to the SEP test of 
Section~\ref{sec:SEP-DS-test}, one now looks for suitable 
binary pulsars with short orbital periods. The best such test comes from 
PSR~J1738+0333 (see Section~\ref{sec:1738}). This binary pulsar is
ideal for this test for several reasons:
\begin{itemize}
\item The orbit has an extremely small, well constrained eccentricity of 
      $\sim 10^{-7}$ \cite{fwe+12}.
\item The (calculated) relativistic precession of periastron is about 
      $1.6^\circ/$yr, and the binary has been observed by now for about 10 years 
      \cite{fwe+12}. Hence, $\theta(t)$ has covered an angle of $16^\circ$ in 
      that time.
\item The 3D velocity with respect to the Solar system is known with good 
      precision from timing and optical observations, meaning that one can 
      compute ${\bf w}$ \cite{avk+12,fwe+12}.
\item The orientation of the system is such, that the unknown angle of the 
      ascending node $\Omega$ has little influence on the $\hat\alpha_1$ limit, 
      hence there is no need for probabilistic considerations to exclude 
      certain values of $\Omega$ \cite{sw12}.
\end{itemize}
Consequently, PSR~J1738+0333 leads to the best constraints of $\alpha_1$-like
violations of the local Lorentz invariance of gravity, giving \cite{sw12} 
\begin{equation}\label{eq:alpha1limit}
  \hat\alpha_1 = -0.4^{+3.7}_{-3.1} \times 10^{-5} 
  \quad \mbox{(95\% confidence)} \;.
\end{equation}
This limit is not only five times better 
than the current most stringent limit on $\alpha_1$ obtained in the Solar 
system (cf.~equation~(\ref{eq:a1LLR})), it is also sensitive to potential 
deviations related to the strong self-gravity of the pulsar. For 
non-perturbative deviations one can, for illustration purposes, do an expansion 
with respect to the fractional binding energy $\epsilon$ of the neutron star,
\begin{equation}\label{eq:alpha1expansion}
  \hat\alpha_1 = \alpha_1 + {\cal C}_1 \epsilon + {\cal O}(\epsilon^2) \;.
\end{equation}
Since $\epsilon \sim -0.1$ for PSR~J1738+0333, we get tight constraints for 
${\cal C}_1$, a parameter that is virtually unconstraint by the LLR experiment, 
since $\epsilon \sim -5 \times 10^{-10}$ for the Earth. 


\subsection{Constraints on $\hat{\alpha}_2$ from binary and solitary pulsars}
\label{sec:a2limits}

In the presence of a non-vanishing $\hat{\alpha}_2$, a small-eccentricity binary 
system experiences a precession of the orbital angular momentum around the fixed 
direction ${\bf w}$ with an angular frequency
\begin{eqnarray}
  \Omega^{\mathrm{prec}}_{\hat{\alpha}_2} 
    &=& -\hat{\alpha}_2 \frac{\pi}{P_b} \left( \frac{w}{c}\right)^2 \cos\psi 
    \nonumber\\
    &\simeq&  -(0.066^\circ/{\rm yr}) \, \hat{\alpha}_2 
             \left(\frac{P_b}{1\,{\rm day}}\right)^{-1}
             \left(\frac{w}{300\,{\rm km/s}}\right)^2 \cos\psi \,,
             \label{eq:alpha2prec}
\end{eqnarray}
where $\psi$ is the angle between the orbital angular momentum and $\mathbf{w}$
\cite{sw12}. In binary pulsars, such a precession should become visible as a
secular change in the projected semi-major axis of the pulsar orbit, $\dot{x}$, 
which is an observable timing parameter. The two binary pulsars PSRs~J1012+5307 
and J1738+0333 turn out to be particularly useful for such a test, since both of 
them have optically bright white dwarf companions, which allowed the
determination of the masses in the system, and the 3D systemic velocity with
respect to the preferred frame \cite{cgk98,lwj+09,avk+12}.

Unfortunately, in general, the orientation of a binary pulsar orbit with
respect to ${\bf w}$ and the line-of-sight cannot be fully determined from 
timing observations. As a consequence, one cannot directly test  
$\hat\alpha_2$ from observed constraints for $\dot{x}$. In fact, since the 
longitude of the ascending node $\Omega$ is not measured, neither for
PSR~J1012+5307 nor for PSR~J1738+0333, the orientation of these systems could in
principle be such, that an $\hat\alpha_2$-induced precession would not lead
to a significant $\dot{x}$. Assuming a random distribution of $\Omega$ in the
interval $[0^\circ,360^\circ)$, one can use probabilistic considerations
to exclude such unfavorable orientations. A detailed discussion of this test can 
be found in \cite{sw12}, where the following 95\% confidence limits are derived
\begin{eqnarray}
  |\hat{\alpha}_2| &<& 3.6\times10^{-4} \quad \mbox{from
    PSR~J1012+5307,}\nonumber\\
  |\hat{\alpha}_2| &<& 2.9\times10^{-4} \quad \mbox{from
    PSR~J1738+0333,}\\
  |\hat{\alpha}_2| &<& 1.8 \times 10^{-4} \quad \mbox{from
    PSRs~J1012+5307 and J1738+0333 combined.}\nonumber
\end{eqnarray}
It is important to note, that for the last limit, based on the statistical 
combination of the two systems, one has to assume that $\hat{\alpha}_2$ has only 
a weak functional dependence on the neutron-star mass in the range of
$1.3$ -- $2.0\,M_\odot$.

The limit for $\hat\alpha_2$ obtained from binary pulsars are still several 
orders of magnitude weaker than the $\alpha_2$ limit which Nordtvedt derived 
in 1987 from the alignment of the Sun's spin with the orbital planes of the 
planets \cite{nor87}. In the same paper, Nordtvedt pointed out that solitary 
fast-rotating pulsars could be used in a similar way to obtain tight constraints 
for 
$\alpha_2$. This can be directly seen from equation~(\ref{eq:alpha2prec}), which
holds for a rotating self-gravitating star if $P_b$ is replaced by the 
rotational period $P$ of the star. While the five-billion-year
base-line for the Solar experiment is typically a factor of $\sim 10^9$ longer 
than the observational time-span $T_{\rm obs}$ of pulsars, for millisecond 
pulsars $P$ is $\sim 10^9$ shorter than the rotational period of the Sun. In 
fact, the first millisecond pulsar PSR~B1937+21, discovered in 1982 
\cite{bkh+82}, by now has a figure of merit $T_{\rm obs}/P$ that is $\sim 10$ 
times larger than that of the Sun.

\begin{figure}[H]
\centering
\includegraphics[height=90mm]{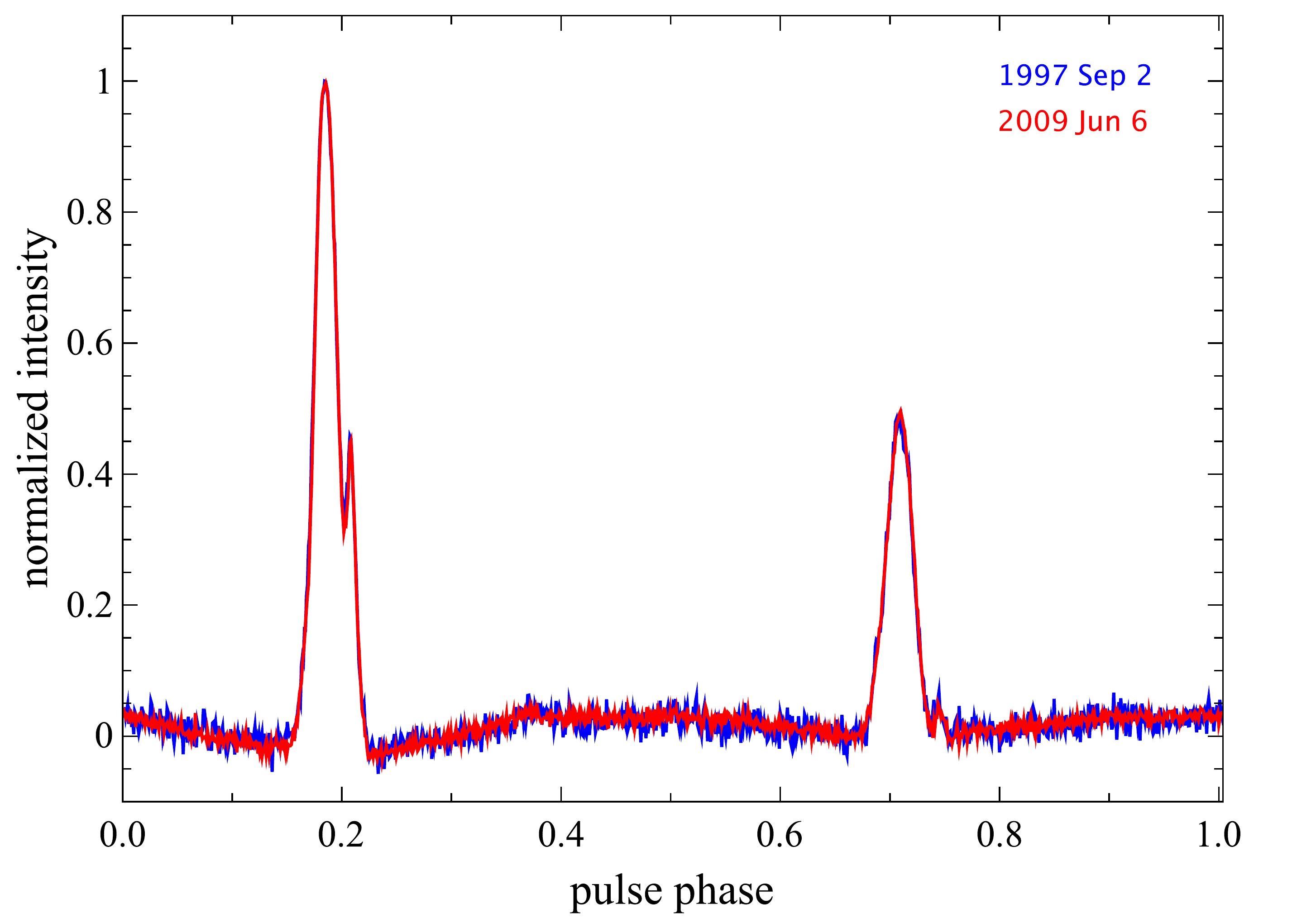}
\caption{Comparison of two pulse profiles of PSR~B1937+21 obtained at two 
different epochs. The blue one was obtained on September 2, 1997, while the red 
one was obtained on June 6, 2009. The main peak is aligned and scaled to have the 
same intensity. There exists no visible difference within the noise level. 
Profiles were taken from \cite{sck+13}.
\label{fig:1937prof}}
\end{figure}

The precession of a solitary pulsar due to a non-vanishing $\hat\alpha_2$ would
lead to characteristic changes in the observed pulse profile over time-scales of
years, just like in the case of binary pulsars that experience geodetic 
precession (cf.\ Section \ref{sec:gp}). Consequently, a non-detection of such 
changes can be converted into constraints for $\hat\alpha_2$. Recently, Shao 
{\it et al.}~\cite{sck+13} used the two solitary millisecond 
pulsars PSRs~B1937+21 and J1744$-$1134 for such an experiment. 
For both pulsars they utilized a consistent set of data, taken over a time span 
of approximately 15 years with the same observing system at the 100-m Effelsberg 
radio telescope. The continuity in the observing system was key for an optimal
comparison of the high signal-to-noise ratio profiles over time. As it turns 
out, both pulsars, PSRs~B1937+21 and J1744$-$1134, do not show any detectable 
profile evolution in the last 15 years. As an example of such a non-detection 
see figure~\ref{fig:1937prof}, which shows two pulse profiles of PSR B1937+21 
obtained at different epochs. 

Similarly to the $\hat\alpha_2$ test with the binary pulsars, there are unknown
angles in the orientation of the pulsar spin, for which certain values have to 
be excluded based on probabilistic considerations. From extensive Monte-Carlo
simulations Shao {\it et al.} found with 95\% confidence 
\begin{eqnarray}
  |\hat{\alpha}_2| &<& 2.5 \times 10^{-8} \quad \mbox{from
    PSR~B1937+21,}\nonumber\\
  |\hat{\alpha}_2| &<& 1.5 \times 10^{-8} \quad \mbox{from
    PSR~J1744$-$1134,}\\
  |\hat{\alpha}_2| &<& 1.6 \times 10^{-9} \quad \mbox{from
    PSRs~B1937+21 and J1744$-$1134 combined.}\nonumber
\end{eqnarray}
These limits are significantly tighter than the $\alpha_2$ limit from the Sun's
spin orientation. Like in the case of the $\hat\alpha_1$ test (previous 
subsection), 
this test also covers potential deviations related to the strong self-gravity of 
the pulsar, and in the combination of the two pulsars, makes the assumption that  
$\hat\alpha_2$ depends only weakly on the neutron-star mass.

An important difference to the aforementioned tests with binary pulsars is, that 
for solitary pulsars one cannot determine the radial velocity. It enters the 
determination of ${\bf w}$ as a free parameter. However, as shown in 
\cite{sck+13}, the unknown radial velocities for PSRs~B1937+21 and J1744$-$1134 
only have a 
marginal effect on the limits. For the limits above it was assumed that both 
pulsars are gravitationally bound in the Galactic potential. But even if one 
relaxes this assumption and allows for unphysically large radial velocities, 
exceeding $1000\,{\rm km/s}$, the limits get weaker by at most 
$\sim 40$\%.

As a final remark, a combined test of $\hat{\alpha}_1$ 
and $\hat{\alpha}_2$ for the gravitational interaction of two strongly 
self-gravitating objects using the Double Pulsar, has been proposed in 
\cite{wk07}. At the time of that publication, however, the observational time 
span was not long enough to disentangle potential preferred-frame effects from 
other orbital contributions. The large correlations with orbital parameters
in the timing solution, lead to rather weak limits on $\hat{\alpha}_1$ 
and $\hat{\alpha}_2$ in the Double Pulsar.


\subsection{Constraints on $\hat{\alpha}_3$ from binary pulsars}
\label{sec:a3limits}

In non-conservative gravity theories, there is a third PPN parameter related to
preferred-frame effects, denoted by $\alpha_3$, which identically vanishes in 
GR. Besides its association with preferred-frame effects, $\alpha_3$ is also 
associated with a violation of the conservation of total momentum, the key
feature of non-conservative gravity theories. Because of this, a non-zero 
$\alpha_3$ leads to a self-acceleration for a self-gravitating rotating body. 
The acceleration is perpendicular to the angular rotation of the body, 
$\mathbf{\Omega}$ and its motion with respect to the preferred frame, 
${\bf w}$. The $\alpha_3$-induced acceleration is given by \cite{will93}
\begin{equation}
  {\bf a}_{\alpha_3} = 
    -\frac{\alpha_3}{3}\,\epsilon\,{\bf w} \times \mathbf{\Omega} \;.
\end{equation}
The quantity $\epsilon$ is the fractional binding energy of the body, as 
defined in equation~(\ref{eq:mGmI}). Due to their high binding energy 
($\epsilon \sim -0.1$) and their fast rotation ($\Omega \sim 10^3$\,rad/s), 
millisecond pulsars are ideal objects to probe for $\alpha_3$ related 
effects. In fact, as shown in \cite{bd96}, millisecond pulsars in binary systems 
with a large orbital period $P_b$ and a small eccentricity $e$ are the best test 
systems for $\hat\alpha_3$, where the hat indicates a strong-field
generalization. As shown in \cite{bd96}, ${\bf a}_{\alpha_3}$ has a polarizing 
effect on the binary orbit, analogous to the one induced by a SEP violation
(cf.~Section~\ref{sec:SEP}). 
Consequently a Damour-Sch\"afer test can be applied to constrain $\alpha_3$.
The same requirements as in the SEP test (cf.~DS1 and DS2 in 
Section~\ref{sec:SEP}) have to be met. DS2 is important, since the binary 
system should not move appreciably in the Galaxy during the build up of the 
polarization induced by ${\bf a}_{\alpha_3}$, otherwise the equations given
in \cite{bd96} do not apply. The latest limit, based on a 
proper statistical analysis of a large sample of known binary pulsars, comes 
from \cite{gsf+11}: $|\hat\alpha_3| < 5.5 \times 10^{-20}$ (95\% confidence). 
PSR~J1711$-$4322, which violates DS1 and DS2, 
has also been included in this analysis. However, due to its slow rotation ($P = 
103$\,ms) it plays only a minor part in that test.  The limit of \cite{gsf+11} 
is more than a factor of $\sim 10^{13}$ better than the best Solar system limit 
\cite{will93}, and by far the tightest limit on any of the PPN parameters. 


\section{Local position invariance of gravity}
\label{sec:LPI}

\index{local position invariance of gravity}

The local position invariance (LPI) of gravity states that the outcome of any 
local gravitational experiment is independent of where and when it is 
performed. If the LPI is violated in the gravitational sector, the gravitational 
interaction of a localized self-gravitating system depends on the direction of 
its acceleration in the gravitational field of an external mass \cite{will93}. 
In such a scenario, the dynamics of the Solar system or a binary system depends 
on the overall matter distribution in our Galaxy, and one would experience a 
directional dependence in the locally measured gravitational constant. Within 
the PPN formalism, such an anisotropy is described by the {\it Whitehead 
parameter} 
\index{Whitehead parameter}
$\xi$ (not to be confused with the parameter $\xi$ in equation~(\ref{eq:LO})) \cite{will93}. It is interesting to note, that even for
fully conservative theories of gravity one may have $\xi \ne 0$. In GR the
gravitational interaction fulfills LPI and therefore GR has $\xi = 0$.

For small-eccentricity binaries, $\xi$ primarily induces a precession of the 
orbital angular momentum around the direction of the external gravitational 
field, ${\bf n}_{\rm G}$, with the angular velocity
\begin{equation}\label{eq:OmegaPrecXi}
  \Omega^{\rm prec} = 
    -\xi \, \frac{2\pi}{P_b} \,\frac{\Phi_{\rm G}}{c^2} \cos\psi \;,
\end{equation}
where $\Phi_{\rm G}$ is the Newtonian Galactic potential at the position of the 
system, and $\psi$ is the angle between the orbital angular momentum and 
${\bf n}_{\rm G}$. Due to the analogy with equation (\ref{eq:alpha2prec}), one 
immediately sees that the same kind of analysis, as outlined in 
Section~\ref{sec:a2limits} for testing $\hat\alpha_2$, can be performed to 
constrain $\hat\xi$. Like in the previous section, the hat indicates the 
strong-field generalization of the PPN parameter. From a combined analysis of 
PSRs~J1012+5307 and J1738+0333 one obtains \cite{swk12}
\begin{equation}
  |\hat\xi| < 3.1 \times 10^{-4} 
  \quad\mbox{(95\% confidence)}\;.
\end{equation}
This limit surpasses the weak-field limit on $\xi$ obtained from the 
non-detection of anomalous Earth tides by about one order of magnitude.

A non-vanishing $\xi$ would also affect an isolated rotating body \cite{nor87}. 
Like for a binary system, the angular momentum, i.e.~the spin, of a 
self-gravitating object with internal equilibrium should precess around 
${\bf n}_{\rm G}$. The precessional frequency is given by 
equation~(\ref{eq:OmegaPrecXi}), if $P_b$ is replaced by the rotational period 
$P$ of the isolated body. Again, one has the analogy to the $\hat\alpha_2$ tests 
of Section~\ref{sec:a2limits}).
Consequently, the same data and method used in the $\hat\alpha_2$ test with
solitary pulsars can be used to constrain $|\hat\xi|$. One obtains with 95\% 
confidence \cite{sw13}:
\begin{eqnarray}
  |\hat{\xi}| &<& 2.2 \times 10^{-8} \quad \mbox{from
    PSR~B1937+21,}\nonumber\\
  |\hat{\xi}| &<& 1.2 \times 10^{-7} \quad \mbox{from
    PSR~J1744$-$1134,} \label{eq:xilimit}\\
  |\hat{\xi}| &<& 3.9 \times 10^{-9} \quad \mbox{from
    PSRs~B1937+21 and J1744$-$1134 combined.}\nonumber
\end{eqnarray}
These limits are significantly (up to three orders of magnitude) better than the 
limit obtained from the Solar spin \cite{nor87,sw13}.

As mentioned above, a violation of the LPI for gravity is directly related to a
directional dependence of the local gravitational constant $G$. Consequently,
the limits (\ref{eq:xilimit}) can straightforwardly be converted into limits on 
an anisotropy of $G$. Corresponding to the combined limit from PSRs~B1937+21 and 
J1744$-$1134 in (\ref{eq:xilimit}), one finds
\begin{equation}\label{eq:deltaG}
  \left| \frac{\Delta G}{G}\right|^{\rm anisotropy} < 
  4 \times 10^{-16} 
  \quad\mbox{(95\% confidence)} \;,
\end{equation}
which is the most constraining limit on the anisotropy of $G$ \cite{sw13}.


\section{A varying gravitational constant}

\index{varying gravitational constant}

The locally measured Newtonian gravitational constant $G$ may vary with time as 
the Universe evolves. In fact, this is expected for most alternatives to GR
that violate the strong equivalence principle \cite{will93}. By now there are 
various tests to constrain a change in the gravitational 
constant on different time scales. Some tests probe a change over the 
cosmological history, i.e.~$G(t)$, others a present change, i.e.~today's 
$\dot{G}$ (see \cite{uza11} for a review). 

In a binary system, a time variation of $G$ changes the orbital period $P_b$. 
If the gravitational binding energy of the masses is small, like for Solar 
system bodies, this change is to first order given by \cite{dgt88}
\begin{equation}\label{eq:PdotGdot}
  \frac{\dot{P}_b}{P_b} = -\frac{\dot{n}_b}{n_b} = -2 \, \frac{\dot{G}}{G} \;,
\end{equation}
and the semi-major axis of the relative motion changes according to
\begin{equation}
  \frac{\dot{a}}{a} = -\frac{\dot{G}}{G} \;.
\end{equation}
In the Solar system, the Lunar Laser Ranging (LLR) experiment gives the best
limit. Based on 39 years of LLR data, \cite{hmb10} derived
a limit of
\begin{equation}\label{eq:GdotLimitLLR}
  \frac{\dot{G}}{G} = (-0.7 \pm 3.8) \times 10^{-13} \, {\rm yr}^{-1} 
                    = (-0.001 \pm 0.005) \, H_0\;.
\end{equation}
The value for the Hubble constant $H_0 = 67.8\,{\rm km/s/Mpc}$ is taken from \cite{PC13}.

\begin{figure}[H]
  \centerline{\includegraphics[height=90mm]{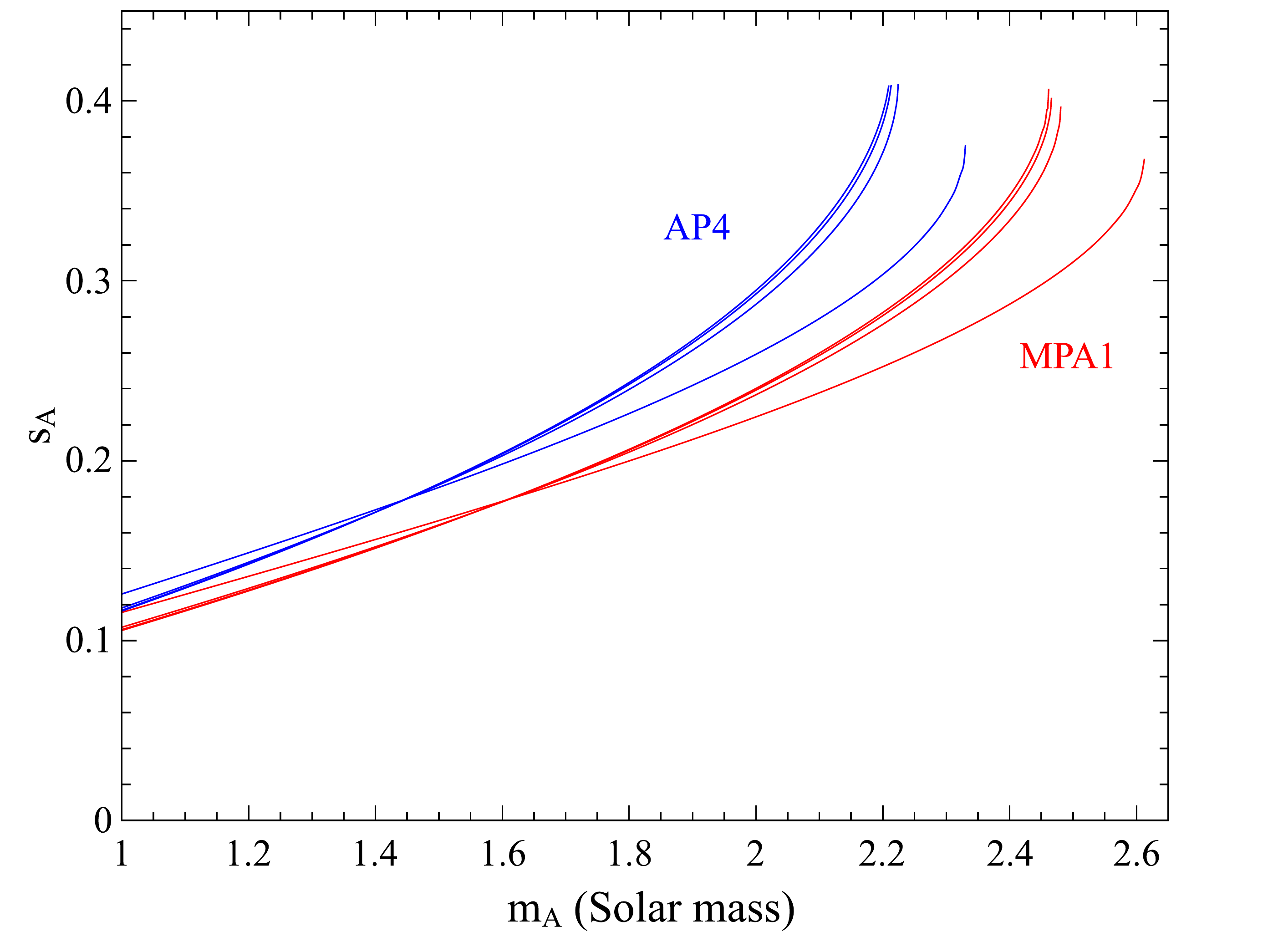}}
  \caption{Sensitivity $s_A$ for Jordan-Fierz-Brans-Dicke theory, i.e.\
  $T_1(\alpha_0,0)$, and for two different
  equations of state (red: MPA1 \cite{mpa87}, blue: AP4 \cite{lp01}).
  For each equation of state four lines have been calculated, corresponding
  to $|\alpha_0| = 0.5,0.2,0.1,0.01$ as the maximum mass decreases.
  For $|\alpha_0| < 0.01$, $s_A$ is practically independent of $\alpha_0$.   
  \label{fig:sJFBD}}
\end{figure}

Equation~(\ref{eq:PdotGdot}) is not applicable to binary pulsars. 
Contrary to weakly self-gravitating bodies, in binary pulsars the dependence on 
the gravitational self-energy cannot be neglected \cite{nor90}. A change in
$G$ changes the gravitational binding energy of a self-gravitating body,
and by this its mass.  While such a change is negligible in the Earth-Moon 
system, since the fractional binding energy is very small for these bodies 
($\epsilon_{\rm Earth} \approx -5 \times 10^{-10}$), it is significant
for neutron stars, where the gravitational self-energy accounts for a 
significant fraction of the mass ($\epsilon_{\rm NS} \sim -0.1$). A detailed
calculation can be found in \cite{nor90}, which shows that 
for a binary pulsar system equation~(\ref{eq:PdotGdot}) has to be modifies to 
\begin{equation}\label{eq:PdotGdotPsr}
  \frac{\dot{P}_b}{P_b} = -2 \, \frac{\dot{G}}{G} 
  \left[1 - \left(1 + \frac{m_c}{2M}\right) s_p 
          - \left(1 + \frac{m_p}{2M}\right) s_c \right] \;,
\end{equation}
where the ``sensitivity'' 
\begin{equation}
  s_A \equiv -\left.\frac{\partial(\ln m_A)}{\partial(\ln G)}\right|_N
\end{equation}
measures how the mass of body $A$ changes with a change of the local 
gravitational constant $G$, for a fixed baryon number $N$ (see \cite{will93}
for details). For a given mass, the sensitivity of a neutron star 
\index{sensitivity of a neutron star}
depends on the 
equation of state and on the specifics of the gravity theory. 
Figure~\ref{fig:sJFBD} shows the sensitivity for Jordan-Fierz-Brans-Dicke 
gravity, for different $\alpha_0$ (i.e.~$\omega_{\rm BD}$) and two different 
equations of state. If the companion of the pulsar is a weakly self-gravitating 
star, like a white dwarf, $s_c$ becomes negligible and 
equation~(\ref{eq:PdotGdotPsr}) simplifies to
\begin{equation}\label{eq:PdotGdotPsrWD}
  \frac{\dot{P}_b}{P_b} \simeq -2 \, \frac{\dot{G}}{G} 
  \left[1 - \left(1 + \frac{m_c}{2M}\right) s_p \right] \;.
\end{equation}

\begin{figure}[H]
\centerline{\includegraphics[height=80mm]{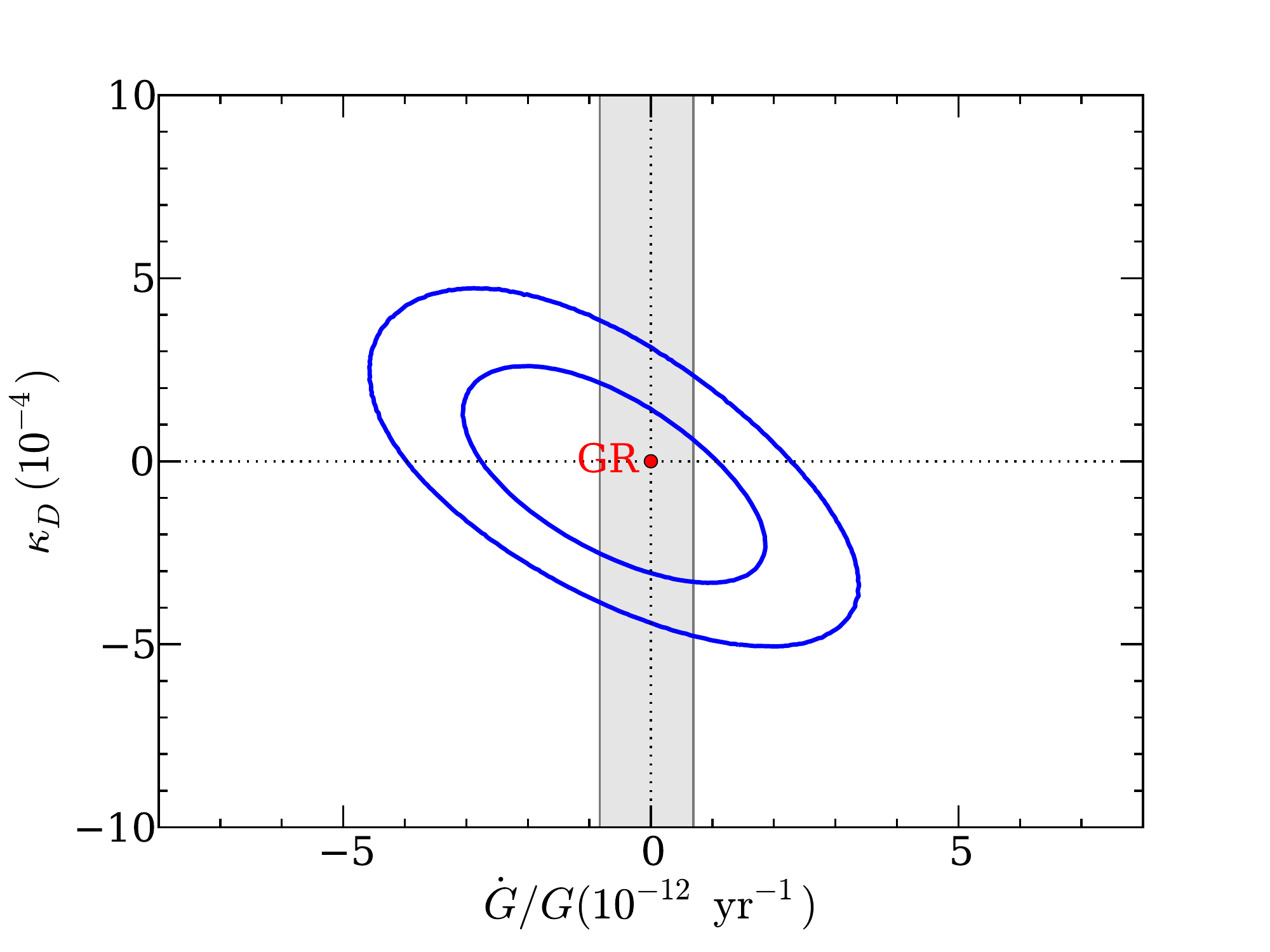}}
\caption{A joint $\dot{G}$-$\kappa_D$ test based on PSRs~J1738+0333 and
  PSR~J0437$-$4715. The inner blue contour includes 68.3\% and the 
  outer contour 95.4\% of all probability. GR ($\dot{G} = \kappa_D = 0$) is 
  well within the inner contour and close to the peak of 
  probability density. The grey band includes regions consistent with the 
  one-sigma constraints for $\dot{G}/G$ from LLR 
  (equation~(\ref{eq:GdotLimitLLR})). Generally only the upper half of the 
  diagram has physical meaning, as the radiation of dipolar gravitational waves 
  is expected to make the system lose orbital energy. Figure is taken from 
  \cite{fwe+12}.
  \label{fig:Gdot_kappaD}}
\end{figure}

The currently best pulsar limit for a change in the gravitational constant
comes from the pulsar-white dwarf system PSR~J0437$-$4715 ($P_b = 5.74\,$d). 
\index{PSR J0437$-$4715} A direct
confrontation of equation~(\ref{eq:PdotGdotPsrWD}) with the timing 
observations of that pulsar yields \cite{dvtb08}
\begin{equation}\label{eq:GdotLimitPsr}
  \frac{\dot{G}}{G} = \frac{(-5 \pm 26) \times 10^{-13}}
                      {1 - 1.1 s_p} \, {\rm yr}^{-1} 
  \qquad \mbox{(95\% confidence)} \;.
\end{equation}
The factor $(1 - 1.1 s_p)$ weakens the limit by typically 30\%, and has been
neglected in \cite{dvtb08}. As pointed out in \cite{lwj+09}, the 
limit~(\ref{eq:GdotLimitPsr}) has the following caveat. It is generally expected 
that a gravity theory with a varying gravitational constant also predicts the
existence of dipolar gravitational waves, that modify $\dot{P}_b$, and could in 
principle even balance a significant part of a decrease in $G$. In fact, just to 
give an example, in Jordan-Fierz-Brans-Dicke theory $\dot{P}_b^{\dot{G}} 
\sim -\dot{P}_b^{\rm dipole}$ for binary pulsar-white dwarf systems
that have orbital periods of $\sim 10$\,d, like PSR~J0437$-$4715. In the absence 
of non-perturbative strong-field effects one finds for the change in the 
orbital period of a pulsar-white dwarf system in a small-eccentricity  
orbit the combined expression (cf.~equation~(\ref{eq:PbdotD}))
\begin{equation}
  \frac{\dot{P}_b - \dot{P}_b^{\rm GR}}{P_b} \simeq -2 \, \frac{\dot{G}}{G} 
  \left[1 - \left(1 + \frac{m_c}{2M}\right) s_p \right] 
  - \frac{4\pi^2}{P_b^2} \, \frac{Gm_pm_c}{c^3M} \, \kappa_D s_p^2 + 
  {\cal O}(s_p^3) \;.
\end{equation}
The constant $\kappa_D$ is a theory dependent constant, which is a priori 
unknown in generic test, where no specific gravity theory is applied. As 
proposed in \cite{lwj+09}, it is now possible to combine two pulsars with a
sufficiently large difference in their orbital periods $P_b$ to constrain 
$\dot{G}$ and $\kappa_D$ simultaneously. In \cite{fwe+12}, the best pulsar
for testing dipolar radiation, PSR~J1738+0333 (see Section~\ref{sec:1738}), 
and the best pulsar for a
$\dot{G}$ test, PSR~J0437$-$4715, have been combined to give joint constraints 
for a variation in $G$ and dipolar radiation (see figure~\ref{fig:Gdot_kappaD}).
In this generic test, one has to make certain reasonable assumptions about 
$s_A$ and how it changes with mass $m_A$, since PSR~J1738+0333 and 
PSR~J0437$-$4715 have different masses (see \cite{fwe+12} for details).
As one can see from figure~\ref{fig:Gdot_kappaD}, the pulsar limit on $\dot{G}$ 
is still somewhat weaker than the one from LLR (\ref{eq:GdotT1_LLR}), but 
obtained with a completely independent method.

\begin{figure}[H]
\centerline{\includegraphics[height=90mm]{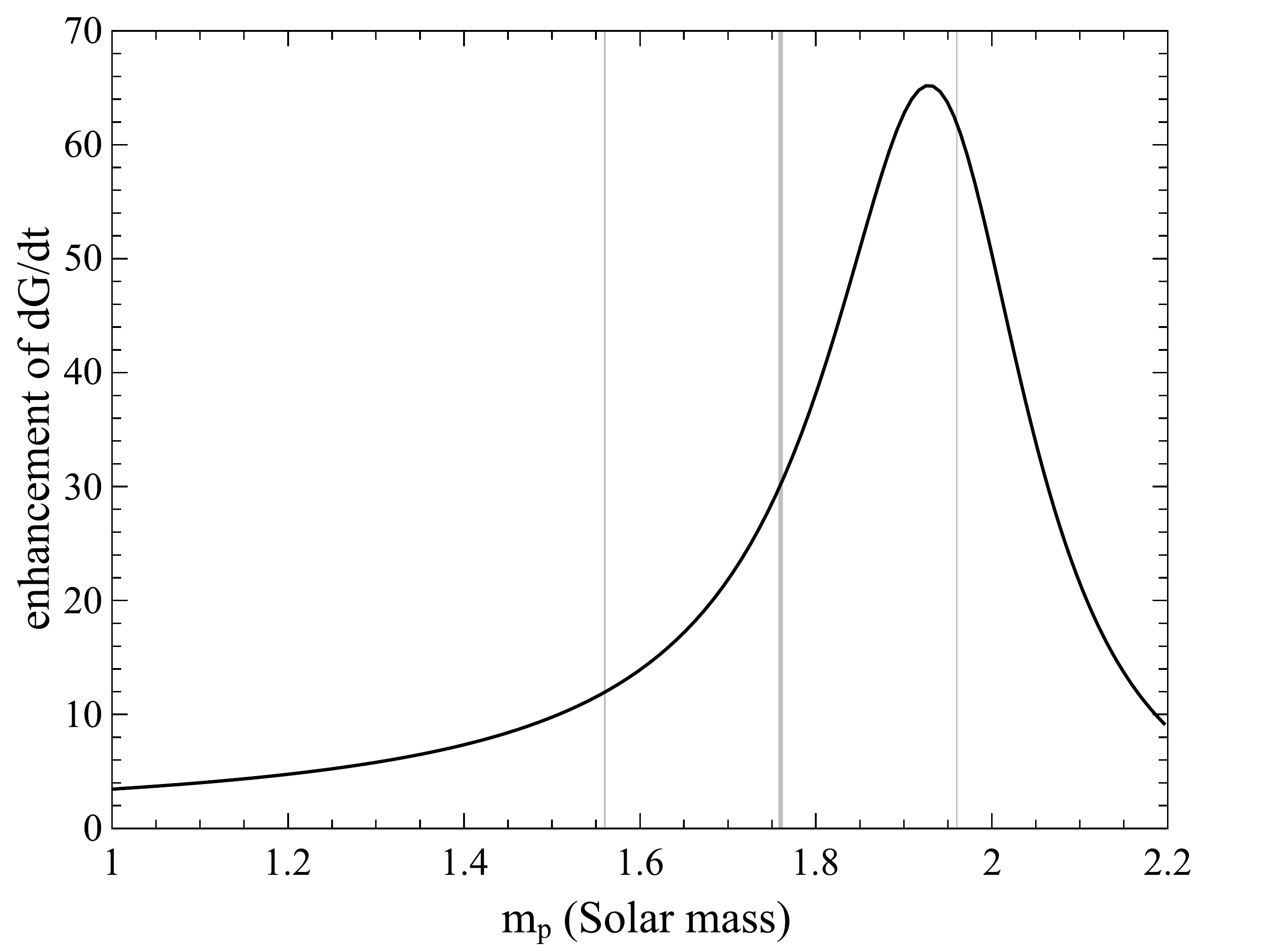}}
\caption{Enhancement of $\dot{G}$ in a pulsar-white dwarf system as a function 
of the pulsar mass $m_p$. Figure shows the ratio $(\dot{\cal G}/{\cal G}) \,/\,
(\dot{G}/G)$ as given in equations~(\ref{eq:GdotT1_LLR}) and 
(\ref{eq:GdotT1_PSR}) for $T_1(10^{-4},-4.3)$ gravity, a theory which still 
passes the PSR~J0348+0432 test (see figure~\ref{fig:0348regime}). The grey 
vertical lines indicate the mass range for PSR~J0437$-$4715 (mean and one-sigma 
uncertainties) \cite{vbv+08}. PSR~J1614$-$2230, with its mass of $1.97 \pm 
0.04\,M_\odot$ and orbital period of 8.7 days \cite{dpr+10}, seems to be a 
promising future test for $\dot{G}$, once tight constraints on the intrinsic 
$\dot{P}_b$ can be derived from the timing observations.
\label{fig:GdotT1}}
\end{figure}

Apart from providing an independent test for a varying gravitational constant,
binary pulsar experiments can test for strong-field enhancements of $\dot{G}$.
To illustrate this, we use scalar-tensor gravity, where a change in the locally 
measured gravitational constant $G$ is the result of a change in the scalar 
field(s). More specifically, in the $T_1(\alpha_0,\beta_0)$ theory of 
\cite{de93,de96a}, LLR tests for a variation of the gravitational
constant that is given by
\begin{equation}\label{eq:GdotT1_LLR}
  \frac{	\dot{G}}{G} = 
    2\left[1 + \frac{\beta_0}{1 + \alpha_0^2}\right]
    \alpha_0\dot{\varphi}_0 
\end{equation}
(see equation~(167) of \cite{uza11}). For the effective gravitational constant between two strongly self-gravitating bodies (as measured in the physical Jordan-frame), equation~(\ref{eq:GdotT1_LLR}) changes to
\begin{equation}\label{eq:GdotT1_PSR}
  \frac{	\dot{\cal G}}{\cal G} = 
    2\left[1 +
    \frac{\alpha_A\beta_B + \alpha_B\beta_A}
         {2\alpha_0(1 + \alpha_A\alpha_B)}\right]
    \alpha_0\dot{\varphi}_0 \;.
\end{equation}
In the presence of significant scalarization effects in the strong gravitational 
fields of neutron stars, the expression in square brackets of 
equation~(\ref{eq:GdotT1_PSR}) can be considerably larger than the corresponding 
one in equation~(\ref{eq:GdotT1_LLR}), even for $\beta_0$ values which are not 
yet excluded by binary pulsar experiments (see figure~\ref{fig:GdotT1}). 
As a conclusion, $\dot{G}$ tests with binary pulsars can be more sensitive than 
LLR tests in situations where a change in the gravitational constant gets 
enhanced by strong-field effects in neutron stars. The details
depend on the specifics of the gravity theory and the mass of the neutron star.
Also, a complete analysis needs to account for corresponding changes in the
masses, i.e.~the analogue to equation~(\ref{eq:PdotGdotPsr}). We will not go 
into these details here.

\section{Summary and Outlook}
\label{sec:summary} 

With their discovery of the first binary pulsar four decades ago, Joseph Taylor 
and Russell Hulse opened a new field of experimental gravity, which has been
an active field of research ever since. Besides the Hulse-Taylor pulsar, which 
led to the first confirmation of the existence of gravitational waves, 
astronomy has seen the discovery of many new binary pulsars suitable for
precision gravity tests. Arguably, the most exiting discovery was the Double 
Pulsar in 2003, which by now provides the best test for GR's quadrupole 
formalism of gravitational wave generation ($< 0.1\%$ uncertainty), and the best 
test for the relativistic spin precession of a strongly self-gravitating body. 
In addition to this, it is the binary pulsar with the most post-Keplerian 
parameters measured, allowing for a number of generic constraints on 
strong-field deviations from GR. For certain aspects of gravity, binary pulsars 
with white dwarf companions have proven to be even better ``test laboratories'' 
than the Double Pulsar. These are gravitational phenomena, predicted by 
alternatives to GR, that depend on the difference in the compactness/binding 
energy of the two components, like gravitational dipolar radiation and a
violation of the strong equivalence principle. By now, pulsar-white dwarf 
systems, like PSR~J1738+0333, set quite stringent limits 
(coupling strength less than about $10^{-3}$)
on the existence of any additional ``gravitational charges'' associated with
light or massless fields. The recent discovery of a massive pulsar in
a relativistic binary system (PSR~J0348+0432), for the first time allowed to 
test the orbital
motion of a neutron star that is significantly more compact than pulsars of
previous gravity tests. For certain aspects of gravity, solitary pulsars 
turned out to be ideal probes. The current best limits on the PPN parameters 
$\alpha_2$, related to the existence of a preferred frame for gravity, and 
$\xi$, related to a violation of local position invariance of the
gravitational interaction, do come from pulse-profile observations of two 
solitary millisecond pulsars. In all these tests, pulsars go beyond Solar
system tests, since they are also sensitive to deviations that occur only 
in the strong-field environment of neutron stars.

So far, GR has passed all these tests with flying colors. Will this continue for 
ever? Is GR our final answer to the macroscopic description of gravity? Pulsar 
astronomy will certainly continue to investigate this question. Many of the 
tests mentioned here will simply improve by continued timing observations of 
the known pulsars. In fact, the measurement precision for some of the 
post-Keplerian parameters increases fast with time. For instance, in regular 
observations (with the same hardware) the uncertainty in the change of the 
orbital period $\dot{P}_b$ decreases with $T_{\rm obs}^{-2.5}$, $T_{\rm obs}$ 
denoting the observing time span. Improvements in the hardware, like new 
broad-band receivers (e.g.~\cite{wbmj09}), will further boost the timing 
precision.
For pulsars like PSR~J1738+0333 and PSR~J0348+0432 soon the modeling of the 
white dwarf will be the limiting factor, while for the Double Pulsar the
corrections of the external contributions to $\dot{P}_b$ will be the challenging 
bit, in particular if one wants to reach the $\sim 10^{-5}$ level at which 
higher oder contributions to $\dot{P}_b$ \cite{bs89,bs90err} and the 
Lense-Thirring contribution to the orbital dynamics \cite{bo75,ds88} become 
relevant (see \cite{kw09} for a detailed discussion). The upcoming next 
generation of radio telescopes, like the Five-hundred-meter Aperture
Spherical radio Telescope (FAST) \cite{nlj+11} and the 
The Square Kilometre Array (SKA) \cite{tay12}, 
\index{Square Kilometre Array (SKA)}
certainly promise a big step towards this goal. With SKA, for many pulsars
one can hope for a factor of 100 improvement in timing precision \cite{sks+09}.
The SKA also promises to provide excellent direct distance measurements to 
pulsars, either directly by utilizing the long baselines of the SKA to form 
high angular resolution images, or by fitting for the timing parallax in the 
arrival times of the pulsar signals \cite{stw+11}. In combination with new
models for the gravitational potential of our Galaxy, in particular after new 
missions like GAIA \cite{pdg+01}, one will be able to accurately determine the 
extrinsic ``contaminations'' of $\dot{P}_b$ via equation (\ref{eq:dPbdotGal}), 
and by this know the intrinsic $\dot{P}_b$. This is key for any high precision 
gravitational wave test with binary pulsars, but also crucial to measure the 
Lense-Thirring drag in the Double Pulsar \cite{kw09}.

Reducing the parameter uncertainties for known pulsars is one way to push 
gravity tests forward, finding new, more relativistic systems is the other.
Presently there are a number of pulsar surveys underway that promise the 
discovery of many new pulsars. New techniques, like acceleration 
searches \cite{rce03} and high performance computing, e.g.~Einstein@Home 
\cite{akc+13}, promise the detection of pulsars in tight orbits, which 
generally cannot be found with traditional methods. There is considerable
hope among pulsar astronomers, that this will finally also lead to the 
discovery of a pulsar-black hole system, 
\index{pulsar-black hole system}
occasionally called the ``holy grail''
of pulsar astronomy. Such a system is expected to provide a superb new probe of 
relativistic gravity and black hole properties, like the dragging of spacetime 
by the rotation of the black hole \cite{wk99,liu12,wle+13}. According to GR, for
an astrophysical black hole (Kerr solution) there is an upper limit for its
spin, given by $S_{\rm max} = GM^2/c$. It would pose an interesting challenge 
to GR, if the timing of a pulsar-black hole system indicates a spin 
$S > S_{\rm max}$. But even for gravity theories that predict the same 
properties for black holes as in GR, a pulsar-black hole system would constitute 
an excellent test system, due to the high grade of asymmetry in the strong-field 
properties of these two components (see \cite{wle+13} for simulations based on
$T_1(\alpha_0,\beta_0)$ scalar-tensor theories). A pulsar in a close orbit 
($P_b < 1$\,yr) around the super-massive black hole  ($m_{\rm BH} \approx 
4 \times 10^6\,M_\odot$) in the center of our Galaxy would be the ultimate test 
system, in that context. According to the mock data analysis in \cite{lwk+12}, 
for such a system a precise 
measurement of the quadrupole moment of the black hole, and therefore a test
of the no-hair theorem, should be possible, provided that the environment of the 
pulsar orbit is sufficiently clean. Finding and timing a pulsar in the 
center of our Galaxy is certainly challenging. A promising result in that
direction is the very recent detection of radio signals from a magnetar 
near the Galactic center black hole \cite{efk+13}, even if this pulsar
is still too far away from the super-massive black hole ($\sim 0.1$\,pc) to 
probe its spacetime. 

Until now, all gravitational wave tests are based on probing the near-zone of a 
binary spacetime by measuring how the back reaction of the gravitational
radiation changes the world lines of the source masses. As outlined above, with 
the Double Pulsar this test has reached a precision of better than 0.1\%. 
Presently there are considerable efforts to achieve a direct detection of 
gravitational waves, i.e.~measure the far-field properties of such 
radiative 
spacetimes by using appropriate test masses. Ground based laser interferometric
gravitational wave observatories, like LIGO and VIRGO, have mirrors
with separations of a few kilometers. Their sensitivity is in the range from 
10\,Hz to few $10^3$\,Hz. Planned space-based detectors, like 
eLISA\footnote{www.elisascience.org/}, will have three 
drag-free satellites as test masses with a typical separation of 
$\sim 10^6$\,km, and should be sensitive to gravitational waves from about 
$10^{-4}$\,Hz to 0.1\,Hz. For the ultra-low frequency band (few nano-Hz)
pulsar timing arrays are currently the most promising detectors \cite{haa+10}.
In these experiments the Earth/Solar system and a collection of very stable
pulsars act as the test masses. A gravitational wave becomes apparent in a
pulsar timing array 
\index{pulsar timing array}
by the changes it causes in the arrival times of the pulsar 
signals. Due to the fitting of the rotational frequency $\nu$ and its 
time derivative $\dot\nu$ for every pulsar, such a detector is only sensitive to 
wavelengths up to $\sim c \, T_{\rm obs}$.\footnote{It has been suggested to use 
the orbital period of binary pulsars to test for gravitational waves of 
considerably longer wavelength \cite{bcr83,kop97}.}
This leads to the special situation that the length of the ``detector arms'' is 
much larger than the wavelength. As a consequence, the observed timing signal 
contains two contributions, the so-called {\it pulsar term}, 
\index{\it pulsar term}
related to the 
impact of the gravitational wave on the pulsar when the radio signal is emitted, 
and the {\it Earth term} 
\index{Earth term}
corresponding to the impact of the gravitational wave 
on the Earth during the arrival of the radio signal at the telescope 
\cite{ew75,det79}. The most promising source in the nano-Hz frequency band is a 
stochastic gravitational wave background, 
\index{stochastic gravitational wave background}
as a result of many mergers of 
super-massive black hole binaries in the past history of the Universe 
\cite{svc08,svv09}. With the large number of ``detector arms'', pulsar timing 
arrays have enough information to explore the properties of the nano-Hz 
gravitational wave background in details, 
once its signal is clearly detected in the data. Are there more than the two 
Einsteinian polarization modes 
\index{polarization modes of gravitational waves}
(alternative metric theories can have up to six)? Is the 
propagation speed of nano-Hz gravitational waves frequency depended? Does the 
graviton carry mass? These are some of the main questions that can be addressed 
with pulsar timing arrays \cite{ljp08,ljp+10}. The isolation of a single source
in the pulsar timing array data would give us a unique opportunity to study
the merger evolution of a super-massive black hole binary, since the signal in
the Earth term and the signal in the pulsar term show two different states of 
the system, which are typically several thousand years apart \cite{jllw04}. For 
these kind of gravity experiments, however, we might have to wait till the 
full SKA has collected a few years of data, which probably brings us close to 
the year 2030. 


\bibliographystyle{numbers}

\end{document}